\documentclass[11pt,a4paper,openany,oneside]{memoir}

\usepackage[OT1]{fontenc}

\usepackage[english]{babel}

\usepackage[utf8]{inputenc}

\usepackage[sc]{mathpazo}

\usepackage{amsmath,amssymb,amsfonts,mathrsfs}

\usepackage[amsmath,thmmarks]{ntheorem}

\usepackage{graphicx}

\usepackage{soul}

\usepackage{pdfpages}

\PassOptionsToPackage{hyphens}{url}
\usepackage{xurl}
\usepackage[linkcolor=black,colorlinks=true,citecolor=black,urlcolor=black,filecolor=black]{hyperref}

\usepackage[capitalise]{cleveref}

\usepackage[german=swiss]{csquotes}

\usepackage{datetime}

\usepackage{mathtools}

\usepackage[h]{esvect}

\usepackage{array}

\usepackage{listings}
\usepackage[ruled,vlined]{algorithm2e}
\usepackage[activate]{microtype}

\usepackage{booktabs}

\usepackage{siunitx}
\DeclareSIUnit\century{century}
\DeclareSIUnit\year{year}
\DeclareSIUnit\flop{flop}
\DeclareSIUnit\bar{bar}
\usepackage{xcolor}

\usepackage{rotating}

\usepackage{placeins} 

\usepackage{listings}
\usepackage{xcolor} 
\usepackage{mdframed} 
\definecolor{light-gray}{gray}{0.95} 

\definecolor{mygreen}{rgb}{0,0.6,0}
\definecolor{mygray}{rgb}{0.96,0.96,0.96}
\definecolor{mymauve}{rgb}{0.58,0,0.82}

\usepackage{inconsolata}

\nonzeroparskip
\defaultlists

\makeatletter

\if@twoside
  \copypagestyle{chapter}{Ruled}
\else
  \copypagestyle{chapter}{ruled}
\fi
\makeoddhead{chapter}{}{}{}
\makeevenhead{chapter}{}{}{}
\makeheadrule{chapter}{\textwidth}{0pt}
\copypagestyle{abstract}{empty}

\makechapterstyle{bianchimod}{%
  \chapterstyle{default}
  \renewcommand*{\chapnamefont}{\normalfont\Large\sffamily}
  
  \renewcommand*{\printchaptername}{%
    \chapnamefont\centering\@chapapp}

  }

\chapterstyle{bianchimod}

\setsecheadstyle{\Large\bfseries\sffamily}
\setsubsecheadstyle{\large\bfseries\sffamily}
\setsubsubsecheadstyle{\bfseries\sffamily}
\setparaheadstyle{\normalsize\bfseries\sffamily}
\setsubparaheadstyle{\normalsize\itshape\sffamily}
\setsubparaindent{0pt}

\captionnamefont{\sffamily\bfseries\footnotesize}
\captiontitlefont{\sffamily\footnotesize}
\setsecnumdepth{subsection}
\settocdepth{subsection}

\pretitle{\vspace{0pt plus 0.7fill}\begin{center}\HUGE\sffamily\bfseries}
\posttitle{\end{center}\par}
\preauthor{\par\begin{center}\let\and\\\Large\sffamily}
\postauthor{\end{center}}
\predate{\par\begin{center}\Large\sffamily}
\postdate{\end{center}}

\def\@advisors{}
\newcommand{\advisors}[1]{\def\@advisors{#1}}
\def\@department{}
\newcommand{\department}[1]{\def\@department{#1}}
\def\@thesistype{}
\newcommand{\thesistype}[1]{\def\@thesistype{#1}}

\renewcommand{\maketitlehookb}{\vspace{1in}%
  \par\begin{center}\Large\sffamily\@thesistype\end{center}}

\renewcommand{\maketitlehookd}{%
  \vfill\par
  \begin{flushright}
    \sffamily
    \@advisors\par
    \@department, ETH Z\"urich
  \end{flushright}
}

\checkandfixthelayout

\makeatother

\theoremstyle{plain}
\numberwithin{equation}{chapter}

\theoremstyle{nonumberplain}
\theorembodyfont{\normalfont}
\theoremsymbol{\ensuremath{\square}}

\renewcommand{\epsilon}{\ensuremath\varepsilon}

\renewcommand{\phi}{\ensuremath{\varphi}}

\title{Pyroclast: A Modular High-Performance Python Solver for Geodynamics}
\author{Marcel Ferrari}
\thesistype{}
\advisors{}
\department{Geophysical Fluid Dynamics Group\\Department of Earth and Planetary Sciences D-EAPS}
\date{}

\begin{document}

\frontmatter

\begin{titlingpage}
  \calccentering{\unitlength}
  \begin{adjustwidth*}{\unitlength-24pt}{-\unitlength-24pt}
    \maketitle
  \end{adjustwidth*}
\end{titlingpage}

\begin{abstract}
This monograph presents the design, implementation, and evaluation of Pyroclast, a modular, high-performance Python framework for large-scale geodynamic simulations. Pyroclast addresses critical limitations of legacy geodynamics solvers, often implemented in monolithic Fortran, C++ or C codebases with limited GPU support and extensibility, by combining modern numerical methods, hardware-accelerated execution, and a flexible, object-oriented software architecture. Designed from the ground up for distributed and GPU-accelerated environments, Pyroclast provides an accessible yet efficient platform for simulating mantle convection and lithospheric deformation using the marker-in-cell method and a matrix-free finite difference discretization.

The work focuses on implementing a robust and scalable two-dimensional viscous mechanical solver that forms the computational core for future visco-elasto-plastic models. This core includes a stress-conservative staggered grid discretization of the incompressible Stokes equations, a matrix-free geometric multigrid solver, support for Krylov and quasi-Newton solvers, and MPI-based domain decomposition. All solver components are designed for GPU-accelerated execution

Preliminary benchmarks were conducted to evaluate performance and scalability. Shared-memory benchmarks show strong scaling of the Stokes solver and demonstrate a 5–10$\times$ speedup on NVIDIA A100 GPUs compared to a multi-core CPU baseline. Distributed advection benchmarks validate the scalability of the domain decomposition strategy, showing near-ideal weak scaling up to 896 CPU cores across 7 compute nodes. The results confirm that Pyroclast achieves high performance while remaining flexible and much more accessible to non-expert users through the high-level Python interface.

This work contributes not only a functional and extensible simulation framework but also a blueprint for modernizing legacy geodynamics codes. Its modular architecture and Python-native implementation lower the barrier to entry for researchers while enabling direct interoperability with modern machine learning libraries. This integration opens the door to hybrid workflows that combine physics-based simulation with data-driven modeling, accelerating scientific discovery in geodynamics and beyond.
\end{abstract}

\cleartorecto
\tableofcontents
\mainmatter

\chapter{Introduction}
\label{chap:introduction}
Understanding the dynamics of the Earth's interior is a fundamental challenge in Earth science, with implications for both scientific research and societal applications. Geodynamic processes such as mantle convection, lithospheric deformation, and plate tectonics shape the structure and evolution of our planet over geological time scales. Numerical modeling has become an indispensable tool to explore these processes, enabling scientists to investigate complex, multi-scale interactions that are otherwise inaccessible to direct observation~\cite{gerya_intro_book}. This work presents the design, implementation, and evaluation of \textit{Pyroclast}, a high performance, modular, and extensible solver for large scale geodynamic simulations, built in the Python programming language. The work bridges geophysical modeling, numerical methods, modern high performance computing techniques, and scientific software engineering.

\section{Why Geodynamics Matters}
Geodynamics is the study of the physical processes that govern the structure, evolution, and dynamics of the Earth's interior and surface. It investigates a wide range of geological phenomena, including the movement of tectonic plates, the formation and deformation of the lithosphere, mantle convection and subduction zones. Geodynamic processes are responsible not only for deep Earth dynamics but also for surface phenomena such as plate tectonics, the evolution of the Earth's topography, earthquakes, and volcanic activity~\cite{turcotte_plate_tectonics, schubert_book}. Geodynamics also contributes to our understanding of the Earth's thermal evolution and how it affects long-term deformation patterns in the lithosphere~\mbox{\cite{turcotte_heat, gerya_heat, turcotte_rheology, gerya_rheology}}. As a result, it plays a central role in Earth sciences, linking geophysics, geology, geochemistry, and planetary science. Understanding these processes is not only of scientific interest but also has direct implications for assessing natural hazards, managing subsurface resources, and reconstructing the Earth's geological history.

\section{Numerical Geodynamic Modeling}
Despite decades of geophysical and geological research, our ability to observe the Earth's interior remains fundamentally limited in both space and time. Direct measurements are largely confined to the surface and uppermost tens of kilometers, while the Earth's interior extends to depths of over 6000 \si{\km}. Similarly, the geological record only preserves sparse information from the last few billion years. As a result, the majority of the Earth's history and deep interior structure are impossible to observe directly~\cite{gerya_intro_book}. This limitation is even more evident when studying other planetary bodies.

Numerical modeling aims to bridge the gap caused by the absence of direct observational data by providing a physically grounded framework to simulate Earth's interior and surface dynamics. Instead of relying on limited measurements, it allows us to describe the geodynamic processes using mathematical tools, particularly partial differential equations that express principles like conservation of mass, momentum, and energy in a deforming medium. By solving these fundamental equations under realistic boundary and material conditions, it is possible to reconstruct the physical mechanisms responsible for large-scale geological phenomena such as mantle convection, plate tectonics, and lithospheric deformation. Particularly, this approach enables researchers to explore the nonlinear and emergent behaviors of a geological system in a controlled and quantitative manner~\cite{gerya_intro_book}.

In this context, the study of geodynamics has become deeply connected with numerical modeling. The increasing complexity of models and the growing availability of computational resources continue to expand the scope of what can be simulated. As a result, modeling has become a foundational component of hypothesis discovery and quantitative study of the Earth's processes, as well as other geological systems~\cite{gerya_intro_book}.

\section{Challenges in Geodynamic Simulations}
Geodynamic simulations face a number of fundamental challenges that stem from both the complexity of the underlying physical processes and the computational demands required to model them at scale.

Firstly, geodynamic systems involve multiple coupled physical processes that interact in strongly nonlinear ways. These include slow, long-term deformation such as mantle convection, as well as faster processes like heat diffusion and seismic events. Accurately capturing these interactions requires advanced multiphysics solvers capable of handling phenomena that span vastly different spatial and temporal scales. Resolving all these scales in a single simulation loop presents significant numerical and algorithmic difficulties~\cite{gerya_seismo_modeling, gerya_hydro_modeling}.

Secondly, the mathematical formulation of geodynamic problems often leads to extremely ill-conditioned systems of equations. This is for example the case when modeling rocks as a continuum. Given the extreme viscosity values, the resulting physical quantities vary by several orders of magnitude. For example, tectonic plate velocities typically lie between 2 and 8~\unit{\centi\metre\per\year}, or approximately \(6 \times 10^{-10}\) to \(2.5 \times 10^{-9}\)~\unit{\metre\per\second}, while internal deviatoric stresses usually range from \(200\) to \(300~\unit{\bar}\) (\(2 \times 10^7\) to \(3 \times 10^7\)~\unit{\newton\per\metre\squared}) and can exceed 1~\unit{\kilo\bar} (\(10^8\)~\unit{\newton\per\metre\squared}) in regions of intense deformation such as volcanic hotspots. The resulting linear systems may have condition numbers exceeding \(10^{25}\), posing a challenge even for state-of-the-art numerical solvers~\cite{ferrari_sparse_solvers}.

Thirdly, sufficient spatial and temporal resolution is essential to accurately capture the dynamics of geodynamic systems. Adaptive mesh refinement strategies, which increase resolution only in regions of interest, have been successfully used to reduce computational cost~\cite{gerya_amr}. However, high-fidelity simulations remain computationally expensive. To make large-scale modeling feasible, it is critical to optimize numerical solvers and take full advantage of modern HPC architectures, including multicore CPUs and GPUs. Exploiting efficient algorithms and hardware acceleration is therefore necessary to enable scientific progress in computational geodynamics.

Finally, there remains a substantial gap between the expertise of geoscientists and the technical engineering knowledge required to implement and run high-performance numerical codes. Many researchers in geology and geophysics lack formal training in numerical methods, parallel computing, or performance optimization. This limits the accessibility of advanced modeling tools and the broader adoption of modern software and hardware capabilities within the geodynamics community.

\section{Motivation for Pyroclast}
While geodynamic modeling has made significant progress over the past decades, many widely used simulation tools remain rooted in legacy codebases written in low-level languages such as Fortran, C, or C++. These monolithic, CPU-centric codes have been essential for advancing the field, but they are often difficult to extend, maintain, or adapt to new computational paradigms. Incorporating new physics, leveraging GPU acceleration, or integrating with modern machine learning workflows is typically difficult or infeasible within these frameworks.

At the same time, the broader Earth system science community has begun to embrace modern, modular approaches to simulation software. In particular, the climate and weather modeling fields have pioneered efforts to transition from traditional high-performance computing (HPC) workflows to Python-based frameworks that prioritize both performance and developer productivity. Projects such as the porting of the ICON weather model to GPU-native Python tools illustrate the potential of this shift~\cite{icon4py}. These initiatives take advantage of the growing ecosystem of AI-optimized software and hardware, which is increasingly built around Python.

Although recent initiatives such as GPU4GEO~\cite{gpu4geo} have explored the use of Julia for GPU-accelerated geophysical modeling. Existing Julia-based work has primarily focused on pseudo-transient solvers and the advection component of the marker-in-cell method~\cite{gpu4geo_software}. However, to the best of our knowledge, no comparable effort has been directed toward Python yet. This distinction is important, as Python occupies a unique position at the intersection of high-performance computing and machine learning. The vast majority of modern AI software infrastructure, including frameworks such as PyTorch~\cite{pytorch} and TensorFlow~\cite{tensorflow}, is written for and deeply integrated with Python. Seamless access to this ecosystem could open new research in geodynamics, enabling hybrid simulation strategies that combine physics-based and data-driven methods. Pyroclast was developed to address this gap. It is designed from the ground up to enable geophysical modeling using modular design, GPU acceleration, and native compatibility with Python-based scientific and AI tools. By rethinking how geodynamic solvers are structured and implemented, Pyroclast aims to bridge the gap between performance and usability, and to enable a new generation of Earth scientists to work at the frontier of HPC and AI-enabled simulation.

The main goal of Pyroclast is to create a modular and maintainable simulation framework that enables geoscientists to run established geodynamic models in a distributed, GPU-accelerated setting without requiring deep expertise in parallel programming or low-level hardware optimization. At the same time, Pyroclast provides a flexible architecture that allows researchers to implement new physics with minimal effort. Instead of forcing users to rewrite solvers from scratch, the framework encourages code reuse and extensibility, making it easier to explore novel physics within a robust simulation environment. Modularity is a key design principle, and by following a strictly object oriented design, Pyroclast improves maintainability and allows domain scientists to interact with the system through high level Python scripting. Heavy numerical workloads are offloaded to optimized libraries and GPU kernels implemented and compiled using specialized tools. The framework does not aim to achieve the absolute peak performance of low-level C++ or Fortran solvers. Instead, it offers a practical compromise that allows non-expert users to reach most of the performance potential, while benefiting from a simpler and more flexible development experience. In this way, Pyroclast lowers the barrier to entry and provides a practical, extensible platform for research.

\section{Goals and Contributions}
The first long-term objective of this project is to port the I2ELVIS and I3ELVIS family of geodynamic models to a modern, modular, Python-based framework. These models have been widely used for thermomechanical simulations in geodynamics. They combine the marker-in-cell method with a finite difference discretization of the Stokes and energy equations and implement a visco-elasto-plastic rheology model~\cite{gerya_i2elvis, gerya_i2elvis_old, Gerya_viscoelastoplastic}.

Given the time constraints of this project, the work focuses on implementing one of the core components required to eventually support the full I2ELVIS and I3ELVIS models. Specifically, this work develops a two-dimensional viscous mechanical model that forms the computational backbone of more complex visco-elasto-plastic simulations. This is not a simplification of the problem but a necessary foundational step. Thanks to the modular design of Pyroclast, the components developed here will be fully reused as the mechanical core of future models. The current implementation includes numerical methods needed to solve the full thermomechanical problem, most notably the solution of the Stokes equations. This component lays the groundwork for the methods and algorithms required to support more complex models. The focus on the mechanical system is deliberate: solving the Stokes equations is particularly challenging, as the discretization results in large, sparse saddle-point systems that are often severely ill-conditioned~\cite{saddle_point}. Efficiently solving such problems requires a specialized approach based on the geometric multigrid method~\cite{gerya_multigrid}.

This work presents the design, algorithms, and implementation strategies that form the foundation of Pyroclast. All solver components are implemented using a matrix-free approach with full GPU support, in order to maximize performance and scalability. Furthermore, MPI-based domain decomposition is introduced to enable distributed-memory parallelism, including an efficient parallel solver for the marker-based advection equation and a distributed multigrid strategy for the Stokes system. The work provides a blueprint for future development and serves as a technical reference for extending the framework to full visco-elasto-plastic models and beyond.

\section{Origin of the Name}
The name \textit{Pyroclast} is inspired by pyroclastic rocks: a class of clastic rocks composed of fragments (called \textit{pyroclasts}) produced and ejected during explosive volcanic eruptions. These rocks, often formed from violent volcanic processes, embody the dynamic and complex behavior of the Earth's interior, which Pyroclast aims to model computationally. The term itself derives from the Greek words \textit{pyr} (meaning ``fire'') and \textit{klastos} (meaning ``broken''), alluding to high-energy, fragmented phenomena. The name also serves as a play on words, adopting the \texttt{Py-} prefix conventionally used for Python libraries, signaling both its programming language foundation and scientific domain.

\section{Manuscript Outline}

The manuscript is structured into nine chapters that follow the conceptual and technical development of the Pyroclast framework.

Chapter~\ref{chap:introduction} introduces the motivation for modernizing geodynamic solvers, outlines the key scientific and technical challenges addressed in this work, and summarizes the main goals and contributions of the project. It also explains the origin of the name \textit{Pyroclast} and closes with this outline.

Chapter~\ref{chap:related_work} reviews related work in computational geodynamics, including existing codes for long-term lithospheric dynamics and seismic wave propagation. It highlights limitations of traditional simulation frameworks and discusses the benefits of the Python ecosystem, recent advances in domain-specific languages, and the growing role of machine learning in scientific computing. The chapter positions Pyroclast within this evolving landscape.

Chapter~\ref{chap:equations} presents the physical and mathematical foundations of the simulations. It covers the conservation of mass and momentum, the advection equation, boundary conditions, and non-dimensionalization.

Chapter~\ref{chap:numerical_methods} describes the numerical implementation of Pyroclast, including the marker-in-cell method, interpolation schemes, Lagrangian advection, staggered grid discretization, and the matrix-free solution of the incompressible Stokes equations. This chapter also introduces the multigrid solver, various smoothers and accelerators, and discusses convergence behavior and solver robustness.

Chapter~\ref{chap:software_architecture} outlines the software architecture and modular design of the framework. It explains the design philosophy, the organization of solver components, the simulation workflow, and the use of high-performance Python tools to enable both flexibility and efficiency.

Chapter~\ref{chap:gpu_support} focuses on GPU support and device-agnostic programming. It describes the integration of CuPy and Numba for GPU acceleration, discusses memory constraints, and outlines the design of stencil operations and marker advection kernels on GPUs.

Chapter~\ref{chap:mpi} describes the distributed parallelization strategy using MPI. It covers domain decomposition, MPI communicator setup, distributed marker exchange, and the implementation of RAS-preconditioned distributed multigrid methods.

Chapter~\ref{chap:benchmarks} presents benchmarking and validation results for shared-memory and distributed executions. It evaluates strong and weak scaling performance, convergence behavior, and GPU acceleration compared to CPU baselines, demonstrating the efficiency and scalability of the framework.

Finally, Chapter~\ref{chap:conclusions} summarizes the core contributions of this work and discusses the scientific impact of Pyroclast. It outlines directions for future work, including support for 3D simulations, coupled thermo-mechanical processes, more advanced numerical features, and integration with machine learning workflows for hybrid modeling.

\chapter{Related Work}
\label{chap:related_work}
In the previous chapter, we argued that numerical modeling plays a central role in geodynamics, where direct observations are limited and physical processes span a wide range of spatial and temporal scales. Over the past several decades, a number of simulation frameworks have been developed to address this challenge, many of which have become foundational tools in the community. However, these codes were often built using low-level programming languages and HPC paradigms that, by today's standards, present limitations in terms of extensibility, maintainability, and compatibility with emerging technologies such as GPU acceleration and machine learning.

This chapter provides a high level overview of the most widely used geodynamic solvers, briefly summarizing their numerical methods, software components, hardware support and reported performance or scaling results on high-performance computing systems. It then examines recent trends in the modernization of scientific software, particularly in climate and weather modeling, where efforts to combine performance with developer productivity have led to the adoption of modular, Python-based approaches. The chapter also explores the growing role of machine learning in scientific computing and highlights its potential to accelerate discovery and augment traditional simulation workflows. These developments motivate the need for a new class of simulation frameworks that combine traditional PDE-base numerical models with a modular design and compatibility with both high-performance computing and AI ecosystems. Pyroclast is introduced in this context as a response to this evolving landscape.

\section{Available Geodynamics Solvers}
This section reviews the main geodynamic simulation frameworks that have shaped current research in mantle convection, lithospheric deformation, seismic wave propagation and related large-scale geophysical processes.
\subsection{Long-Term Lithospheric Dynamics}
Over the past decades, a number of high-performance computing codes have been developed to simulate mantle convection, lithospheric deformation, and long-term geodynamic evolution. These solvers typically implement finite difference, finite volume, or finite element methods to solve thermo-mechanical equations, and many support parallel execution via MPI. Some have been extended or adapted to exploit GPU acceleration, although this remains limited. Below is a selection of widely used solvers and their key characteristics.

\subsubsection*{ASPECT}
ASPECT is a finite-element code developed under the Computational Infrastructure for Geodynamics (CIG). It is designed for 2D and 3D mantle convection and planetary thermal evolution problems. ASPECT supports advanced features such as adaptive mesh refinement (AMR), nonlinear rheologies, and compressibility. It is highly scalable and has been tested on thousands of CPU cores using MPI. However, GPU acceleration is not currently available in the standard version~\cite{heister:etal:2017,kronbichler:etal:2012,aspect-doi-v3.0.0,aspectmanual}.

\subsubsection*{CitcomS}
CitcomS is another CIG-supported finite-element code focused on 3D mantle convection, particularly in spherical shell geometry. The code supports domain decomposition via MPI and has been tested in HPC environments. While an experimental CUDA port exists, CitcomS remains primarily a CPU-based MPI code~\cite{citcoms_software, citcoms_article_1, citcoms_article_2, citcoms_article_3, citcoms_gpu}.

\subsubsection*{I2ELVIS/I3ELVIS}
I2ELVIS and its 3D extension I3ELVIS are finite-difference solvers based on the marker-in-cell method. These models implement visco-elasto-plastic rheology and support additional physics such as self-gravitation and free-surface evolution. The numerical strategy combines conservative staggered-grid finite differences with a fully nonlinear rheological formulation that is solved iteratively at every time step. The I2/I3ELVIS codes do not currently support MPI or GPU acceleration, which limits their scalability on modern parallel architectures~\cite{gerya_i2elvis, gerya_i2elvis_old}.

\subsubsection*{GAIA}
GAIA is a finite-volume code developed at DLR for simulating thermo-chemical evolution of planetary mantles. It supports creeping flow with strongly variable viscosity and models thermochemical convection. GAIA is designed for distributed-memory parallelism and includes GPU-enabled solvers using CUDA and hybrid CPU-GPU strategies, making it one of the few geodynamics codes with native GPU support~\cite{gaia_software, gaia_article_1, gaia_article_2, gaia_article_3}.

\subsubsection*{LaMEM}
LaMEM (Lithosphere and Mantle Evolution Model) is a finite-difference code from the University of Mainz designed for 3D lithosphere–mantle simulations with visco-elasto-plastic rheology. It uses a staggered grid and supports marker-in-cell techniques. Built on PETSc, LaMEM scales to hundreds of thousands of CPU cores. Recent efforts have explored GPU acceleration via Julia bindings, although these developments are still experimental and not part of the main release.

\subsubsection*{pTatin3D}
pTatin3D is a geodynamic modeling package designed for long-term lithospheric deformation problems, such as subduction and continental rifting. It combines the material point method for tracking material composition with a finite-element discretization of the incompressible visco-plastic Stokes equations. The solver uses a matrix-free geometric multigrid preconditioner, which significantly reduces memory bandwidth requirements and improves scalability on modern architectures. Performance studies have demonstrated more than 2× speedups compared to matrix-based methods, with sustained weak scaling on systems like the Cray XC-30. pTatin3D supports MPI-based parallelism and may benefit from GPU acceleration via its PETSc back end~\cite{ptatin3d}.

\subsubsection*{StagYY}
StagYY is a finite-volume code developed at ETH Zürich for 3D spherical shell convection, including compressibility, phase changes, and thermochemical effects. While not open-source, it is widely used in the mantle dynamics community. It supports MPI parallelism and is being modernized with the StagBL abstraction layer which is based on PETSc~\cite{tackley_stagyy, stagbl}.

\subsubsection*{Underworld2}
Underworld2 is a particle-in-cell finite-element code developed at the University of Melbourne for simulating lithosphere and mantle deformation, including visco-elasto-plastic rheologies. It provides a Python interface for problem setup while relying on a PETSc-based C solver backend. Underworld2 scales well in CPU-based MPI environments but does not include native GPU support. PETSc-based components may benefit from external GPU-accelerated libraries, though this is not the focus of the code~\cite{underworld_1, underworld_2, underworld_3, underworld_4, underworld_5}.

While these solvers represent major milestones in the development of computational geodynamics, several common limitations remain. Most are implemented in low-level languages such as Fortran, C or C++, making them difficult to extend or interface with the modern AI/ML software ecosystems. GPU support, when available, is often limited to specific modules or achieved indirectly through PETSc~\cite{petsc-web-page, petsc-user-ref, petsc-efficient, petsc-gpu} or other external libraries rather than through native GPU implementations. As a result, few of these frameworks can fully exploit current heterogeneous HPC architectures. Furthermore, monolithic design and complex build systems pose challenges for modularity and rapid experimentation, limiting accessibility for researchers who wish to modify or extend the models. Among the codes reviewed, LaMEM is the most similar to I3ELVIS in both its numerical methods and physical modeling approach, and therefore also closely aligned with the goals of Pyroclast. Like Pyroclast, it uses finite difference discretization combined with the marker-in-cell method. 

\subsection{Seismic Wave Propagation}
In addition to long-term geodynamic and mantle convection codes, the geophysics community has developed a number of high-performance simulation frameworks targeting other classes of physical problems. Notably, many solvers focus on seismic wave propagation, dynamic earthquake rupture, and crustal deformation. These codes typically solve hyperbolic or elastodynamic equations and are optimized for high-resolution, short-timescale simulations. While they address a different set of physical processes, they share many of the same computational challenges, including the need for scalable parallelism and efficient use of accelerator hardware. Some of the most widely used tools in this space are briefly described below.

\subsubsection*{SeisSol}
SeisSol is a high-order discontinuous Galerkin code for seismic wave propagation and dynamic rupture. It is designed for simulating complex 3D earthquake scenarios and supports both CPU and GPU acceleration. The code uses hybrid MPI and OpenMP parallelization and has demonstrated strong scaling on large HPC systems~\cite{seissol_1, seissol_2, seissol_3, seissol_4}.

\subsubsection*{SPECFEM3D}
SPECFEM3D is a spectral-element code for global and regional 3D seismic wave propagation. It is widely used in earthquake ground motion modeling and seismic tomography. The solver supports distributed-memory parallelism via MPI and offers GPU acceleration through both CUDA and HIP backends~\cite{specfem_1, specfem_2, specfem_3}.

\subsubsection*{AWP-ODC}
AWP-ODC is a finite-difference code for anelastic wave propagation and earthquake rupture simulations. It has been optimized for large-scale ground motion modeling and has demonstrated excellent scaling on supercomputers. GPU-enabled versions of the code have achieved substantial speedups in critical kernels, enabling high-frequency simulations~\cite{awp_odc_1, awp_odc_2}.

\subsubsection*{SW4}
SW4 (Seismic Waves 4th-order) is a finite-difference solver for regional 3D seismic wave propagation that supports complex topography and material heterogeneities. It supports MPI-based parallelism and has been ported to GPUs using the RAJA~\cite{raja_1, raja_2} performance portability framework, resulting in strong speedups on GPU platforms~\cite{sw4_1, sw4_2, sw4_3}.

\subsubsection*{PyLith}
PyLith is a finite-element solver for quasi-static and dynamic crustal deformation, developed for earthquake cycle modeling. It is built on top of PETSc and supports MPI parallelism. While PyLith does not currently support GPU acceleration, its use of PETSc may allow future integration with GPU capable solvers~\cite{pylith}.

The development of high-performance solvers for seismic wave propagation and dynamic rupture has led to significant advances in earthquake modeling. These codes are often highly optimized for specific use cases and have demonstrated excellent scalability on modern HPC systems. Many now support GPU acceleration, either through native implementations or performance portability frameworks. Although they target different physical systems than long-term geodynamics models, they share many of the same computational challenges, including the need for efficient parallelization, flexible solver design, and integration with evolving hardware architectures. Together with the mantle convection solvers discussed earlier, they illustrate the broader landscape of geophysical simulation tools and help motivate the design goals behind Pyroclast.

\section{Limitations of Traditional Simulation Frameworks}
Despite their significant contributions to geodynamics research, many traditional simulation frameworks face structural and architectural limitations that hinder  development and broader adoption. Most existing codes were designed for CPU-centric hardware architectures and implemented in low-level programming languages such as C, C++, or Fortran. Their structure is often monolithic, with tightly coupled components that make it difficult to extend or modify the software without deep knowledge of both the underlying physics and the technical implementation. Extending these solvers or introducing new physical models often requires expertise from multiple domains: geophysicists must not only understand the governing equations and numerical methods, but also have the engineering background needed to deal with low-level programming, performance tuning, and parallelization on heterogeneous architectures. This includes knowledge of topics such as cache optimization, shared versus distributed memory models, MPI communication, and GPU programming. Few researchers have this combined skill set, which greatly limits the accessibility of existing tools to the broader scientific community.

This added complexity also makes it difficult for many solvers to adopt modern hardware and software technologies. While several seismic and wave-propagation codes now include GPU support, few of the models for long-term lithospheric dynamics presented in the previous section feature stable and fully integrated GPU acceleration. This may be due to the added implementation difficulty of using a marker based method, which requires dealing with both particle and grid operations. This makes them harder to accelerate on GPUs compared to purely grid-based methods, and as a result, GPU support is often incomplete, experimental, or limited to specific components.

Many existing codes rely on PETSc~\cite{petsc-web-page, petsc-user-ref, petsc-efficient, petsc-gpu} to enable distributed parallelism and, in some cases, partial GPU acceleration. PETSc is a robust and widely used library, but it is nevertheless a large monolithic framework written in C that was conceived before the era of GPU programming for scientific computing, with the first stable version released to the public in 1996~\cite{petsc_changes}. GPU acceleration within PETSc applies to selected operations and is still under development, mainly targeting algebraic solvers and BLAS routines~\cite{petsc_gpu_roadmap}. Custom kernels, such as stencil operations used in matrix-free geometric multigrid solvers, still require manual implementation in CUDA or similar low-level languages. Alternatively, developers must rely on a performance portability framework such as Kokkos~\cite{kokkos_1, kokkos_2} or RAJA~\cite{raja_1, raja_2}, or adopt a directive-based approach such as OpenACC~\cite{openacc-spec}. These models are typically implemented in C or C++ and require significant technical expertise to use effectively. 

\section{Benefits of the Python Ecosystem}
In contrast, the modern Python ecosystem, influenced by the needs of AI and machine learning, treats GPU computing as a first-class concern. Array-based programming models built around \texttt{ndarray} or tensor abstractions naturally allow mathematical operations to be offloaded to GPUs, automatically targeting optimized vendor libraries such as cuBLAS~\cite{cublas}, cuTensor~\cite{cutensor}, or equivalently hipBLAS~\cite{hipblas} and hipTensor~\cite{hiptensor}. This design enables the same high-level code to run efficiently on CPUs or GPUs, depending on the available hardware context.

For custom operations that cannot be easily expressed as tensor operations, such as stencil kernels, Python provides several just-in-time (JIT) compilation frameworks that allow users to write high-performance CPU and GPU kernels directly in Python. Projects like Numba~\cite{numba}, Numba-CUDA~\cite{numba_cuda}, Numba-HIP~\cite{numba_hip}, and DaCe~\cite{dace} make it possible to generate optimized machine code without leaving the Python environment, achieving performance comparable to manually written CUDA or HIP kernels. In many cases, the same kernel can be compiled for multiple hardware backends with little or no modification.

When direct interfacing with low-level languages is necessary, either to access specialized external libraries or to implement highly optimized kernels, bindings can be written easily using tools such as Nanobind~\cite{nanobind}. These bindings integrate seamlessly with Python ndarray and tensor objects, allowing zero-copy data sharing with libraries like NumPy~\cite{numpy}, CuPy~\cite{cupy}, and PyTorch~\cite{pytorch}.

The benefits of the Python ecosystem extend well beyond GPU programming and performance portability. Python provides access to a vast collection of mature libraries for data analysis, visualization, and numerical computing, which work together natively. Deployment of Python-based tools is also straightforward thanks to the AI-driven software infrastructure that has emerged in recent years. Container technologies such as Apptainer~\cite{apptainer_1, apptainer_2} and Enroot~\cite{enroot} simplify deployment on HPC systems and provide access to optimized GPU-ready environments, including prebuilt images like the NVIDIA PyTorch container~\cite{nvidia_pytorch_container}. These environments include vendor-optimized math libraries, GPU-aware MPI, and advanced communication layers such as NCCL~\cite{nccl} and NVSHMEM~\cite{nvshmem}, simplifying scaling across distributed GPU clusters.

Beyond performance and interoperability, Python’s simple yet expressive syntax plays a key role in improving developer productivity. The language allows complex numerical operations to be written in a clear and compact way, often resembling the underlying mathematical formulation. This reduces the amount of boilerplate code required and makes implementations easier to read, modify, and maintain. Features such as dynamic typing, automatic memory management, and high-level data structures further simplify development by removing much of the overhead associated with low-level programming. In this respect, Python shares many of the same advantages as Julia, offering a high-level interface that enables researchers to focus on the scientific and algorithmic aspects of their models rather than the technical details of implementation. In addition, Python has strong support for object-oriented programming, which makes it easier to build modular, extensible simulation frameworks. This contributes to shorter development cycles, better code organization, and long-term maintainability.

It is important to note that Python is not without trade-offs. As an interpreted language, it introduces runtime overhead compared to compiled languages like C or Fortran, and poorly structured code can lead to significant performance bottlenecks. However, these limitations are largely mitigated when performance-critical components are offloaded to optimized libraries or JIT compiled. With careful design, Python-based simulations can achieve performance comparable to low-level implementations while retaining a much higher level of abstraction and developer productivity. Recent results from large-scale scientific computing projects have demonstrated that, when carefully designed, Python-based frameworks can achieve performance on par with traditional low-level implementations, even at the highest end of HPC. For example, recent work on large-scale quantum transport simulations achieved sustained exascale performance using a Python-based framework built with NumPy, CuPy, mpi4py, Numba, and SciPy. The team ran simulations on over 37,000 GPUs, reaching 1.15~\si{\exa\flop\per\second} at 82\% weak-scaling efficiency, and the project was nominated for the 2025 ACM Gordon Bell Prize~\cite{quantum_transport, Russo2025NanoribbonTransistor}. This result demonstrates that Python, when used with the right tools and design, can scale to extreme levels of performance.

Overall, while traditional frameworks have served the geodynamics community for decades, their reliance on low-level languages, monolithic design, and limited support for heterogeneous computing make them increasingly difficult to extend and maintain. The Python ecosystem, by contrast, offers a flexible, modular, and hardware-aware platform that aligns naturally with the current direction of high-performance and AI-accelerated computing. Adopting such modern software paradigms represents a necessary step toward bridging the gap between performance, accessibility, and scientific innovation in geodynamics.

\section{Domain-Specific Languages in Scientific Python}
In recent years, several projects have emerged with the goal of abstracting away the complexity of performance optimization from domain scientists, specifically targeting the Python ecosystem. These efforts aim to make it easier to write high-performance simulation code without requiring expertise in low-level programming, memory hierarchies, or hardware-specific tuning. One of the most prominent examples is ICON~\cite{ICON2025.04}, a global climate and weather prediction model, which is currently being ported from Fortran to Python with the ICON4Py initiative~\cite{icon4py}.

A central component of this project is GT4Py~\cite{gt4py}, a domain-specific language (DSL) for stencil computations embedded directly in Python. GT4Py allows scientists to express stencil operations using a high-level, backend-agnostic syntax that is compiled into optimized low-level kernels targeting CPUs, GPUs, or FPGAs. It supports multiple backends, including a native kernel generation backend and one based on the DaCe data-centric compiler~\cite{dace}. Although primarily developed for use in ICON, GT4Py is open source and can be used in any scientific application that involves stencil-based computations.

Another example is Devito~\cite{devito-api, devito-compiler}, a domain-specific language for solving partial differential equations embedded in Python. Devito adopts a higher-level approach: instead of manually writing stencil code, users define their physical models symbolically using the SymPy~\cite{sympy} package. From these symbolic expressions, Devito automatically derives discretizations and generates optimized low-level stencil kernels. The compiler supports a range of advanced optimizations and most notably, recent developments also enable the automatic generation of MPI-parallel code, allowing Devito to scale across distributed-memory systems~\cite{bisbas2024automatedmpixcodegeneration}. The framework has been successfully used for applications such as wave propagation and seismic inversion, and shows an extreme level of abstraction compared to traditional simulation codes.

\section{Machine Learning for Scientific Computing}

In recent years, machine learning has emerged as a powerful tool in scientific computing, extending far beyond traditional tasks such as data analysis or classification. It is now actively used to accelerate numerical solvers, build surrogate models, and solve inverse problems across a wide range of scientific domains. The growing field of scientific machine learning spans a very broad set of different models and techniques, far too many to review comprehensively here. Instead, we highlight a few representative approaches that illustrate the increasing role of machine learning in modern physics-based simulation workflows.

A major class of models are physics-informed neural networks (PINNs), which incorporate differential equation constraints directly into the training objective by embedding PDE residuals into the loss function~\cite{pinn_original_paper_1, pinn_original_paper_2}. This enables the model to approximate PDE solutions without requiring labeled data. PINNs have been successfully applied to problems in fluid mechanics, weather and climate modeling, and geophysics~\cite{pinn_fluid_mechanics, pinn_geoscience}.

Neural operators take a more general approach by learning mappings between function spaces, such as input and solution fields of PDEs. Architectures like Fourier Neural Operators (FNOs) learn these mappings in Fourier space and have been applied to simulate turbulent flows and other complex fluid dynamics problems~\cite{fno_paper}. Extensions such as spherical FNOs have been used to forecast global atmospheric dynamics~\cite{sfno}. Other variants, including convolutional neural operators (CNOs), use CNN-based architectures and have shown success in similar settings~\cite{cno}. These models enable fast surrogate training and especially inference and are well suited for learning unknown physics using a data-driven approach.

Closely related are differentiable physics methods, which implement simulation pipelines using automatic differentiation to enable end-to-end optimization through physical models. This allows, for instance, inverse material design by backpropagating through simulation steps~\cite{diff_phys_materials}, or learning control policies by optimizing through differentiable Newtonian solvers~\cite{diff_taichi}.

More recently, research into large foundation models for PDEs has shown that transformer-based architectures can generalize across different equation types and domains. Poseidon~\cite{poseidon}, a multiscale operator transformer trained on diverse PDE datasets, has demonstrated strong performance on a variety of downstream tasks, outperforming domain-specific models and pointing toward general-purpose PDE solvers. Another recent example of a large-scale foundation model is Aurora~\cite{aurora}. Aurora is a transformer-based model of the Earth system trained on more than one million hours of heterogeneous geophysical data, covering atmospheric, oceanic, and climatic processes. It has demonstrated performance surpassing that of several operational forecasting systems in tasks such as air quality prediction, ocean wave modeling, tropical cyclone tracking, and high-resolution weather forecasting, while operating at orders of magnitude lower computational cost. Aurora can also be fine-tuned for specialized applications with relatively modest computational resources, making it an important step toward accessible, efficient, and unified Earth system prediction. Its success highlights the transformative potential of machine learning for large-scale environmental modeling and further reinforces the need for simulation frameworks that can interface seamlessly with AI-driven tools.

Together, these examples illustrate that machine learning is becoming a central component of scientific computing, offering powerful tools to accelerate simulations, infer hidden, unknown or intractable physics, and solve problems that in general remain out of reach for traditional numerical methods alone.

\section{Positioning Pyroclast}
The landscape of geodynamic simulation tools has been shaped by decades of high-performance computing development. As discussed in the previous sections, most established codes rely on monolithic architectures written in low-level languages. While these tools have enabled significant scientific advances, their structure often limits extensibility, complicates maintenance, and hinders the adoption of emerging technologies such as GPU acceleration and machine learning. This creates a growing disconnect between modern hardware/software capabilities and what domain scientists can readily access.

Pyroclast is designed to bridge this gap. It combines the features of a modern and scalable HPC solver, including distributed-memory parallelism and GPU acceleration, with a simple modular architecture written entirely in Python. Unlike frameworks that treat Python as a thin wrapper around low-level code, Pyroclast uses Python as the primary development environment. Performance-critical operations are offloaded to JIT-compiled kernels, vectorized backends, or GPU libraries, delivering speed and portability without compromising on simplicity and developer productivity.

In contrast to traditional solvers for long-term lithospheric dynamics, which require deep familiarity with C/Fortran or PETSc internals to extend, Pyroclast offers a clean and accessible interface that allows scientists to implement or adapt models with minimal technical overhead. Its object-oriented structure encourages modularity and reuse, making it straightforward to extend or replace components independently. This is especially valuable for developing hybrid models that combine mechanistic simulation with machine learning, a growing trend across computational science.

Pyroclast is also tightly integrated with the Python scientific ecosystem, including libraries for data analysis, visualization, and deep learning. This enables seamless deployment of simulation workflows in AI-accelerated environments and supports direct interfacing with differentiable solvers, neural surrogates, and foundation models for physical systems.

By embracing the broader modernization trends in scientific computing, Pyroclast defines a new class of geodynamics simulation framework, one that is high-performance yet accessible, modular yet robust, and natively compatible with both HPC and AI ecosystems. It builds on the strengths of past solvers while removing key barriers to innovation and experimentation.

\section{Scope of This Work}
As discussed in Chapter~\ref{chap:introduction}, the long-term goal of this project is to port the I2ELVIS/I3ELVIS family of geodynamic models to a modern, high-performance Python framework. Since fully implementing the complete thermomechanical model is beyond the scope of this first manuscript, we focus instead on a necessary and manageable first step: building a distributed, GPU-accelerated viscous mechanical solver.

To keep the problem tractable, we assume purely viscous deformation and fixed material properties that are advected by the velocity field. The implementation is limited to two dimensions, which simplifies development and visualization. However, all algorithms are designed to extend naturally to 3D, where larger problem sizes are expected to yield better performance. Pyroclast’s modular design supports this by allowing straightforward reuse of components across different models and dimensions.

Developing a simulation framework presents a variety of mathematical, engineering, and HPC challenges. While not all aspects can be covered in detail, this manuscript focuses on the key equations, numerical methods, algorithms, and optimization strategies that form the basis of the current implementation.

\chapter{Governing Equations}
\label{chap:equations}

This chapter outlines the governing equations used in the current implementation of Pyroclast, which models incompressible viscous deformation relevant to lithospheric and mantle dynamics. The model is based on simple principles of continuum mechanics and describes slow, creeping flow dominated by viscous forces. We focus exclusively on the mechanical component of the problem, as this forms the computational core of the more complex visco-elasto-plastic thermomechanical models used in I2ELVIS and I3ELVIS.

The material presented in this chapter is largely based on the work proposed in~\cite{gerya_book}, which provides a detailed overview of the governing principles and rheological models used in numerical geodynamic modeling. The purpose here is not to introduce new theory, but rather to summarize the essential equations and assumptions that constitute the current implementation in Pyroclast. These equations define the conservation of mass and momentum under the assumptions of incompressible, viscous flow, and provide the basis for the numerical models discussed in subsequent chapters.

\section{Physical Assumptions}

The mechanical model describes slow, buoyancy-driven flow in the Earth's lithosphere and upper mantle under a set of simplifying physical assumptions. These approximations preserve the dominant dynamics of the system while allowing for a tractable continuum formulation.

\subsubsection*{Creeping Flow}
The flow regime is characterized by extremely low Reynolds numbers, such that inertial forces are negligible compared to viscous forces. The governing equations therefore reduce to the steady-state Stokes equations. This assumption is well justified for long-term lithospheric deformation, where typical plate velocities are on the order of $10^{-9}~\si{\centi\meter\per\second}$ and evolve over geological timescales of about $10^{13}~\si{\second}$. The resulting accelerations are around $10^{-22}~\si{\meter\per\second\squared}$, many orders of magnitude smaller than gravity ($g \approx 10~\si{\meter\per\second\squared}$). The momentum balance thus expresses a quasi-static equilibrium between viscous, pressure, and gravitational forces.

\subsubsection*{Incompressibility}
The material is assumed to be incompressible, implying that volume changes are negligible. The velocity field $v = (v_x, v_y)$ therefore satisfies the divergence-free condition:
\begin{equation}
\frac{\partial v_x}{\partial x} + \frac{\partial v_y}{\partial y} = 0.
\label{eq:incomp}
\end{equation}

\subsubsection*{Viscous Rheology}
The material behaves as a linear, Newtonian viscous fluid. The deviatoric stress tensor $\sigma'_{ij}$ is related to the strain rate tensor $\dot{\epsilon}_{ij}$ by
\begin{equation}
\sigma'_{ij} = 2 \eta \dot{\epsilon}_{ij}
= \eta \left( \frac{\partial v_i}{\partial x_j} +  \frac{\partial v_j}{\partial x_i} \right),
\label{eq:stress}
\end{equation}
where $\eta$ denotes the dynamic viscosity. Both viscosity and density are treated as spatially variable quantities that are advected with the flow but do not evolve internally within a time step.

\subsubsection*{Stress Decomposition}
The Cauchy stress tensor $\sigma_{ij}$ is commonly decomposed into an isotropic pressure component and a deviatoric stress component:
\begin{equation}
\sigma_{ij} = -p\,\delta_{ij} + \sigma'_{ij},
\label{eq:stress_decomp}
\end{equation}
where $p$ represents the mean normal stress (pressure) and $\delta_{ij}$ is the Kronecker delta. The deviatoric part $\sigma'_{ij}$ carries the information about shear deformation, while $p$ accounts for volumetric effects.  
For incompressible flow, $\nabla \cdot v = 0$ implies that the volumetric strain vanishes, so the pressure term acts solely as a Lagrange multiplier enforcing incompressibility. The governing equations can therefore be written directly in terms of $\sigma'_{ij}$ and $p$.

\section{Conservation of Mass and Momentum}

Under these assumptions, the system is governed by the two-dimensional incompressible Stokes equations, consisting of the $x$-momentum equation, the $y$-momentum equation, and the incompressibility condition. The Cauchy formulation of the Stokes problem expresses the momentum balance in terms of deviatoric stresses and pressure as follows:
\begin{equation}
\frac{\partial \sigma_{xx}'}{\partial x} + \frac{\partial \sigma_{xy}'}{\partial y} - \frac{\partial p}{\partial x} = 0,
\label{eq:xmom}
\end{equation}
\begin{equation}
\frac{\partial \sigma_{yx}'}{\partial x} + \frac{\partial \sigma_{yy}'}{\partial y} - \frac{\partial p}{\partial y} = \rho g_y,
\label{eq:ymom}
\end{equation}
\begin{equation}
\frac{\partial v_x}{\partial x} + \frac{\partial v_y}{\partial y} = 0.
\label{eq:mass}
\end{equation}

Equations~\eqref{eq:xmom} and~\eqref{eq:ymom} represent the balance of momentum in the $x$ and $y$ directions, respectively, while Equation~\eqref{eq:mass} enforces incompressibility. The constitutive relationship~\eqref{eq:stress} links the deviatoric stresses to the velocity field through the viscosity.

The physical quantities appearing in the above equations are:
\begin{itemize}
    \item $\sigma'_{ij}$: deviatoric stress on the $i$-plane acting in the $j$-direction
    \item $p$: pressure
    \item $v_x, v_y$: velocity components in the $x$ and $y$ directions
    \item $\rho$: density
    \item $g_y$: gravitational acceleration in the $y$-direction
    \item $\eta$: dynamic viscosity
    \item $\dot{\epsilon}_{ij}$: strain rate on the $i$-plane acting in the $j$-direction
\end{itemize}

Together, Equations~\eqref{eq:xmom}--\eqref{eq:mass} and the constitutive law~\eqref{eq:stress} define the incompressible Stokes problem used in Pyroclast.

\section{Advection Equation}

In addition to the conservation of mass and momentum, the evolution of material properties such as viscosity and density is governed by the advection equation. These quantities are transported by the velocity field but do not diffuse or evolve through internal processes within a single time step.

In Pyroclast, advection is treated from a Lagrangian point of view. A set of discrete particles, commonly referred to as \emph{markers}, is used to represent the material domain. Each marker carries physical properties, such as $\eta$ or $\rho$, which are advected according to the local velocity field $v = (v_x, v_y)$. The advection of a scalar property $\phi$ follows
\begin{equation}
\frac{D \phi}{D t} = \frac{\partial \phi}{\partial t} + v_x \frac{\partial \phi}{\partial x} + v_y \frac{\partial \phi}{\partial y} = 0,
\label{eq:advection}
\end{equation}
where $D / D t$ denotes the material derivative. This equation expresses that the quantity $\phi$ remains constant along the trajectory of a material parcel.

From a Lagrangian point of view, we follow material trajectories $\mathbf{x}(t) = (x(t), y(t))$ defined by the characteristic ordinary differential equation
\begin{equation}
\frac{d\mathbf{x}}{dt} = v\!\left(\mathbf{x}(t), t\right), 
\qquad 
\mathbf{x}(t_0) = \mathbf{x}_0,
\label{eq:char_ode}
\end{equation}
where $\mathbf{x}_0$ denotes the initial position of the parcel at time $t_0$. Along any such trajectory,
\begin{equation}
\frac{d}{dt}\,\phi\!\left(\mathbf{x}(t),t\right) 
= \frac{D\phi}{Dt} 
= 0 
\quad\Longrightarrow\quad 
\phi\!\left(\mathbf{x}(t),t\right) 
= \phi_0(\mathbf{x}_0),
\label{eq:phi_const_along_char}
\end{equation}
meaning that $\phi$ remains constant along particle paths. Therefore, the solution of~\eqref{eq:advection} is obtained by (i) integrating the trajectory equation~\eqref{eq:char_ode} and (ii) assigning to each parcel the same property it had initially.

The use of Lagrangian markers allows for accurate tracking of material interfaces and sharp property contrasts while minimizing numerical diffusion. The mechanical equations, however, are solved on a fixed Eulerian grid. Coupling between the Lagrangian and Eulerian descriptions is achieved by interpolating marker properties to the grid before solving the Stokes equations, and by mapping the resulting velocity field back to the markers for advection.

The specific numerical schemes used for interpolation and time integration, as well as their implementation in Pyroclast, are described in detail in the following chapters.

\section{Boundary Conditions}

The system must be supplemented by appropriate boundary conditions that constrain the velocity field along the domain boundaries. These conditions determine how the material can move or deform at the edges of the computational domain and play an important role in defining the physical behavior of the model.

Common types of mechanical boundary conditions include~\cite{gerya_boundary_conditions}:
\begin{itemize}
    \item \textbf{Free-slip boundaries}: the normal velocity component vanishes, and the shear stress along the boundary is zero. In terms of velocity derivatives, this is expressed as
    \begin{equation}
        v_n = 0, 
        \qquad 
        \frac{\partial v_t}{\partial n} = 0,
        \label{eq:free_slip_bc}
    \end{equation}
    where $v_n$ and $v_t$ denote the normal and tangential velocity components, respectively, and $n$ indicates differentiation in the direction normal to the boundary.

    \item \textbf{No-slip boundaries}: both the normal and tangential velocity components are set to zero:
    \begin{equation}
        v_n = 0, 
        \qquad 
        v_t = 0.
        \label{eq:no_slip_bc}
    \end{equation}

    \item \textbf{Periodic boundaries}: velocity and pressure fields are periodic across opposite boundaries, ensuring continuity of both velocities and fluxes:
    \begin{equation}
        v_i(x_\mathrm{min}) = v_i(x_\mathrm{max}), 
        \qquad 
        p(x_\mathrm{min}) = p(x_\mathrm{max}).
        \label{eq:periodic_bc}
    \end{equation}

    \item \textbf{Prescribed velocity (Dirichlet) boundaries}: specific values of $v_i$ or of the normal velocity gradient $\partial v_i / \partial n$ may be imposed to drive the flow or approximate open boundaries.
\end{itemize}

This is not a comprehensive overview, but the important takeaway is that boundary conditions are an essential part of the Stokes problem and must be treated with particular care, especially when the domain is distributed across multiple processes. In a parallel setting, each subdomain must correctly represent its local boundaries as well as the shared interfaces with neighboring domains. Ensuring consistency and continuity across these boundaries is crucial for both numerical stability and physical accuracy. The implementation of boundary conditions within the MPI domain decomposition strategy used by Pyroclast is discussed in detail in Chapter~\ref{chap:mpi}.

\section{Non-dimensionalization and Unit Scaling}

Pyroclast does not currently support automatic non-dimensionalization of the Stokes equations. In our limited testing, solving the non-dimensional form of the equations did not lead to a significant improvement in convergence or numerical stability. However, the same numerical benefits can be achieved manually through unit scaling of the physical parameters of the model. This method keeps all quantities dimensional while ensuring that numerical values remain well balanced in magnitude.

The idea of unit scaling is to rescale all physical quantities in the system relative to a set of reference values, chosen to represent typical magnitudes for the problem. This effectively normalizes the equations without changing their physical dimensions, thereby improving numerical conditioning while keeping results expressed in physical units.

The procedure can be summarized as follows:

\begin{enumerate}
    \item \textbf{Select reference values.}  
    Choose one reference value for each independent physical dimension present in the problem.  
    For the Stokes system, these are time~$T$, length~$L$, and mass~$M$.  
    These values serve as the scaling anchors for the system and define what is considered a “unit” of each base dimension.  
    For example:
    \[
    t_0 = 1~\si{\second}, 
    \qquad 
    x_0 = 100~\si{\kilo\meter} = 10^5~\si{\meter},
    \qquad 
    \eta_0 = 10^{18}~\si{\pascal\second}.
    \]
    The first two values set the temporal and spatial scales of the problem, while $\eta_0$ defines the characteristic viscosity (and thus indirectly fixes mass scale).

    \item \textbf{Express reference values in terms of base reference dimensions.}  
    Each reference value can be written as a combination of the fundamental dimensions of time, length, and mass.  
    For instance, viscosity has physical dimensions $[\eta] = M/(L\,T)$.  
    This step ensures that all chosen reference quantities are dimensionally consistent:
    \[
    t_0 = T_0, 
    \qquad 
    x_0 = L_0, 
    \qquad 
    \eta_0 = \frac{M_0}{L_0\,T_0}.
    \]

    \item \textbf{Solve for the dimensional scaling coefficients.}  
    By substituting the known reference values into the dimensional relations above, one can solve for the reference coefficients $T_0$, $L_0$, and $M_0$.
    \[
    T_0 = t_0, 
    \qquad 
    L_0 = x_0, 
    \qquad 
    M_0 = \eta_0\,x_0\,t_0.
    \]
    In this example, the mass scale $M_0$ follows directly from the definition of viscosity, since $\eta_0$ already contains $M$, $L$, and $T$.

    \item \textbf{Express all other quantities relative to these coefficients.}  
    Once the reference scales are defined, any physical variable can be expressed as a dimensionless quantity $\tilde{q}$ multiplied by its reference scale $q_0$.  
    \begin{gather*}
    \tilde{\eta} = \frac{\eta}{\eta_0}, \qquad
    \tilde{x} = \frac{x}{x_0}, \qquad
    \tilde{t} = \frac{t}{t_0}, \\
    \rho_0 = \frac{M_0}{L_0^3} \;\Longrightarrow\;
    \tilde{\rho} = \frac{\rho}{\rho_0}, \qquad
    g_0 = \frac{L_0}{T_0^2} \;\Longrightarrow\;
    \tilde{g} = \frac{g}{g_0}.
    \end{gather*}

    These relations define the scaling for all remaining physical quantities in the Stokes system. Density, gravity, velocity, and pressure can all be derived analogously.

\end{enumerate}

Numerically, this procedure achieves the same effect as full non-dimensionalization: it rescales all terms in the equations to comparable magnitudes, improving numerical conditioning. The difference is that physical units are preserved throughout the process, making it easier to interpret input parameters and results.

\chapter{Numerical Methods}
\label{chap:numerical_methods}
This chapter describes the numerical algorithms that form the core of the Pyroclast solver. Pyroclast is based on the marker-in-cell (MIC) method, also known as the material point method, which combines a fixed Eulerian grid with Lagrangian markers used to represent and advect material properties. The chapter introduces the overall simulation pipeline used in MIC methods and then focuses on the specific numerical components implemented in Pyroclast. These include the finite-difference discretization of the governing equations on a staggered grid, the interpolation procedures used to exchange information between markers and grid, the advection scheme for material transport, and the iterative solver for the incompressible Stokes system with variable viscosity. Special emphasis is placed on the matrix-free formulation of the operators and the use of geometric multigrid methods for efficient velocity smoothing. Details related to parallelization, GPU acceleration, and MPI support are deferred to the following chapters.

\section{The Marker-in-Cell Method}

Pyroclast is based on the marker-in-cell (MIC) method, which combines an Eulerian discretization of the governing equations with a Lagrangian representation of material properties. The governing PDEs are solved on a fixed rectilinear grid, while quantities such as viscosity, density, temperature, etc. are stored on Lagrangian markers that move with the flow. Marker values are interpolated to the grid before each solve, and the resulting velocity field is interpolated back to the markers to update their positions.

This method is essentially a form of operator splitting, specifically Lie-Trotter splitting, of the advection and PDE solve steps, and yields a first-order accurate time integration scheme~\cite{splitting_methods}. Material advection is handled separately from the solution of the governing equations, such as the incompressible Stokes system or the energy equation. Grid-based solves use material properties interpolated from the markers, and the resulting velocity field is used to update marker positions independently. In principle, a second-order Strang splitting scheme could be used to improve time accuracy to second-order~\cite{splitting_methods}, but Pyroclast currently employs the simpler first-order variant.

\begin{figure}[h]
\centering
\includegraphics[width=0.8\textwidth]{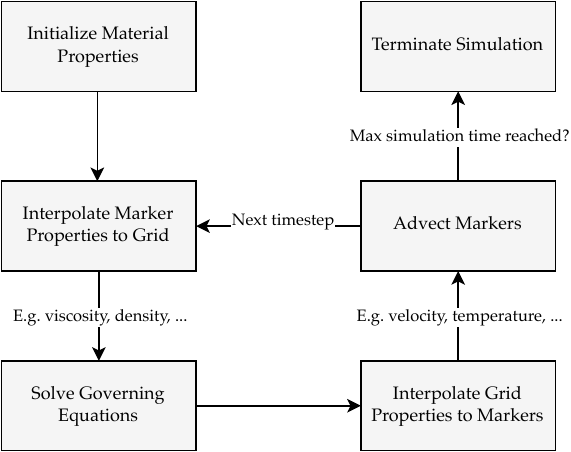}
\caption{High-level simulation loop used in the marker-in-cell method. Material properties such as viscosity and density are initialized on Lagrangian markers and interpolated to the Eulerian grid. The governing equations are solved on the grid, and the resulting properties (e.g., velocity, temperature) are interpolated back to the markers. The markers are then advected, and the loop continues until the maximum simulation time is reached.}
\label{fig:mic-loop-diagram}
\end{figure}

A typical simulation cycle proceeds through the following steps: (1) marker properties are interpolated to the grid to obtain nodal fields such as density and viscosity; (2) the discretized governing equations are solved on the Eulerian grid to compute solution fields, for example velocity and pressure in the Stokes problem; (3) relevant fields such as velocity and temperature are interpolated back to the markers; and (4) the markers are advected forward in time using an explicit time integration scheme. This loop is repeated at each time step, ensuring that the marker properties evolve consistently with the underlying flow field. Figure~\ref{fig:mic-loop-diagram} provides a schematic overview of this simulation pipeline.

The MIC method is well suited for geodynamic simulations because it enables accurate advection of complex, heterogeneous material distributions with minimal numerical diffusion~\cite{gerya_mic}. By storing material properties on Lagrangian markers, the method preserves sharp material interfaces over long timescales. Decoupling advection from the solution of the governing equations also improves modularity, allowing the advection solver and the PDE solver to be implemented and optimized independently. These features make MIC an effective approach for modeling lithospheric and mantle dynamics involving strong spatial variations in viscosity, density, and composition.

\section{Interpolation Between Markers and Grid}
\label{sec:marker-interpolation}
\begin{figure}[h]
\centering
\includegraphics[trim={0.2cm 0 -1.0cm 0.5cm}, clip, width=0.8\linewidth]{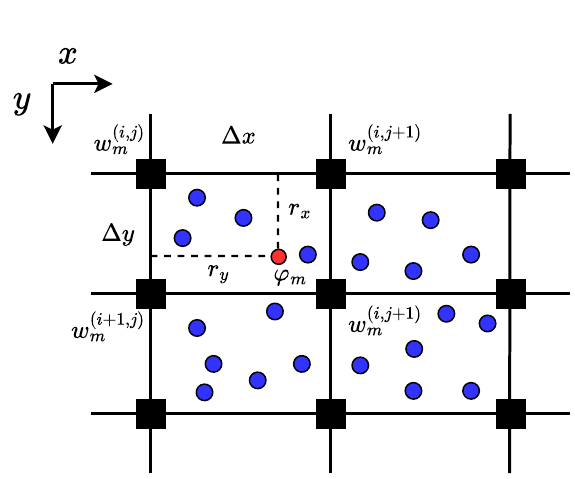}
\caption{Schematic of linear interpolation between markers (blue circles) and grid nodes (black squares) in two dimensions. The red marker is located inside a grid cell bounded by four nodes. Distances $r_x$ and $r_y$ are used to compute interpolation weights.}
\label{fig:interp-schematic}
\end{figure}

Figure~\ref{fig:interp-schematic} illustrates the interpolation scheme used to transfer quantities between markers and grid nodes. Pyroclast uses a weighted linear (bilinear in 2D, trilinear in 3D) interpolation kernel to perform both marker-to-grid and grid-to-marker transfers. Let $\phi$ be a scalar field stored on markers, such as density, viscosity, or temperature.

\subsubsection*{Marker-to-grid interpolation}
To transfer a quantity from markers to the grid:
\begin{enumerate}
    \item For each marker, identify the reference grid node at the top-left corner of the surrounding cell.
    \item Compute the distances $r_x$ and $r_y$ from the marker to the reference node.
    \item Compute interpolation weights for the four surrounding grid nodes:
    \begin{align}
        w_m^{(i,j)}     &= (1 - r_x/\Delta x)(1 - r_y/\Delta y), \\
        w_m^{(i+1,j)}   &= (r_x/\Delta x)(1 - r_y/\Delta y), \\
        w_m^{(i,j+1)}   &= (1 - r_x/\Delta x)(r_y/\Delta y), \\
        w_m^{(i+1,j+1)} &= (r_x/\Delta x)(r_y/\Delta y).
    \end{align}
    \item For each neighboring node, accumulate $w_m^{(i,j)}  \, \phi_m$ and $w_m^{(i,j)} $ into temporary arrays for weighted sum and normalization.
    \item After processing all markers, compute the final grid value at each node as the weighted average:
    \begin{equation}
        \phi^{(i,j)} = \frac{\sum_{m=1}^{n_m} w_m^{(i,j)} \phi_m}  {\sum_{m=1}^{n_m} w_m^{(i,j)}},
    \end{equation}
    where the sums are taken over all $n_m$ markers that contribute to \mbox{node \((i,j)\)}.
\end{enumerate}
To avoid race conditions during accumulation, the summation is typically implemented using atomic operations, scatter-add primitives, or MPI parallelism, depending on the target architecture. Although this type of interpolation is standard in marker-in-cell and particle-in-cell methods~\cite{gerya_mic}, it is important to present it here in detail because it plays a central role in the design of the distributed advection step. A detailed treatment of boundary and ghost node handling is given in Chapter~\ref{chap:mpi}, which covers MPI domain decomposition and distributed marker advection.

\subsubsection*{Grid-to-marker interpolation}
To interpolate a grid field back to markers:
\begin{enumerate}
    \item For each marker, identify the four surrounding nodes.
    \item Compute the same bilinear weights $w^{(i,j)}$ as above.
    \item Interpolate the field to the marker location as:
    \begin{equation}
        \phi_m = \sum_{(i, j)} w_m^{(i,j)} \phi^{(i,j)}
    \end{equation}
    where the sum over $(i,j)$ takes into account the four surrounding nodes.
\end{enumerate}
Since the weights naturally sum to one, no normalization is required. Parallel execution of this step is straightforward, as each marker can be processed independently by summing values from its four neighboring nodes with no risk of race conditions.

\section{Advection of Lagrangian Markers}

The advection of material properties in Pyroclast is performed by numerically integrating the characteristic trajectories defined by the velocity field. As introduced in Chapter~\ref{chap:equations}, the motion of a material parcel located at \mbox{$\mathbf{x}_0 = (x(t_0), y(t_0))$} is governed by the ordinary differential equation
\begin{equation}
\frac{d\mathbf{x}}{dt} = v(\mathbf{x}), \quad \mathbf{x}(t_0) = \mathbf{x}_0,
\label{eq:char_ode_repeat}
\end{equation}
where $v = (v_x, v_y)$ is the velocity field obtained from the solution of the incompressible Stokes system. Since Pyroclast employs operator splitting to decouple advection from the Stokes solve, the velocity field is treated as constant in time during each advection step.

Each Lagrangian marker in the simulation represents such a parcel and carries a set of physical properties, including for example viscosity, density, and temperature. These quantities may in general evolve over time due to mechanical or thermal processes. However, because advection is performed separately from the other governing equations, marker properties are held fixed while integrating~\eqref{eq:char_ode_repeat}. In other words, each property $\phi$ is treated as constant along the trajectory during the advection step, and any intrinsic change in the material is handled in a separate update step.

\subsection{Time Step Selection and Stability}

Pyroclast uses adaptive time stepping to control the size of each advection step. This is essential in multiphysics simulations, where the range of relevant time scales can vary significantly during the simulation. In practice, it is not feasible to choose a fixed global time step that satisfies all physical and numerical constraints.

Two main classes of criteria are used to restrict the advection time step. The first consists of \emph{physics-based constraints}, which ensure accurate resolution of physical processes. These include limits such as the maximum allowable particle displacement or temperature change within a single time step. The second consists of \emph{grid-based constraints}, which arise from numerical stability requirements and implementation details. For example, the advection step must be limited so that markers remain within the local ghost region in MPI-parallel simulations, avoiding the need for intermediate communication. Both types of constraints are evaluated at runtime and used to compute the maximal admissible step size based on a CFL-like condition.

\subsection{Advection Equation Integration Schemes}
To accommodate different accuracy and performance requirements, Pyroclast supports three types of integration schemes for marker advection: (i) the forward Euler method, (ii) explicit Runge-Kutta methods, and (iii) a locally polynomial integration scheme based on a Taylor expansion of the velocity field around each marker. These methods differ in cost, accuracy, and implementation complexity, especially in parallel environments. Each approach is described in detail in the following sections.

\subsubsection{Euler Integration}
The simplest integration scheme available in Pyroclast is the explicit forward Euler method. This method provides a first-order accurate solution of the trajectory equation~\eqref{eq:char_ode_repeat} and is given by
\begin{equation}
\mathbf{x}^{n+1} = \mathbf{x}^{n} + \Delta t \, v(\mathbf{x}^{n}),
\label{eq:euler_advection}
\end{equation}
where $\mathbf{x}^{n}$ denotes the marker position at the beginning of the time step, and $\Delta t$ is the time increment. The velocity $v(\mathbf{x}^{n})$ is evaluated at the marker position by trilinear interpolation of the grid-based velocity field.

The forward Euler method is computationally inexpensive since it requires only one velocity evaluation per time step. Its low order of accuracy makes it sensitive to time step size and may lead to significant integration error over long simulations. To improve accuracy while preserving simplicity, the Euler method can be used with subcycling, where the global time step is divided into several smaller substeps. This composite update approach can be useful in distributed-memory implementations, as it simplifies the logic for marker exchange and velocity interpolation compared to multi-stage Runge-Kutta methods.

\subsubsection{Runge-Kutta Integration}

To achieve higher spatial integration accuracy, Pyroclast supports explicit Runge-Kutta (RK) methods of second and fourth order. These methods integrate the trajectory equation~\eqref{eq:char_ode_repeat} by evaluating the velocity field at multiple intermediate points within each time step~\cite{butcher_runge_kutta}. The general form of an explicit Runge-Kutta method is
\begin{equation}
\mathbf{x}^{n+1} = \mathbf{x}^{n} + \Delta t \sum_{i=1}^{s} b_i k_i,
\qquad
k_i = v\!\left(\mathbf{x}^{n} + \Delta t \sum_{j=1}^{i-1} a_{ij} k_j \right),
\label{eq:rk_general}
\end{equation}
where $s$ is the number of stages and $(a_{ij}, b_i)$ are the coefficients defining the specific RK scheme~\cite{butcher_runge_kutta}.

In practice, Pyroclast implements the second-order Heun method and the classical fourth-order Runge-Kutta (RK4) method.  
The second-order scheme evaluates the velocity twice per step, once at the beginning and once at a predicted end position:
\begin{align}
\mathbf{x}^{*} &= \mathbf{x}^{n} + \Delta t \, v(\mathbf{x}^{n}), \\
\mathbf{x}^{n+1} &= \mathbf{x}^{n} + \tfrac{1}{2}\Delta t \, [v(\mathbf{x}^{n}) + v(\mathbf{x}^{*})].
\label{eq:heun_method}
\end{align}

The classical fourth-order RK scheme performs four velocity evaluations per step:
\begin{align}
k_1 &= v(\mathbf{x}^{n}), \\
k_2 &= v\!\left(\mathbf{x}^{n} + \tfrac{1}{2}\Delta t\,k_1\right), \\
k_3 &= v\!\left(\mathbf{x}^{n} + \tfrac{1}{2}\Delta t\,k_2\right), \\
k_4 &= v\!\left(\mathbf{x}^{n} + \Delta t\,k_3\right), \\
\mathbf{x}^{n+1} &= \mathbf{x}^{n} + \tfrac{\Delta t}{6}\,(k_1 + 2k_2 + 2k_3 + k_4).
\label{eq:rk4_method}
\end{align}
This method achieves fourth-order accuracy in space while maintaining good stability properties for smooth velocity fields~\cite{butcher_runge_kutta}. Although RK4 requires four velocity evaluations per step, its higher accuracy reduces cumulative trajectory drift and interpolation errors.

All Runge-Kutta schemes used here are explicit and therefore straightforward to parallelize, as each marker trajectory can be advanced independently. The choice of integration order represents a trade-off between computational efficiency and accuracy in capturing particle trajectories, particularly in regions of strong spatial velocity gradients.

\subsubsection{Locally Polynomial Integration}

An alternative approach is what we termed the locally polynomial integration scheme.  
Rather than evaluating the velocity field at multiple intermediate positions as in Runge-Kutta methods, this technique approximates the velocity field around each marker by a local Taylor expansion and integrates it analytically over the time step.  
Expanding the velocity around the marker position $\mathbf{x}^{n}$ up to quadratic order gives
\begin{equation}
v(\mathbf{x}) \approx v_0
+ \mathbf{J}(\mathbf{x} - \mathbf{x}^{n})
+ \tfrac{1}{2}(\mathbf{x} - \mathbf{x}^{n})^{\!\mathsf{T}} \mathbf{H}(\mathbf{x} - \mathbf{x}^{n}),
\label{eq:velocity_taylor}
\end{equation}
where $v_0 = v(\mathbf{x}^{n})$, $\mathbf{J} = \nabla v(\mathbf{x}^{n})$ is the velocity gradient tensor, and $\mathbf{H}$ is the tensor of second derivatives of $v$.  
This representation is consistent with the stress-conservative discretization of the stokes equations discussed in the following section, which enables stable computation of first and second order velocity derivatives.

The marker trajectory satisfies the characteristic relation
\begin{equation}
\mathbf{x}^{n+1} = \mathbf{x}^{n} + \int_{0}^{\Delta t} v(\mathbf{x}(t))\,dt.
\end{equation}
To evaluate this integral, the locally expanded velocity field~\eqref{eq:velocity_taylor} is substituted into the expression above.  
Since the velocity field is assumed steady during each advection step, only its spatial variation influences the trajectory curvature.  
Within the short time interval $\Delta t$, the marker path can therefore be approximated as a straight segment with constant velocity,
\begin{equation}
\mathbf{x}(t) \approx \mathbf{x}^{n} + t\,v_0.
\end{equation}
Inserting this trajectory into the Taylor expansion yields an explicit expression for the velocity experienced by the marker as it moves:
\begin{equation}
v(\mathbf{x}(t)) \approx v_0 + t\,\mathbf{J}v_0 + \tfrac{1}{2}t^2(\mathbf{H} : v_0 v_0),
\end{equation}
where $(\mathbf{H} : v_0 v_0)_i = H_{ijk}v_{0,j}v_{0,k}$ denotes the contraction of the second-derivative tensor with the velocity vector twice. 
Integrating the above expression over $t \in [0, \Delta t]$ gives the analytic update
\begin{equation}
\mathbf{x}^{n+1} = \mathbf{x}^{n}
+ \Delta t\,v_0
+ \tfrac{1}{2}\Delta t^{2}\,\mathbf{J}v_0
+ \tfrac{1}{6}\Delta t^{3}\,(\mathbf{H} : v_0 v_0)
+ \mathcal{O}(\Delta t^{4}).
\label{eq:lpi_update}
\end{equation}

Truncating after the second term gives a second-order accurate update, while including the quadratic term improves accuracy further at negligible cost when $\mathbf{H}$ is already available. 

This locally polynomial approach offers two main advantages.  
First, it increases the accuracy of the single-stage Euler update without the additional velocity interpolations required by multi-stage schemes.  
Second, it is particularly advantageous in distributed-memory (MPI) simulations: since the intermediate velocity evaluations are replaced by a local analytic expansion, markers that temporarily leave the local domain during the integration step can be advanced correctly without communication of remote velocity values.  

\section{Discretization of the Governing Equations}
\label{sec:discretization}
Pyroclast discretizes the incompressible Stokes equations using a finite-difference scheme on a fully staggered Cartesian grid. The discretization is stress-conservative and follows an established strategy used in other MIC geodynamic solvers, including I2ELVIS and I3ELVIS~\cite{gerya_i2elvis, gerya_numerical_solution_of_stokes}. Although the approach is not novel, it provides a robust and accurate foundation for simulating slow viscous flows in strongly heterogeneous media. The use of a staggered grid ensures that all operators are discretized using central-differences, drastically improving stability of the model.

The remainder of this section describes the structure of the staggered grid and outlines the stress-conservative discretization of the governing equations. These two components form the basis for the mechanical solver implemented in Pyroclast.

\subsection{Staggered Grid Layout}
\label{sec:staggered-grid-layout}
\begin{figure*}[h!]
\centering
\includegraphics[width=\textwidth, trim={0.2cm 1.3cm 0.3cm 0.4cm}, clip]{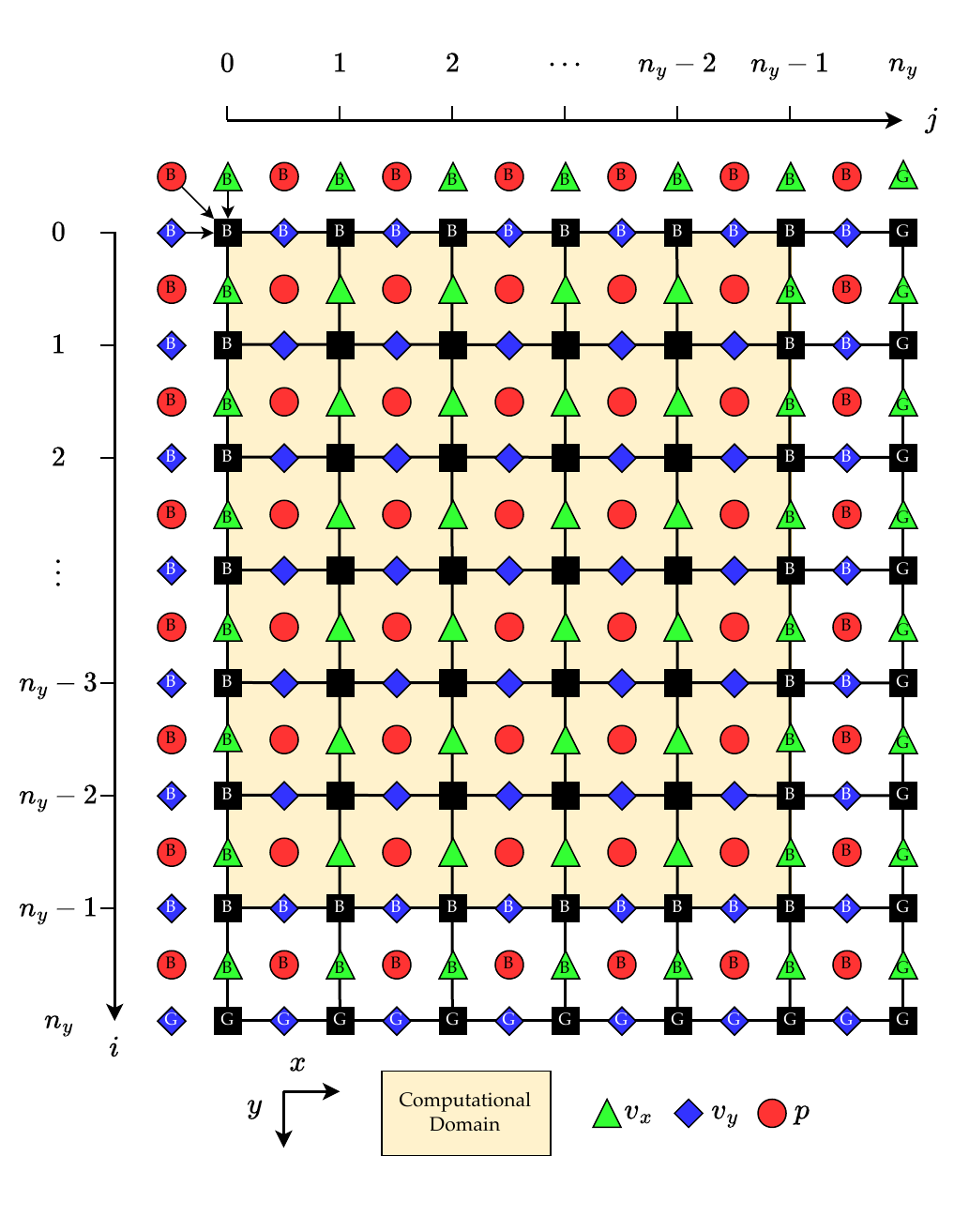}
\caption{Staggered grid arrangement used for the finite-difference discretization of the incompressible Stokes equations. The velocity components $v_x$ (green triangles) and $v_y$ (blue diamonds) are defined at cell faces, while pressure $p$ (red circles) is stored at cell centers. Black squares mark the locations of basic nodes that carry auxiliary quantities such as viscosity and density. Boundary nodes (\textbf{B}) extend the computational domain to enforce boundary conditions. Ghost nodes (\textbf{G}) hold no physical information, but ensure that all quantities can be represented by arrays of size $n_y + 1 \times n_x + 1$, greatly simplifying stencil-based operations and interpolation.
This layout ensures a stress-conservative discretization where all spatial derivatives are approximated using central differences. The figure is adapted from~\cite{Gerya_stokes_flow}.}
\label{fig:staggered-grid-diagram}
\end{figure*}

The spatial arrangement of the primary variables for the Stokes problem is shown in \Cref{fig:staggered-grid-diagram}. Each pressure value $p_{i,j}$ is surrounded by four velocity components $v_x$ and $v_y$, positioned on adjacent cell faces. Black squares denote basic grid nodes, which store auxiliary quantities such as viscosity and density. Boundary nodes (marked \textbf{B}) are used to enforce conditions at the physical domain boundaries. For example, imposing $v_y = 0$ on the west and east boundaries is achieved by mirroring the adjacent interior value with a negative sign, resulting in a zero average at the true boundary location. Ghost nodes (marked \textbf{G}) are appended on the East and South sides of the domain and do not store physical values. Their purpose is to simplify marker-to-grid interpolation and prevent out-of-bounds accesses when applying finite-difference operators.

Pyroclast adopts a row-major indexing convention. Grid indices $i$ and $j$ range from $0$ to $n_y$ and $0$ to $n_x$, respectively. Each staggered quantity shares the index of the lower-right nodal point it belongs to, as illustrated in the top-left corner of \Cref{fig:staggered-grid-diagram}.

\FloatBarrier

\subsection{Stress-Conservative Finite Differences}
\label{sec:stress_conservative_fd}
The discretization of the momentum equations in Pyroclast follows a stress-conservative formulation. Instead of directly expressing the viscous terms in terms of velocity gradients, the equations are first written in terms of the deviatoric stress components, as shown in Equations~\eqref{eq:xmom}–\eqref{eq:ymom} from Chapter~\ref{chap:equations}. Each stress component is then discretized independently using finite differences and expressed in terms of velocity through the constitutive relationship:
\begin{equation}
\sigma'_{ij} = 2 \eta \dot{\epsilon}_{ij} = \eta
\left( 
\frac{\partial v_i}{\partial x_j} + \frac{\partial v_j}{\partial x_i}
\right),
\end{equation}
with the viscosity $\eta$ evaluated at the location where the corresponding stress component is defined. This ensures that the discrete divergence of the stress tensor is computed consistently with the spatial placement of the velocity and pressure unknowns on the staggered grid.

\begin{figure}[h!]
\centering
\includegraphics[width=0.6\textwidth, trim={0cm 1.0cm 0cm 0cm}, clip]{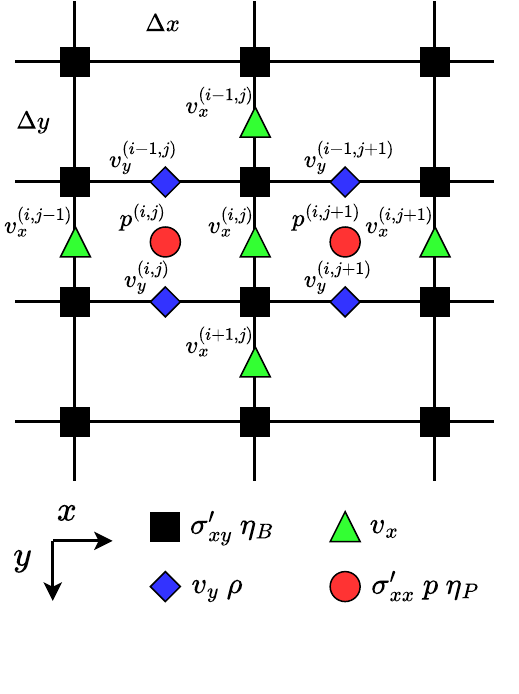}
\caption{Stencil for the discretization of the $x$‑momentum equation on a staggered grid. The figure shows the indexing of the primary solution variables ($v_x$, $v_y$, and $p$); all other quantities stored at the same nodal locations (e.g., $\sigma'_{xx}$, $\sigma'_{xy}$, $\eta_P$, $\eta_B$, $\rho$) follow the same indexing convention. Velocity components $v_x$ and $v_y$ are defined on cell faces, pressure $p$ at cell centers, and the deviatoric stress components $\sigma'_{xx}$ and $\sigma'_{xy}$ at their respective natural positions. Viscosity is stored at the same nodes as the corresponding stress components ($\eta_P$ at pressure nodes and $\eta_B$ at basic nodes), ensuring a consistent stress‑conservative discretization. The figure is adapted from~\cite{gerya_numerical_solution_of_stokes, ferrari_3d_blocking}.}
\label{fig:xmomentum-stencil}
\end{figure}

To illustrate this approach, consider the $x$‑momentum equation~\eqref{eq:xmom}, which can be written in discrete form as:
\begin{equation}
\frac{\partial \sigma_{xx}'}{\partial x} + \frac{\partial \sigma_{xy}'}{\partial y}
- \frac{\partial p}{\partial x} = 0.
\end{equation}
On the staggered grid, this equation is centered at the location of $v_x$, defined on vertical cell faces. The stencil for this discretization is shown in Figure~\ref{fig:xmomentum-stencil}. The first term is approximated using finite differences of the normal stresses $\sigma_{xx}'$, defined on pressure nodes, while the second term uses the shear stresses $\sigma_{xy}'$, defined at basic nodes:
\begin{align}
\left(\frac{\partial \sigma_{xx}'}{\partial x}\right)_{i,j}
&\approx
\frac{\sigma_{xx}'^{(i,j+1)} - \sigma_{xx}'^{(i,j)}}{\Delta x}, \\
\left(\frac{\partial \sigma_{xy}'}{\partial y}\right)_{i,j}
&\approx
\frac{\sigma_{xy}'^{(i,j)} - \sigma_{xy}'^{(i-1,j)}}{\Delta y}.
\end{align}
Each stress component is then expressed in terms of the local velocity gradients using the viscosity defined at its corresponding location:
\begin{align}
\sigma_{xx}'^{(i,j)} &= 2\,\eta_P^{(i,j)}
\left(\frac{v_x^{(i,j)} - v_x^{(i,j-1)}}{\Delta x}\right), \\
\sigma_{xy}'^{(i,j)} &= \eta_B^{(i,j)}
\left(
\frac{v_x^{(i+1,j)} - v_x^{(i,j)}}{\Delta y}
+
\frac{v_y^{(i,j+1)} - v_y^{(i,j)}}{\Delta x}
\right).
\end{align}
Here, $\eta_P$ denotes the viscosity stored at pressure points and $\eta_B$ the viscosity defined at basic grid nodes. The same procedure is applied to the $y$‑momentum equation, using the appropriate stress components and their natural grid locations.

This formulation ensures that each component of the stress tensor is discretized using local velocity differences and the viscosity evaluated at the same position, thereby conserving stress across cell interfaces and maintaining a physically consistent discretization.

\section{Fast Solver for the Stokes System}
\label{sec:stokes_solver}

The discretization of the Stokes equations described in the previous section gives rise to a large, sparse linear system with a saddle-point structure. This type of system is particularly ill-conditioned and presents a major computational challenge in geodynamic modeling. Since the objective of Pyroclast is to provide a scalable simulation framework capable of running efficiently on distributed and GPU-accelerated architectures, direct methods based on sparse LU or QR factorization~\cite{ferrari_sparse_solvers} are not practical. Instead, we rely on a fast, fully iterative solver specifically designed for the Stokes system.

A \textit{saddle-point} system refers to a class of block-structured linear systems of the general form~\cite{saddle_point}:
\begin{equation}
\label{eq:saddle_point_def}
\begin{bmatrix}
A & B_1^\top \\
B_2 & -C
\end{bmatrix}
\begin{bmatrix}
x \\
y
\end{bmatrix}
=
\begin{bmatrix}
f \\
g
\end{bmatrix},
\quad \text{or compactly} \quad
\mathcal{A}u = b,
\end{equation}
where $A$, $B_1$, and $B_2$ are nonzero submatrices and one or more of the following properties hold:
\begin{enumerate}
    \item $A$ is symmetric ($A = A^\top$);
    \item The symmetric part $H = \tfrac{1}{2}(A + A^\top)$ is positive semidefinite;
    \item $B_1 = B_2 = B$;
    \item $C$ is symmetric positive semidefinite;
    \item $C = 0$.
\end{enumerate}
Such systems typically arise from the first-order optimality conditions of constrained minimization problems. For example, consider the quadratic program:
\begin{equation}
\min_x \; J(x) = \frac{1}{2} x^\top A x - f^\top x \quad \text{subject to} \quad Bx = g,
\end{equation}
with associated Lagrangian:
\begin{equation}
\mathcal{L}(x, y) = \frac{1}{2} x^\top A x - f^\top x + y^\top (Bx - g),
\end{equation}
where $y$ is a vector of Lagrange multipliers. The saddle-point of $\mathcal{L}$ satisfies the linear system~\eqref{eq:saddle_point_def}~\cite{saddle_point}.

In the context of the incompressible Stokes equations, the velocity field $v$ and pressure $p$ form the unknown vector $u = [v, p]^\top$, with the pressure acting as a Lagrange multiplier that enforces incompressibility. The discrete Stokes system can therefore be written in the canonical saddle-point form:
\begin{equation}
\label{eq:stokes_saddle}
\begin{bmatrix}
L & G \\
D & 0
\end{bmatrix}
\begin{bmatrix}
v \\
p
\end{bmatrix}
=
\begin{bmatrix}
f \\
0
\end{bmatrix},
\end{equation}
where $L$ is the discrete viscosity-weighted Laplacian operator, and $G$ and $D$ are the discrete gradient and divergence operators, respectively. The right-hand side $f$ represents the body force vector. The zero block on the pressure diagonal encodes the incompressibility constraint and is the primary source of indefiniteness in the system. This coupling between velocity and pressure makes the system very difficult to solve without a preconditioner specifically designed for this problem.

Since the system in Equation~\eqref{eq:stokes_saddle} is the discrete equivalent of the incompressible Stokes equations, each block in the system corresponds to a finite-difference approximation of a continuous differential operator acting on the velocity or pressure fields. In operator notation, the discrete terms correspond to:
\begin{align}
L v &= \nabla_h \cdot \sigma' = \nabla_h \cdot \left( \eta \left( \nabla_h v + \nabla_h v^\top \right) \right), \label{eq:laplacian_discrete} \\
G p &= - \nabla_h p, \label{eq:gradient_discrete} \\
D v &= \nabla_h \cdot v = 0, \label{eq:divergence_discrete}
\end{align}
where \( \nabla_h \) denotes the discrete gradient operator. The matrix \( L \) corresponds to the discrete viscosity-weighted Laplacian, written here in terms of velocity derivatives and variable viscosity. The matrix \( G \) represents the discrete gradient, and \( D \) is the discrete divergence. All operators are implemented using central finite differences on the staggered grid described previously. This discretization preserves the symmetry and coupling structure of the continuous Stokes system at the discrete level.

The remainder of this chapter presents the numerical strategy implemented in Pyroclast for the efficient solution of this saddle-point system. We first introduce Uzawa iteration, which decouples the velocity and pressure subproblems through a Schur complement reduction. We then describe the matrix-free geometric multigrid solver used for the velocity update and discuss the smoother and transfer operators. Finally, we outline the use of black-box accelerators to improve convergence of the overall iteration.

\subsection{Uzawa Iteration}
\label{sec:uzawa_iteration}

The derivation of the Uzawa iteration presented here is taken directly from the work presented in ~\cite{ferrari_3d_blocking}, which also describes part of the numerical framework implemented in Pyroclast. Originally developed in the context of concave programming~\cite{uzawa_book}, the Uzawa algorithm has become a widely used strategy for solving saddle-point systems, including those arising from the discretization of the incompressible Stokes equations~\cite{saddle_point}.

The core idea of Uzawa iteration is to alternate between updating the velocity and pressure fields. At each iteration, the velocity is updated by solving the discretized momentum equations using the current estimate of pressure. The updated velocity is then used to compute a pressure correction derived from the Schur complement of the system, enforcing the incompressibility constraint. This decoupling of velocity and pressure makes the method especially attractive for matrix-free implementations and naturally compatible with the geometric multigrid solvers described in the next section.

Starting from the discrete Stokes system in block form:
\begin{equation}
\label{eq:stokes_block_system}
\begin{bmatrix}
L & G \\
D & 0
\end{bmatrix}
\begin{bmatrix}
v \\
p
\end{bmatrix}
=
\begin{bmatrix}
f \\
0
\end{bmatrix},
\end{equation}
we extract the first row, corresponding to the discrete momentum balance:
\begin{equation}
\label{eq:momentum_eq}
L v + G p = f.
\end{equation}
Solving this equation for \( v \) gives:
\begin{equation}
\label{eq:velocity_elimination}
v = L^{-1} (f - G p).
\end{equation}
Substituting this expression into the incompressibility constraint \( D v = 0 \) yields the Schur complement equation for pressure:
\begin{equation}
\label{eq:schur_system}
D L^{-1} G p = D L^{-1} f,
\end{equation}
where the matrix \( S = D L^{-1} G \) is known as the Schur complement of the Stokes system.

Solving the system~\eqref{eq:schur_system} directly is not tractable, and in fact even forming \( S \) or \( L^{-1} \) explicitly is impossible. Instead, we apply a Richardson fixed-point iteration to the pressure equation:
\begin{equation}
\label{eq:richardson_pressure}
p^{k+1} = p^k + \alpha \left( D L^{-1} (f - G p^k) \right),
\end{equation}
where \( \alpha \) is a relaxation parameter. 

Next, we define:

\begin{equation}
\label{eq:velocity_update}
v^{k+1} := L^{-1}(f - G p^k),
\end{equation}
and use it to simplify the pressure update:
\begin{equation}
\label{eq:pressure_update_precond}
p^{k+1} = p^k + \alpha\, D v^{k+1}.
\end{equation}

Preconditioning the pressure update in Eq.~\eqref{eq:pressure_update_precond} is essential to achieve fast convergence of the Uzawa iteration. The goal is to approximate the inverse of the Schur complement,
\begin{equation}
S = D L^{-1} G,
\end{equation}
with a computationally inexpensive surrogate \( \widetilde{S}^{-1} \approx S^{-1} \). To guide this construction, we examine the dimensional scaling of the discrete operators defined in Eqs.~\eqref{eq:laplacian_discrete}-\eqref{eq:divergence_discrete}. The discrete gradient and divergence operators scale as \( G \sim \mathcal{O}(1/\Delta x) \) and \( D \sim \mathcal{O}(1/\Delta x) \), while the inverse of the discrete viscosity weighted Laplacian operator scales as \( L^{-1} \sim \mathcal{O}(\Delta x^{2} / \eta) \). Combining these relations yields
\begin{equation}
S = D L^{-1} G \sim \frac{1}{\Delta x} \cdot \frac{\Delta x^{2}}{\eta} \cdot \frac{1}{\Delta x} = \mathcal{O}\!\left(\frac{1}{\eta}\right).
\end{equation}
This dimensional analysis indicates that the Schur complement is primarily governed by the inverse viscosity, suggesting that an effective preconditioner should reproduce this scaling. Accordingly, a simple diagonal approximation
\begin{equation}
\widetilde{S} = \frac{1}{\eta} \quad \Longrightarrow \quad \widetilde{S}^{-1} = \eta
\end{equation}
This simple preconditioner captures the dominant behavior of the Schur complement at minimal cost. Although it neglects coupling between pressure degrees of freedom, it preserves the correct coefficient scaling and has proven effective in practice for stabilizing Uzawa iteration. This strategy is not new and has been widely used in the context of Stokes solvers with variable viscosity~\cite{gerya_multigrid, ptatin3d}.

Applying this preconditioner to the pressure update yields the full Uzawa iteration in its final form:
\begin{equation}
\label{eq:uzawa_iteration}
\begin{cases}
v^{k+1} := L^{-1}(f - G p^k), \\
p^{k+1} := p^k + \alpha\, \eta\, D v^{k+1},
\end{cases}
\end{equation}
where \( \eta \) is evaluated pointwise at pressure nodes. The divergence residual is scaled by the local viscosity to ensure consistent coupling between the momentum and continuity equations, even in the presence of strong viscosity variations.

This formulation defines the Uzawa iteration: at each step, the velocity is updated by solving the discrete momentum equation using the current pressure iterate, and the pressure is then corrected using the divergence of the updated velocity field.

The Uzawa algorithm assumes that the inverse of the velocity block \( L^{-1} \) can be applied exactly. In practice, this is not possible, particularly for large systems. Pyroclast approximates the action of \( L^{-1} \) using an efficient geometric multigrid solver tailored to the velocity subsystem. This results in an inexact Uzawa iteration scheme, where the velocity update is solved approximately at each step. Despite this, the method remains effective and scalable, provided that the velocity solver reduces the residual sufficiently. The next section describes the mathematcial aspects of this multigrid solver in detail.

\subsection{Pressure Nullspace Elimination}
\label{sec:pressure_nullspace}

The incompressible Stokes equations, as introduced in Chapter~\ref{chap:equations}, include only the pressure gradient term in the momentum balance. As a result, the absolute value of pressure is not physically constrained, only pressure differences influence the flow field. Mathematically, this implies that the discrete Stokes system has a one-dimensional nullspace corresponding to the addition of an arbitrary constant to the pressure field:
\begin{equation}
p \;\rightarrow\; p + C, \quad C \in \mathbb{R}.
\end{equation}
The presence of this nullspace renders the coefficient matrix singular, since any constant offset in pressure leaves the velocity field unchanged. While the physical solution remains unique up to this constant, the singularity must be handled explicitly to ensure numerical stability and convergence of the iterative solver.

To eliminate this pressure nullspace, a single constraint must be imposed to fix the pressure reference level. In Pyroclast, we enforce a zero-mean pressure condition:
\begin{equation}
\langle p \rangle = \frac{1}{V_\Omega} \int_\Omega p\, dV = 0,
\end{equation}
where \(V_\Omega\) is the total domain volume and \(\langle p \rangle\) denotes the volume-weighted mean pressure.

For a discrete grid with cell volumes \(V_i = \Delta x \Delta y\) or respectively \(V_i = \Delta x \Delta y \Delta z\), the condition is expressed as
\begin{equation}
\langle p \rangle_h = 
\frac{\sum_{i=1}^{n_p} V_i\, p_i}{\sum_{i=1}^{n_p} V_i} = 0.
\end{equation}
Accordingly, after each Uzawa pressure correction step, the mean pressure is subtracted from the field:
\begin{equation}
p^{k+1} \;\leftarrow\;
p^{k+1} - 
\frac{\sum_{i=1}^{n_p} V_i\, p_i^{k+1}}
     {\sum_{i=1}^{n_p} V_i}.
\label{eq:pressure_normalization}
\end{equation}
Here \(V_i\) represents the local control volume associated with pressure node \(i\), and \(n_p\) is the total number of pressure unknowns.

This normalization removes the nullspace component without affecting the physical pressure gradients that drive the flow. It ensures that the pressure correction step in Eq.~\eqref{eq:uzawa_iteration} remains well-conditioned and that the iterative solver converges reliably. The zero-mean condition is mathematically equivalent to fixing the pressure at a single node, but it preserves full symmetry of the system and avoids introducing artificial boundary effects. On a uniform grid, Eq.~\eqref{eq:pressure_normalization} simplifies to a simple arithmetic mean subtraction,
\begin{equation}
p^{k+1} \leftarrow p^{k+1} - 
\frac{1}{n_p}\sum_{i=1}^{n_p} p_i^{k+1}.
\end{equation}

\section{The Multigrid Method}
\label{sec:multigrid_method}

As previously stated, the Uzawa iteration introduced in the previous section requires repeated application of the inverse of the velocity operator \( L^{-1} \), which corresponds to a discretized, variable-viscosity Laplacian. Computing this inverse explicitly is infeasible in large-scale or matrix-free settings, and classical iterative methods such as Jacobi or Gauss-Seidel~\cite{yousef_saad_iterative_methods_linear_systems} converge too slowly to be practical. These methods efficiently damp high-frequency errors but are ineffective at removing smooth, low-frequency error modes, which dominate the residual after only a few iterations. As a result, achieving global convergence with standard smoothers alone would require an impractically large number of iterations.

Multigrid (MG) methods overcome this limitation by addressing error components at their natural spatial scales. The central idea is to accelerate convergence by combining smoothing on fine grids, which eliminates high-frequency error, with corrections on coarser grids to handle the low-frequency components. As noted by Hackbusch in~\cite{hackbusch_book_intro}, “the characteristic feature of multigrid iteration is its fast convergence; its convergence speed does not deteriorate as the discretisation is refined”. Consequently, the computational work required grows only in proportion to the number of unknowns, achieving not only optimal complexity but also an exceptionally small constant of proportionality, making multigrid methods among the most efficient solvers for large-scale elliptic problems.

In the context of this work, the multigrid method serves as an efficient way to approximate the action of \( L^{-1} \) and provides the computational backbone of the inexact Uzawa iteration. The method is implemented in a fully matrix-free, geometric form, exploiting the structured nature of the computational grid. This allows the solver to represent transfer operators, smoothers, and coarse-grid corrections directly in terms of stencils rather than assembled matrices, improving performance and scalability on both CPUs and GPUs.

The remainder of this section is dedicated to describing the multigrid method in more detail and how it is applied in Pyroclast to solve the velocity subsystem of the Stokes equations. We begin by illustrating the limitations of classical smoothers and the motivation for multi-level approaches. We then outline the structure of multigrid cycles, compare geometric and algebraic MG algorithms, and describe the implementation of restriction, prolongation, and smoothing operators. The section concludes by discussing the integration of the multigrid solver within the Uzawa iteration and the use of global acceleration techniques related to Krylov methods~\cite{yousef_saad_iterative_methods_linear_systems}.

\subsection{Smoothing and Information Propagation}
\label{sec:info_propagation}

Classical iterative methods such as Jacobi and Gauss-Seidel update each point in the solution grid using a small local stencil. These stencil operations are computationally efficient and straightforward to implement, but they are inherently limited by their spatial locality. Each update modifies only a single point based on its immediate neighbors, which restricts the rate at which information can propagate across the domain.

\begin{figure}[h!]
\centering
\includegraphics[width=0.8\textwidth, trim={0cm 0cm 1.5cm 0cm}, clip]{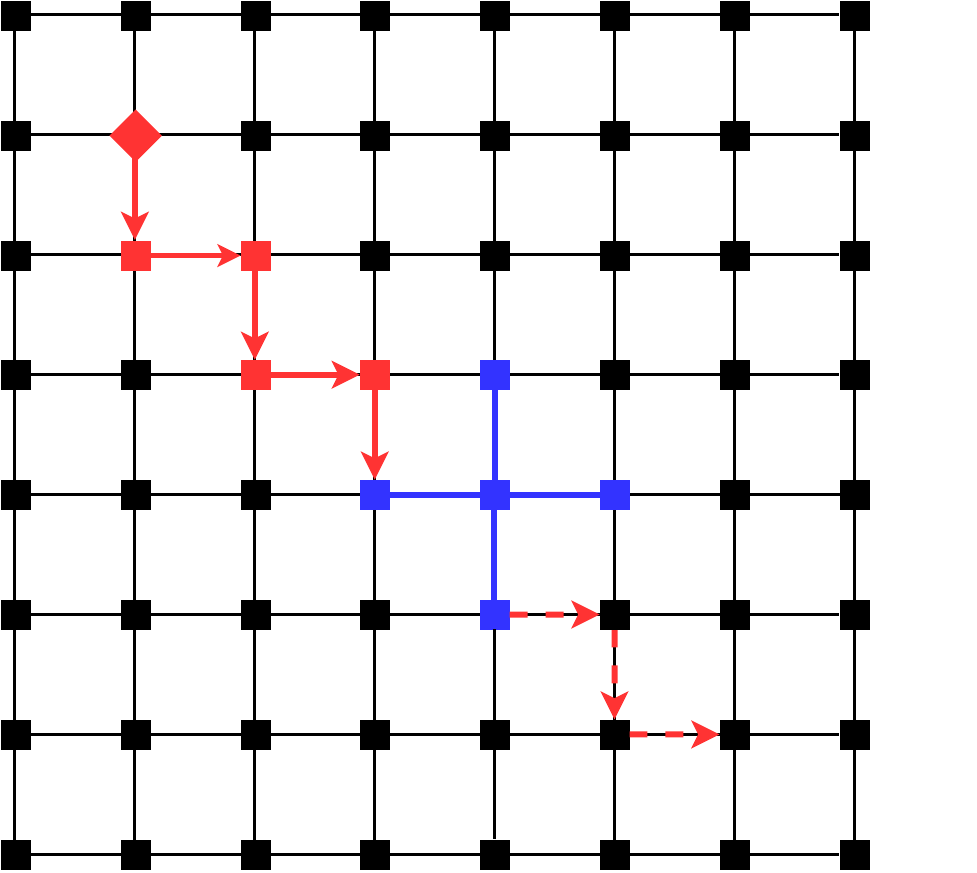}
\caption{
Illustration of information propagation using a local 5-point stencil (in blue) on a structured grid. The red diamond marks a source of information in the top-left corner. Each stencil application updates a node based on its immediate North, South, East, and West neighbors. In Jacobi iteration, all updates use values from the previous sweep, meaning information can only advance by one grid edge per iteration (red arrows). Propagating information across the domain therefore requires a number of sweeps proportional to the grid diameter. Gauss-Seidel iteration improves on this by immediately using updated values, allowing information to chain through the grid within a single sweep. However, this chaining leads to gradual contamination of the signal, and the method is inherently serial, making it unsuitable for parallel execution. Red-Black Gauss-Seidel (RBGS) enables parallelism by splitting the grid and processing it in two alternating sweeps, but this also limits information flow. As a result, RBGS behaves similarly to Jacobi, requiring approximately half as many sweeps (i.e., diameter~\( /~2 \)) to propagate information across the grid.
}
\label{fig:stencil_propagation}
\end{figure}

To illustrate this, consider a simple 5-point stencil on a regular Cartesian grid, as shown in Figure~\ref{fig:stencil_propagation}. The stencil updates a central node using its North, South, East, and West neighbors, and this process is repeated across the grid at each iteration. Suppose a perturbation or correction originates at a point in the top-left corner (marked by the red diamond). Since each iteration of the Jacobi method uses only values from the previous sweep, information from this source can travel at most one grid edge per iteration. Reaching a distant part of the domain therefore requires a number of iterations proportional to the Manhattan distance between the source and the target node.

In theory, Gauss-Seidel iteration improves on this by using updated values as soon as they are available within the same sweep. This allows information to chain through the grid more rapidly, potentially reaching the entire domain in a single pass. However, this daisy-chaining process is inherently serial and the propagated values are incrementally "contaminated" at each step, degrading the signal quality over long distances. Moreover, the sequential nature of Gauss-Seidel makes it unsuitable for modern parallel architectures.

To address this, Red-Black Gauss-Seidel (RBGS) introduces a two-color ordering that enables parallel updates. By splitting the grid into two interleaved sets, each color can be updated independently and in parallel. However, this also reintroduces a barrier to information flow: only alternating nodes are connected at each step. The result is a propagation behavior similar to Jacobi, with information spreading at most two grid edge every sweep. Consequently, RBGS still requires a number of iterations proportional to the grid diameter to propagate global corrections.

These limitations are fundamental to point-wise iterative methods and illustrate why such smoothers are effective only at reducing high-frequency errors. Low-frequency (long-wavelength) errors, which dominate the solution at later stages, require global information to resolve and cannot be efficiently corrected through local stencil updates alone. This motivates the use of multigrid methods, which accelerate convergence by targeting different error scales on different grid resolutions.

\subsection{Multigrid Cycles}
\label{sec:multigrid_cycles}
The core idea behind multigrid methods is to accelerate convergence by solving for error corrections across a hierarchy of grid resolutions.

Let \( A x = b \) be the linear system arising from discretization on the finest grid. Suppose \( x^\ell \) is an approximate solution on level \( \ell \). We define the error \( \delta x^\ell = x^* - x^\ell \), where \( x^* \) is the exact solution at that level. Substituting this into the system gives:
\begin{equation}
A^\ell (x^\ell + \delta x^\ell) = b^\ell
\quad \Rightarrow \quad
A^\ell \delta x^\ell = b^\ell - A^\ell x^\ell = r^\ell,
\label{eq:mg_residual}
\end{equation}
where \( A^\ell \) is the system matrix at level \( \ell \), \( b^\ell \) is the right-hand side, and \( r^\ell \) is the residual. Solving this residual equation yields a correction that can be added to the current solution:
\begin{equation}
x^\ell \gets x^\ell + \delta x^\ell.
\end{equation}

The multigrid method constructs a hierarchy of grids indexed by levels \( \ell = 1, \dots, N \), with \( \ell = 1 \) denoting the finest grid and \( \ell = N \) the coarsest. Residuals are transferred from fine to coarse levels (\( \ell \to \ell+1 \)) using the restriction operator \( R^\ell \), while corrections are interpolated back from coarse to fine levels (\( \ell+1 \to \ell \)) using the prolongation operator \( P^\ell \).

At each level, the correction equation is defined and solved approximately by transferring residuals to coarser levels and propagating corrections back to finer levels. The operations at each level can be expressed as:
\begin{equation}
\begin{aligned}
&\text{(1) Pre-smoothing:} & x^\ell &\gets S^\ell(x^\ell, A^\ell, b^\ell) \\
&\text{(2) Compute residual:} & r^\ell &= b^\ell - A^\ell x^\ell \\
&\text{(3) Restrict to coarse grid:} & b^{\ell+1} &= R^\ell r^\ell \\
&\text{(4) Solve coarse-grid system with MG:} & A^{\ell+1} \delta x^{\ell+1} &= b^{\ell+1} = R^\ell r^\ell \\
&\text{(5) Prolongate and apply correction:} & x^\ell &\gets x^\ell + P^\ell \delta x^{\ell+1} \\
&\text{(6) Post-smoothing:} & x^\ell &\gets S^\ell(x^\ell, A^\ell, b^\ell)
\end{aligned}
\label{eq:multigrid_levels}
\end{equation}
Here, \( R^\ell \) and \( P^\ell \) denote the restriction and prolongation operators between levels \( \ell \) and \( \ell+1 \), respectively. 

On the coarsest level \( N \), the correction equation
\begin{equation}
A^N \delta x^N = b^N = R^{N-1} r^{N-1}
\end{equation}
is either solved with a direct method~\cite{tim_davis_direct_methods_linear_systems, ferrari_sparse_solvers} or approximated by a few smoothing iterations. This terminates the MG coarse-grid solve recursion of step (4) and allows the prolongation phase to begin.

\begin{figure}[h!]
\centering
\includegraphics[width=0.87\textwidth]{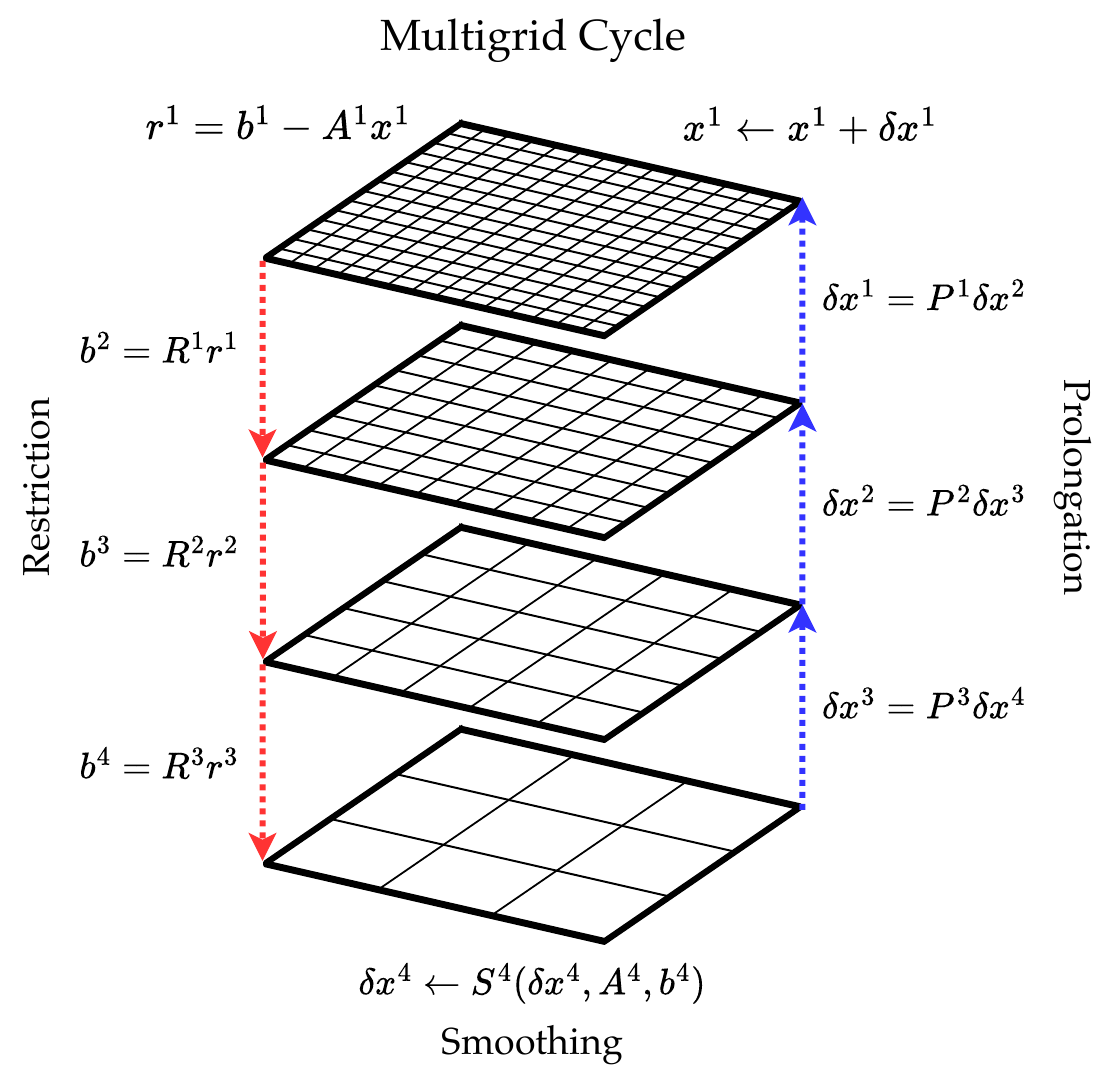}
\caption{Schematic representation of the multigrid V-cycle.
Residuals are successively restricted from fine to coarse grids (red arrows) to eliminate low-frequency errors efficiently. On the coarsest grid, the system is smoothed or solved with a direct solver. The correction is then prolongated back to finer grids (blue arrows), where it is used to correct the current solution. This hierarchical process accelerates convergence by combining smoothing of high-frequency errors on fine levels with coarse-grid corrections for low-frequency components.}
\label{fig:mg_hierarchy}
\end{figure}

The combination of these steps defines a \emph{V-cycle}, which recursively traverses down to coarser levels before returning to the fine grid. This process ensures that error components at all frequencies are efficiently reduced: high-frequency errors are smoothed on fine grids, while low-frequency errors are corrected on coarse grids. The structure of this recursive process is illustrated in Figure~\ref{fig:mg_hierarchy}, where each level represents a different grid resolution used to accelerate convergence.

This recursive formulation yields a solver whose convergence rate is independent of grid resolution. When implemented efficiently, multigrid achieves linear computational complexity \( \mathcal{O}(N) \), making it one of the most effective methods for solving large-scale elliptic systems.

\subsection{Geometric vs Algebraic Multigrid}
\label{sec:geom_vs_amg}
Multigrid methods can be broadly classified into two categories: geometric multigrid (GMG) and algebraic multigrid (AMG). The key distinction lies in how the hierarchy of coarse grids and the corresponding transfer operators are constructed.

In geometric multigrid, the grid structure is defined explicitly using the underlying geometry of the domain. Coarser grids are created by rediscretizing the underlying PDE at a lower spatial resolution, and restriction and prolongation operators are derived from this structure using interpolation stencils or averaging schemes. This approach requires knowledge of the grid layout and is best suited to problems with regular or structured meshes, where grid levels can be generated efficiently. Because the method operates directly on the geometric hierarchy, it is well-suited for matrix-free implementations and offers high performance and low memory usage~\cite{hackbusch_book}. 

Algebraic multigrid, by contrast, builds the multigrid hierarchy directly from the assembled system matrix without relying on any geometric information. The coarsening process and interpolation operators are inferred from the sparsity pattern and the matrix entries. This makes AMG more general and widely applicable to unstructured meshes or complex geometries where no obvious grid hierarchy exists. However, this generality comes at the cost of increased setup time, memory usage, and implementation complexity. AMG also typically requires access to the matrix in assembled form, making it less attractive for matrix-free solvers~\cite{amg_book}.

The main benefit of AMG lies in its black-box usability. Once a system matrix is provided, the solver can automatically construct a suitable hierarchy and apply the multigrid cycle without user intervention. This has led to the development of many different AMG libraries, such as PyAMG~\cite{pyamg2023}, AMGCL~\cite{amgcl}, and NVIDIA’s AmgX~\cite{amgx}, which are commonly integrated into finite element and finite volume codes with assembled matrices and irregular meshes.

In Pyroclast, we adopt a geometric multigrid approach for three reasons. First, the use of a regular Cartesian grid allows for straightforward generation of coarser levels and enables the definition of transfer operators directly in stencil form. Second, GMG integrates naturally with our matrix-free implementation, avoiding the need to assemble or store large sparse matrices. Third, and critically, it provides full control over how physical fields are transferred across grid levels. This is essential in our context, where variables such as velocity and pressure differ significantly in magnitude. By treating each field independently, GMG ensures that coarse-level representations preserve the structure and scaling of the underlying physics. These advantages make geometric multigrid the optimal choice for Pyroclast.

\subsection{Restriction and Prolongation Operators}
\label{sec:restriction_prolongation}

The transfer of residuals and corrections between adjacent levels in the multigrid hierarchy is performed using the restriction and prolongation operators, \( R^\ell \) and \( P^\ell \), respectively. Both rely on the same weighted bilinear interpolation scheme introduced in Section~\ref{sec:marker-interpolation}.

\subsubsection*{Prolongation}

The prolongation operator \( P^\ell \) interpolates coarse-grid corrections \( \delta x^{\ell+1} \) to the finer level \( \ell \), generating a fine-grid correction field \( \delta x^\ell \). In Pyroclast, this operation is implemented identically to the grid-to-marker interpolation routine: each fine node receives contributions from the four surrounding coarse nodes, weighted bilinearly based on their relative positions.

In compact form:
\begin{equation}
\delta x^\ell = P^\ell \delta x^{\ell+1}.
\end{equation}
And in pointwise form:
\begin{equation}
\delta x^{\ell,(p,q)} = \sum_{(i,j)} w^{(i,j)}_{(p,q)} \, \delta x^{\ell+1,(i,j)},
\end{equation}
where \( w^{(i,j)}_{(p,q)} \) are the bilinear interpolation weights. Since these weights always sum to one, no normalization is required. The operation is fully local, depends only on neighboring coarse nodes, and is trivially parallelizable.

\subsubsection*{Restriction}

The restriction operator \( R^\ell \) transfers the fine-grid residual \( r^\ell \) to the coarser level:
\begin{equation}
b^{\ell+1} = R^\ell r^\ell.
\end{equation}
This operation is structurally analogous to marker-to-grid interpolation, as it reduces data from many fine-grid points to fewer coarse-grid targets using bilinear weighting. Unlike prolongation, restriction presents a concurrency challenge: multiple fine-grid nodes may contribute to the same coarse node, leading to potential race conditions during parallel execution.

To avoid costly atomic operations, Pyroclast exploits the regular Cartesian structure of the grid hierarchy. Since all levels use structured layouts, even with Swiss-cross adaptive refinement~\cite{gerya_amr}, nodal coordinates \( (x^\ell, y^\ell) \) can be represented as separable 1D vectors. This enables efficient computation of index bounds via coordinate bisection and can be exploited to parallelize this operation.

Restriction proceeds in two stages. First, each coarse node \((i,j)\) accumulates contributions from nearby fine-grid residuals using bilinear weights. Second, the accumulated values are normalized:
\begin{equation}
r^{\ell+1,(i,j)} = 
\frac{
\displaystyle \sum_{(p,q)} w^{(i,j)}_{(p,q)} \, r^{\ell,(p,q)}
}{
\displaystyle \sum_{(p,q)} w^{(i,j)}_{(p,q)}
}.
\end{equation}

\begin{figure}[ht!]
\centering
\includegraphics[width=0.62\textwidth]{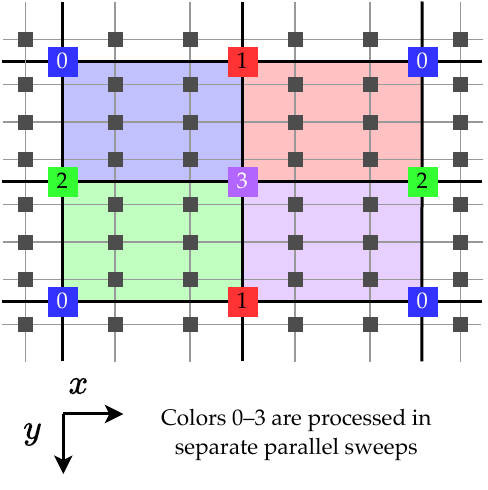}
\caption{Four-color decomposition of the coarse grid used for the parallel restriction operation. Each color (0--3) represents a group of coarse cells that can be processed simultaneously without race conditions, since no two cells of the same color share a node. Fine-grid values within each colored region are restricted to the coarse grid using weighted bilinear interpolation, with one separate parallel sweep per color. This partitioning enables a deterministic and thread-safe reduction without atomic operations.}
\label{fig:restriction-parallel}
\end{figure}

To ensure thread safety without atomics, the coarse grid is partitioned into four disjoint color groups, as shown in Figure~\ref{fig:restriction-parallel}. Each group contains a set of coarse nodes that do not share contributing fine nodes with each other, allowing for independent, race-free processing. This four-color decomposition enables parallel restriction using separate parallel sweeps per color group.

Figure~\ref{fig:restriction-parallel} illustrates this strategy, while Algorithm~\ref{alg:restriction_checkerboard} provides the complete implementation. The algorithm begins by bisecting the fine-grid coordinates to compute bounding indices for each coarse cell. During each color sweep, fine nodes within each bounded region are used to compute and accumulate weighted contributions. A final normalization step ensures the result is scaled correctly.

\begin{algorithm}[hb]
\caption{Bisect Bound}
\label{alg:find_lower_index}
\SetKwInOut{Input}{Input}
\SetKwInOut{Output}{Output}

\Input{Monotonic array $x[0 \ldots n{-}1]$, target value $x_t$.}
\Output{Smallest index $i$ such that $x[i] \ge x_t$.}

\BlankLine
$l := 0$, \quad $r := n - 1$\;
\While{$l < r$}{
  $m := \lfloor (l + r)/2 \rfloor$\;
  \eIf{$x[m] < x_t$}{
    $l := m + 1$\;
  }{
    $r := m$\;
  }
}
\Return{$l$}
\end{algorithm}
\newpage

\begin{algorithm}[ht]
\caption{4-Color Parallel Restriction}
\label{alg:restriction_checkerboard}
\SetKwInOut{Input}{Input}
\SetKwInOut{Output}{Output}
\SetKw{KwBy}{by}

\Input{Fine-grid field $u^h$, coordinates $(x^h, y^h)$;\\
       Coarse-grid buffers $u^H$, $w^H$, coordinates $(x^H, y^H)$.}
\Output{Restricted coarse-grid field $u^H$.}

\BlankLine
\textbf{Initialize:} $u^H := 0$, \quad $w^H := 0$\;
$\Delta x^H := x^H[1] - x^H[0]$, \quad $\Delta y^H := y^H[1] - y^H[0]$\;

\BlankLine
\tcp{Compute fine index bounds}
Allocate $i_{\text{bounds}}[0 \ldots n_y^H]$, $j_{\text{bounds}}[0 \ldots n_x^H]$\;
\ForPar{$i^H = 0$ \KwTo $n_y^H - 1$}{
  $i_{\text{bounds}}[i^H] := \textsc{BisectBound}(y^h,\, y^H[i^H])$\;
}
\ForPar{$j^H = 0$ \KwTo $n_x^H - 1$}{
  $j_{\text{bounds}}[j^H] := \textsc{BisectBound}(x^h,\, x^H[j^H])$\;
}
$i_{\text{bounds}}[n_y^H] := n_x^h - 1$, \quad $j_{\text{bounds}}[n_x^H] := n_y^h - 1$\;

\BlankLine
\tcp{Four checkerboard sweeps over coarse cells}
\For{$c_y = 0$ \KwTo $1$}{
  \For{$c_x = 0$ \KwTo $1$}{
    \ForPar{$i^H = c_y$ \KwTo $n_y^H - 1$ \KwBy $2$}{
      \ForPar{$j^H = c_x$ \KwTo $n_x^H - 1$ \KwBy $2$}{

        \tcp{Loop over contributing fine nodes}
        \For{$i^h = i_{\text{bounds}}[i^H]$ \KwTo $i_{\text{bounds}}[i^H{+}1]$}{
          \For{$j^h = j_{\text{bounds}}[j^H]$ \KwTo $j_{\text{bounds}}[j^H{+}1]$}{

            \tcp{Compute and apply bilinear weights}
            \For{$(\delta_i, \delta_j) \in \{(0,0), (1,0), (0,1), (1,1)\}$}{
              $r_x := |x^h[j^h] - x^H[j^H + \delta_j]| / \Delta x^H$\;
              $r_y := |y^h[i^h] - y^H[i^H + \delta_i]| / \Delta y^H$\;
              $w := (1 - |r_x|) (1 - |r_y|)$\;
              $u^H[i^H+\delta_i, j^H+\delta_j] \mathrel{+}= w \, u^h[i^h, j^h]$\;
              $w^H[i^H+\delta_i, j^H+\delta_j] \mathrel{+}= w$\;
            }

          }
        }

      }
    }
  }
}

\BlankLine
\tcp{Normalize accumulated values}
\ForPar{each coarse node $(i^H, j^H)$}{
  $u^H[i^H, j^H] := u^H[i^H, j^H] / w^H[i^H, j^H]$\;
}
\end{algorithm}
\newpage

\FloatBarrier

\subsection{Smoothers}
\label{sec:smoothers}
Smoothers are essential on all grid levels. On fine grids, they reduce high-frequency errors, while on coarser grids they damp the remaining low-frequency components. By operating through local updates, they improve the solution without requiring global communication. Pyroclast supports several smoothing algorithms, which are briefly discussed in this section. A more detailed discussion of their mathematical formulation, along with other iterative methods for sparse linear systems, can be found in~\cite{yousef_saad_iterative_methods_linear_systems}.

We consider a linear system
\begin{equation}
A x = b,
\label{eq:linear_system}
\end{equation}
where
\begin{equation}
A =
\begin{bmatrix}
a_{11} & a_{12} & \dots & a_{1n} \\
a_{21} & a_{22} & \dots & a_{2n} \\
\vdots & \vdots & \ddots & \vdots \\
a_{n1} & a_{n2} & \dots & a_{nn}
\end{bmatrix},
\quad
x =
\begin{bmatrix}
x_1 \\ x_2 \\ \vdots \\ x_n
\end{bmatrix},
\quad
b =
\begin{bmatrix}
b_1 \\ b_2 \\ \vdots \\ b_n
\end{bmatrix}.
\end{equation}
The goal is to iteratively improve the approximation \(x^{(k)}\) to the true solution \(x^*\) by reducing the residual \(r^{(k)} = b - A x^{(k)}\).

\subsubsection*{Jacobi Smoother}

The classical Jacobi method updates each element of \(x\) using only values from the previous iteration:
\begin{equation}
x_i^{(k+1)} = \frac{1}{a_{ii}} \left( b_i - \sum_{j \neq i} a_{ij} x_j^{(k)} \right),
\quad i = 1, 2, \dots, n.
\label{eq:jacobi_update}
\end{equation}
Because every update depends solely on the old iterate, all entries of \(x\) can be computed in parallel. This makes the Jacobi smoother ideal for massively parallel architectures such as GPUs.  

In practice, to improve numerical stability and prevent overcorrection, the Jacobi method is applied in its damped form:
\begin{equation}
x_i^{(k+1)} = (1 - \omega) x_i^{(k)} + \omega
\frac{1}{a_{ii}} \left( b_i - \sum_{j \neq i} a_{ij} x_j^{(k)} \right),
\quad \omega \in (0,1].
\label{eq:damped_jacobi}
\end{equation}
The relaxation parameter \(\omega\) controls the damping strength. For \(\omega = 1\), we recover the standard Jacobi iteration. In Pyroclast, the damped Jacobi smoother serves as a robust and fully parallel relaxation method, well suited for matrix-free implementations.

\subsubsection*{Red-Black Gauss-Seidel Smoother}

The Gauss-Seidel method improves upon Jacobi by using the most recently updated values within the same iteration:
\begin{equation}
x_i^{(k+1)} = \frac{1}{a_{ii}} \left( b_i - \sum_{j < i} a_{ij} x_j^{(k+1)} - \sum_{j > i} a_{ij} x_j^{(k)} \right),
\quad i = 1, 2, \dots, n.
\label{eq:gs_update}
\end{equation}
Because each update immediately uses the latest available information, Gauss-Seidel may converge faster than Jacobi, depending on the problem. However, this sequential data dependency makes it difficult to parallelize. 

To enable efficient parallel execution, Pyroclast implements a Red-Black Gauss-Seidel (RBGS) smoother~\cite{yousef_saad_iterative_methods_linear_systems}. In this variant, the grid is partitioned into two interleaved subsets (``red'' and ``black'' nodes) such that the Gauss-Seidel update on each red node depends only on black neighbors, and vice versa. During each sweep, all red nodes are updated concurrently using the most recent black values, followed by a parallel update of the black nodes. This two-color approach removes direct data dependencies within each color group, allowing Gauss-Seidel to be executed in parallel while retaining much of its convergence advantage over Jacobi.

To retain stability and improve control over convergence, we again employ a damped Gauss-Seidel variant:
\begin{equation}
x_i^{(k+1)} = (1 - \omega) x_i^{(k)} + \omega
\frac{1}{a_{ii}} \left( b_i - \sum_{j < i} a_{ij} x_j^{(k+1)} - \sum_{j > i} a_{ij} x_j^{(k)} \right),
\label{eq:sor_update}
\end{equation}
where \(\omega \in (0,1]\) is the relaxation parameter. This formulation is also known as Successive Over-Relaxation (SOR)~\cite{yousef_saad_iterative_methods_linear_systems}.

\subsubsection*{RAS-Type Jacobi Smoother}

In matrix-free solvers such as Pyroclast, the unknown vector \(x\) corresponds to one or more physical fields stored as multidimensional arrays, e.g. \(u[i, j]\). For large 2D or 3D grids, stencil operations are typically limited by memory bandwidth and cache reuse rather than arithmetic cost. To improve memory locality, we employ a blocking strategy, where the computational domain is divided into tiles that fit within cache and are processed independently. This is known as spatial blocking~\cite{ferrari_3d_blocking}.

However in practice, pure spatial blocking in two dimensions provides limited performance improvement. To overcome this, Pyroclast implements a RAS-type temporal blocking scheme, introduced in~\cite{ferrari_3d_blocking}. In this approach, each tile performs several local Jacobi iterations before moving to the next tile, reusing data that remains in cache. Tiles are defined with small overlaps to ensure continuity between subdomains. This is similar in spirit to a Restricted Additive Schwarz (RAS) or domain decomposition method~\cite{yousef_saad_iterative_methods_linear_systems, hackbusch_book}.

Each iteration updates the interior of a tile while treating its outer boundaries as fixed, and after a few local iterations the results are written back to the global field. In order to avoid minimize artifacts, at the next outer iteration, the tiling grid is shifted randomly in both spatial directions, ensuring that the boundaries of the tiles vary over time and preventing the accumulation of errors along fixed seams. The pseudocode for this algorithm is shown in Algorithm~\ref{alg:ras_temporal}. This RAS-type Jacobi smoother achieves significantly better cache efficiency and scalability than standard Jacobi or spatially blocked variants, while maintaining the same mathematical structure and numerical stability~\cite{ferrari_3d_blocking}.

\begin{algorithm}[h]
\caption{RAS-Type Temporal Blocking (adapted from~\cite{ferrari_3d_blocking})}
\label{alg:ras_temporal}
\SetKwInOut{Input}{Input}
\SetKwInOut{Output}{Output}

\Input{Field buffers $u_{\text{read}}, u_{\text{write}}$, tile size $(T_I,T_J)$,
Even number of maximum iterations $N_{\max}$,
Even number of inner iterations $T_{\text{inner}}$.}
\Output{Updated field $u$.}

$N_{\text{outer}} := \left\lceil N_{\max}/T_{\text{inner}} \right\rceil$, \quad $N_{\text{outer}} \mathrel{+}= N_{\text{outer}} \% 2$\;

\For{$t := 1$ \KwTo $N_{\text{outer}}$}{
  Draw random shifts $s_i \in [0,T_I)$, $s_j \in [0,T_J)$ with periodic wrap-around\;
  Partition the interior into overlapping tiles of size $(T_I,T_J)$ using $(s_i,s_j)$\;
  \ForEach{tile $\mathcal{T}$ \textnormal{(in parallel)}}{
    \For{$\tau := 1$ \KwTo $T_{\text{inner}}$}{
      \For{$(i,j) \in \mathcal{T}$}{
        $u_{\text{write}}(i,j) := \text{JacobiStencil}\big(u_{\text{read}}, i, j\big)$\;
      }
      \tcp{Local buffer ping-pong}
      Swap $u_{\text{read}} \leftrightarrow u_{\text{write}}$ 
    }
    \tcp{Each cell has a single writer but may be read by multiple tiles (benign races)}
  }
}
\end{algorithm}

Together, these three smoothers provide a flexible toolkit for different hardware and problem scales. The damped Jacobi and RBGS smoothers offer simple, stable, and easily parallelized options for general use, while the RAS-type Jacobi smoother provides a high-performance alternative optimized for modern cache hierarchies.

\subsection{Multigrid Solution of the Stokes System}
\label{sec:mg_stokes_solution}

As established in Section~\ref{sec:uzawa_iteration}, the Uzawa algorithm for solving the Stokes system proceeds by alternating between a velocity update and a pressure correction:
\begin{equation}
\label{eq:uzawa_recap}
\begin{cases}
v^{k+1} := L^{-1}(f - G p^k), \\
p^{k+1} := p^k + \alpha\, \eta\, D v^{k+1},
\end{cases}
\end{equation}
where \(L\) denotes the discrete variable-viscosity Laplacian, and \(G\), \(D\) are the discrete gradient and divergence operators, respectively. The pressure update ensures compliance with the incompressibility condition, and the local viscosity field \(\eta\) provides consistent physical scaling.

A key challenge in this scheme lies in the application of the inverse velocity operator \(L^{-1}\). Since explicitly inverting \(L\) is computationally infeasible for large problems, we instead approximate the solution to the velocity subproblem:
\begin{equation}
\label{eq:velocity_subproblem}
L v = f - G p^k,
\end{equation}
using a small number (typically one or two) of multigrid V-cycles. The pressure is held fixed during this solve, and the right-hand side represents the momentum residual under the current pressure field. 

After obtaining the approximate velocity \(v^{k+1}\), the divergence of the velocity field is computed and used to update the pressure using the preconditioned formula in Equation~\eqref{eq:uzawa_recap}. Since only a few multigrid iterations are used per step, the overall method remains computationally efficient while still delivering robust convergence.

\subsection{Viscosity Rescaling and Initial Guess}
\label{sec:viscosity_rescaling}

Solving the Stokes equations in the presence of strong viscosity contrasts poses a significant challenge for iterative solvers. In such cases, poor initial guesses can cause the pressure field to converge slowly or stagnate entirely. To improve robustness, Pyroclast implements a viscosity rescaling strategy, similar to what is proposed in~\cite{gerya_multigrid}, that gradually introduces the full viscosity contrast in stages, allowing the solver to build successively better approximations to the solution.

Let \( \eta_{\min} = \min \eta \) denote the minimum viscosity in the domain. At each stage, we define a rescaled computational viscosity field:
\begin{equation}
\eta_{\text{comp}} = (1 - \theta)\,\eta_{\min} + \theta\,\eta,
\end{equation}
where \( \theta \in [0,1] \) is a blending factor that controls the amount of contrast. Initially, the system is solved using a uniform viscosity field \( \eta_{\text{comp}} = \eta_{\min} \), which leads to a smoother pressure solution. The value of \( \theta \) is then incremented in fixed steps (e.g., \( \theta \in \{0, 0.25, 0.5, 0.75, 1.0\} \)), gradually restoring the full physical contrast. At each stage, the pressure and velocity solutions from the previous iteration are reused as initial guesses, guiding the solver towards the true solution as the system becomes increasingly ill-conditioned.

In addition to viscosity rescaling, Pyroclast supports physically informed initialization of the pressure field. For problems involving gravity, we initialize pressure using the lithostatic distribution:
\begin{equation}
p_{\text{litho}}(x, y) = \int_0^y \rho(x, y')\,g_y\,dy',
\end{equation}
which represents the hydrostatic pressure required to balance gravitational forces in a column of fluid with density \( \rho \). This provides a reasonable first approximation to the pressure field.

\section{Accelerators}
\label{sec:accelerators}
While the Uzawa iteration described in the previous sections provides an efficient and scalable solver for the Stokes system, it can be further enhanced by global acceleration techniques. In the numerical linear algebra literature, it is common to treat Uzawa-type fixed-point iterations as a preconditioner for a Krylov subspace method~\cite{tackley_fast_stokes, ptatin3d}. However, in Pyroclast, we adopt a different perspective: since Uzawa iteration is the core solver responsible for most of the computation, we refer to these outer Krylov or quasi-Newton techniques as \emph{accelerators}.

This distinction is motivated by practical considerations in high-performance computing. Stencil-based fixed-point methods such as Uzawa are well suited for distributed-memory parallelism and GPU execution. Their communication patterns are inherently local, requiring only halo exchange between neighboring domains. Krylov methods, by contrast, involve global reductions to compute inner products and often require more complex orthogonalization procedures~\cite{yousef_saad_iterative_methods_linear_systems}. Although pipelined and communication-avoiding Krylov methods have been developed and implemented in libraries like PETSc~\cite{petsc-web-page}, they remain sensitive to global communication overheads and are more difficult to scale.

Another limiting factor is memory consumption. Krylov methods typically require storing multiple past search directions or solution vectors, often between 10 and 30 for problems of this kind~\cite{tackley_fast_stokes, ptatin3d}. On GPU-accelerated systems where memory is a critical bottleneck, this increased memory footprint severely limits the size of physical problems that can be solved efficiently. In our experience, memory usage increases by a factor of 10x–30x when using standard Krylov solvers as accelerators.

Nevertheless, Krylov methods offer improved robustness and are valuable when solving particularly challenging systems. In Pyroclast, we find that the viscosity rescaling strategy described in Section~\ref{sec:viscosity_rescaling} is often sufficient to ensure convergence without the need for Krylov acceleration. However, for very difficult systems or extreme viscosity contrasts, acceleration methods can substantially reduce the number of Uzawa iterations required.

Because Pyroclast implements an \emph{inexact} Uzawa iteration, where the action of \( L^{-1} \) is only approximated using a multigrid solver, the underlying fixed-point operator changes at each iteration. As a result, standard Krylov methods are not directly applicable. Instead, we must use \emph{flexible} Krylov solvers, which accommodate variable preconditioners~\cite{yousef_saad_iterative_methods_linear_systems}.

In this section, we describe two such acceleration techniques that are compatible with Pyroclast's iterative framework:
\begin{itemize}
    \item The Generalized Conjugate Residual (GCR) method, a flexible Krylov solver applicable to linear fixed-point problems.
    \item Anderson Acceleration, a quasi-Newton method that applies to both linear and nonlinear fixed-point iterations and can also be used to accelerate other components of the solver, such as plastic iterations~\cite{Gerya_viscoelastoplastic}.
\end{itemize}
These methods offer complementary strengths and are both supported as optional accelerators within the Pyroclast framework.

\subsection{Uzawa Iteration as a Fixed Point Method}
\label{subsec:uzawa_fixed_point}

We recall the discrete Stokes system introduced in Equation~\eqref{eq:stokes_saddle},
\begin{equation}
\label{eq:stokes_saddle_recall}
\underbrace{
\begin{bmatrix}
L & G \\
D & 0
\end{bmatrix}}_{A}
\underbrace{
\begin{bmatrix}
v \\ p
\end{bmatrix}}_{x}
=
\underbrace{
\begin{bmatrix}
f \\ 0
\end{bmatrix}}_{b},
\end{equation}
where \(L\) is the discrete vector Laplacian, \(G\) and \(D\) are the discrete gradient
and divergence operators, and \(x = [v, p]^\top\) is the full velocity–pressure vector.

Again, we recall the Uzawa iteration, defined in Section~\ref{sec:uzawa_iteration},
which updates velocity and pressure sequentially according to
\begin{equation}
\label{eq:uzawa_iteration_recall}
\begin{cases}
v^{k+1} := L^{-1}\big(f - G p^{k}\big), \\[3pt]
p^{k+1} := p^{k} + \alpha\, \eta\, D\,v^{k+1},
\end{cases}
\end{equation}
where \(\alpha>0\) is a relaxation parameter and \(\eta\) is a diagonal matrix
containing the local viscosities. The first step approximately inverts the viscous
operator \(L\), while the second applies a pressure correction based on the velocity divergence.

\subsubsection*{Matrix-Splitting Representation}

Following the approach discussed in~\cite{ho2016acceleratinguzawaalgorithm},
the Uzawa iteration~\eqref{eq:uzawa_iteration_recall} can be reformulated
as a linear fixed-point iteration on the full velocity–pressure space.
This is achieved by identifying a suitable matrix splitting of the Stokes operator \(A\)
such that the iteration can be expressed as
\begin{equation}
\label{eq:uzawa_split_compact}
M x^{k+1} = N x^{k} + b,
\end{equation}
for appropriately defined matrices \(M\) and \(N\).

Starting from~\eqref{eq:uzawa_iteration_recall}, we write the two update equations in linearized form:
\begin{align*}
L v^{k+1} &= f - G p^{k}, \\
D v^{k+1} - (\alpha\eta)^{-1} p^{k+1} &= -(\alpha\eta)^{-1} p^{k}.
\end{align*}
These two relations can be combined into the block system
\begin{equation}
\label{eq:uzawa_split_system}
\underbrace{
\begin{bmatrix}
L & 0 \\
D & -(\alpha\eta)^{-1}
\end{bmatrix}}_{M}
\begin{bmatrix}
v^{k+1} \\[2pt] p^{k+1}
\end{bmatrix}
=
\underbrace{
\begin{bmatrix}
0 & -G \\
0 & -(\alpha\eta)^{-1}
\end{bmatrix}}_{N}
\begin{bmatrix}
v^{k} \\[2pt] p^{k}
\end{bmatrix}
+
\begin{bmatrix}
f \\ 0
\end{bmatrix}.
\end{equation}
The matrices \(M\) and \(N\) define a splitting of the full Stokes operator,
\begin{equation}
\label{eq:uzawa_splitting}
A = M - N =
\begin{bmatrix}
L & G \\ D & 0
\end{bmatrix}.
\end{equation}

\subsubsection*{Fixed-Point Formulation}
The matrix relation~\eqref{eq:uzawa_split_compact} defines a stationary linear iteration, which can be equivalently written as a preconditioned Richardson iteration. This follows directly from the split form:
\begin{align}
M x^{k+1} &= N x^{k} + b \\
x^{k+1} &= M^{-1} N x^{k} + M^{-1} b \\
x^{k+1} &= x^{k} + (M^{-1} N - I) x^{k} + M^{-1} b
\end{align}
Since \(A = M - N\), we have \(M^{-1} N - I = -M^{-1} A\),
and therefore
\begin{equation}
x^{k+1} = x^{k} - M^{-1} A x^{k} + M^{-1} b
= x^{k} + M^{-1}\big(b - A x^{k}\big).
\end{equation}
Defining the residual \(r^k = b - A x^k\),
this iteration can be written compactly as
\begin{equation}
x^{k+1} = x^{k} + M^{-1}\big(b - A x^{k}\big) = x^{k} + M^{-1} r^k,
\end{equation}
which is the standard preconditioned Richardson form.

In practice, the action of \(M^{-1}\) is approximated by first applying a multigrid solver
to the viscous operator \(L\) and then performing a viscosity-weighted pressure correction.
Because the multigrid solve is inexact and may vary between iterations,
the effective preconditioner changes over time, requiring the use of \emph{flexible} methods~\cite{yousef_saad_iterative_methods_linear_systems}.
This interpretation views the Uzawa algorithm as a linear fixed-point iteration
on the full saddle-point system and forms the basis for the Krylov and quasi-Newton accelerators introduced in the following sections.

\subsection{Generalized Conjugate Residual Method}
\label{subsec:gcr_accelerator}

The Generalized Conjugate Residual (GCR) method~\cite{eisenstat_GCR, yousef_saad_iterative_methods_linear_systems}
is a Krylov subspace algorithm that minimizes the residual norm at each iteration while supporting flexible preconditioning.
It extends the Conjugate Residual method to nonsymmetric systems
and is particularly suitable for cases where the preconditioner varies between iterations.

At each step, the method computes a preconditioned direction
\(
z^{k} = M_{k}^{-1} r^{k}
\)
and its image under the operator
\(
w^{k} = A z^{k},
\)
then orthogonalizes \(w^{k}\) against previously computed vectors to ensure that
the set \(\{w^{j} \}\) remains mutually orthogonal.
The solution is updated such that the residual \(r^{k+1}\)
is minimized over the subspace spanned by \(\{z^{j}\}_{j \leq k}\),
guaranteeing monotonic reduction of the residual norm.
Because the preconditioner \(M_{k}^{-1}\) may change at each iteration,
GCR naturally accommodates the inexact Uzawa preconditioner described in the previous section.

The complete procedure is summarized in Algorithm~\ref{alg:gcr},
which presents the restarted flexible GCR(\(m\)) variant implemented in Pyroclast.

\begin{algorithm}[ht]
\caption{Flexible GCR(\(m\)) for \(A x = b\)}
\label{alg:gcr}
\SetKwInOut{Input}{Input}
\SetKwInOut{Output}{Output}

\Input{Implicit operator $y \mapsto A y$, preconditioner $r \mapsto z = M_k^{-1} r$, initial guess $x^0$, restart length $m$.}
\Output{Approximate solution $x$.}

\BlankLine
\textbf{Initialize:} $r^0 := b - A x^0$, \quad $k := 0$\;

\BlankLine
\While{not converged}{
  \For{$i = 0$ \KwTo $m{-}1$}{

    \BlankLine
    \tcp{Apply variable preconditioner and compute search direction}
    $z^{k+i} := M_{k+i}^{-1} r^{k+i}$\;
    $w^{k+i} := A z^{k+i}$\;

    \BlankLine
    \tcp{Modified Gram-Schmidt orthogonalization}
    \For{$j = 0$ \KwTo $i{-}1$}{
      $\gamma := \dfrac{w^{k+i} \cdot w^{k+j}}{w^{k+j} \cdot w^{k+j}}$\;
      $w^{k+i} := w^{k+i} - \gamma\, w^{k+j}$\;
      $z^{k+i} := z^{k+i} - \gamma\, z^{k+j}$\;
    }

    \BlankLine
    \tcp{Normalize new basis vectors}
    $\nu := \|w^{k+i}\|_2$\;
    $w^{k+i} := w^{k+i} / \nu$\;
    $z^{k+i} := z^{k+i} / \nu$\;

    \BlankLine
    \tcp{Update solution and residual}
    $\alpha^{k+i} := r^{k+i} \cdot w^{k+i}$ \tcp*{since $\|w^{k+i}\|=1$}
    $x^{k+i+1} := x^{k+i} + \alpha^{k+i} z^{k+i}$\;
    $r^{k+i+1} := r^{k+i} - \alpha^{k+i} w^{k+i}$\;

    \BlankLine
    \If{converged}{\textbf{break}}
  }

  \BlankLine
  \tcp{Restart after $m$ steps}
  $k := k + i + 1$\;
}
\end{algorithm}

The formulation follows directly from the fixed-point representation of the Uzawa iteration
introduced in Section~\ref{subsec:uzawa_fixed_point},
\begin{equation}
x^{k+1} = x^{k} + M_{k}^{-1}\big(b - A x^{k}\big),
\end{equation}
where \(M_{k}^{-1}\) represents the action of one Uzawa sweep,
consisting of a multigrid velocity solve and a viscosity-scaled pressure correction.
By interpreting Uzawa as a preconditioned Richardson iteration with variable \(M_{k}^{-1}\),
it can be directly embedded into the GCR framework by defining
\begin{equation}
z^{k} = M_{k}^{-1} r^{k}, \qquad w^{k} = A z^{k}.
\end{equation}
The GCR update
\begin{equation}
x^{k+1} = x^{k} + \alpha^{k} z^{k},
\end{equation}
therefore performs a residual-minimizing combination of successive Uzawa iterates.
Hence, GCR acts as a global linear accelerator
built on top of the local fixed-point iterations defined by the matrix splitting \(A = M - N\), improving its convergence rate.

\subsubsection*{Comparison with FGMRES}
Among flexible Krylov methods, both GCR and FGMRES~\cite{yousef_saad_iterative_methods_linear_systems}
are capable of handling variable preconditioners and minimizing the residual norm at each iteration.
FGMRES achieves an optimal residual reduction at every restart,
but requires storing a complete orthonormal basis of Krylov vectors
and solving a small least-squares problem at each step.
In contrast, GCR maintains a short-recurrence formulation that produces similar convergence
for linear problems while being simpler to implement and more memory efficient. Moreover, the literature suggests that GCR has been successfully applied to very similar saddle-point problems in geophysical
and fluid dynamics applications~\cite{ptatin3d, tackley_fast_stokes}.
This makes GCR particularly attractive for large-scale or GPU-accelerated simulations,
where memory and communication costs dominate.

\subsection{Anderson Acceleration}
\label{subsec:anderson_accelerator}

Anderson Acceleration (AA)~\cite{anderson_original, anderson_for_fixed_point}
is a nonlinear acceleration technique that improves the convergence of fixed-point iterations
by constructing an optimized linear combination of previous iterates.
Unlike Krylov methods, which require explicit access to the operator \(A\),
AA operates purely in terms of the fixed-point map
\(
x^{k+1} = G(x^{k}),
\)
making it applicable to both linear and nonlinear systems.
It can therefore be viewed as a quasi-Newton method
that implicitly approximates the inverse Jacobian of \(G(x)\)
using information from recent iterates.

Given a sequence of iterates and their residuals
\(
f^{k} = G(x^{k}) - x^{k},
\)
the method seeks an updated iterate
as a combination of the most recent \(m\) residual histories.
The coefficients are determined by minimizing the residual norm
under the constraint that their sum equals one.
This least-squares formulation allows AA to accelerate convergence
without explicitly forming or applying the system matrix.

\begin{algorithm}[ht]
\caption{Anderson Acceleration(\(m\)) with Mixing for Inexact Uzawa Iteration}
\label{alg:anderson}
\SetKwInOut{Input}{Input}
\SetKwInOut{Output}{Output}

\Input{Fixed-point map $G(x)$, initial guess $x^{0} = [v^{0}, p^{0}]^\top$, depth $m$, mixing $\beta \in (0,1]$.}
\Output{Accelerated solution $x = [v, p]^\top$.}

\BlankLine
\textbf{Initialize:} $x^{1} := G(x^{0})$ \tcp*{Perform one Uzawa iteration}

\BlankLine
\For{$k := 1, 2, \ldots$ until convergence}{
    $m_k := \min(m, k)$\;
    $r^{k} := G(x^{k}) - x^{k}$\;
    Form $R^{k} := [\, r^{\,k - m_k} \ \ldots \ r^{\,k} \,]$\;
    Find $\alpha^{k} := \arg \min_{\alpha \in A_k} \| R^{k} \alpha \|_2$\;
    where $A_k := \{\, \alpha \in \mathbb{R}^{m_k + 1} \,:\, \mathbf{1}^\top \alpha = 1 \,\}$\;

    \BlankLine
    \tcp{Compute mixed Anderson update}
    $x^{k+1} := (1 - \beta)\sum_{i=0}^{m_k} (\alpha^{k})_i \, x^{\,k - m_k + i}
                + \beta\, \sum_{i=0}^{m_k} (\alpha^{k})_i \, G(x^{\,k - m_k + i})$\;
}
\end{algorithm}

\clearpage
In the linear case, where \(G(x) = x + M^{-1}(b - A x)\),
Anderson Acceleration can be shown to be mathematically equivalent
to a generalized minimal residual (GMRES) method applied to the preconditioned system~\cite{ho2016acceleratinguzawaalgorithm}.
However, its formulation purely in terms of the fixed-point map
makes it far easier to apply in complex or nonlinear contexts,
where the operator \(A\) is not explicitly available.

\subsubsection*{Application to the Uzawa Iteration}

The Uzawa algorithm can be written in fixed-point form as
\begin{equation}
x^{k+1} = G(x^{k}) = x^{k} + M_{k}^{-1}\big(b - A x^{k}\big),
\end{equation}
where the implicit action of \(M_{k}^{-1}\)
represents one sweep of the inexact preconditioner
consisting of a multigrid velocity solve and a viscosity-scaled pressure correction.
Following the approach proposed in~\cite{ho2016acceleratinguzawaalgorithm},
Anderson Acceleration can be directly applied to this fixed-point formulation
to improve convergence without requiring explicit access to the operator \(A\).

At each Uzawa cycle, Pyroclast computes the next iterate
\(
x^{k+1} = G(x^{k})
\)
and stores the most recent \(m\) pairs \(\{x^{i}, G(x^{i})\}\)
to form the least-squares system described in Algorithm~\ref{alg:anderson}.
The accelerated update is then used to overwrite
the velocity and pressure fields before the next Uzawa sweep.
Because Anderson Acceleration operates independently of the underlying discretization,
it can be seamlessly enabled for both CPU and GPU solvers.
In practice, moderate depths (\(m \approx 5{-}15\)) and damping factors (\(\beta \approx 0.5{-}0.8\))
provide robust and consistent acceleration across a wide range of viscosity contrasts.

\section{Assessment of Convergence and Accuracy}
\label{sec:convergence_accuracy}

Accurate evaluation of solver convergence requires residual norms that are consistent with
the physical scaling of the incompressible Stokes equations.
Because the magnitudes of velocity and pressure differ substantially,
a simple relative $\ell^2$ norm is not meaningful in dimensional form.
Instead, Pyroclast employs surrogate \emph{energy-based} norms
derived from the continuous operators of the Stokes system.
The following formulation is adapted from the approach presented in~\cite{ferrari_3d_blocking}.

\subsection{Residual Norms}
\label{sec:residuals}

The residual norms are defined in terms of energy inner products
associated with the discrete momentum and continuity equations.
They provide a physically interpretable measure of convergence,
quantifying the fraction of mechanical work that remains unbalanced in the current iterate.

\subsubsection*{Velocity Residual}

The discrete viscosity-weighted Laplacian \(L\) is symmetric and \emph{negative definite}
under our sign convention for the Stokes equations.
To obtain a positive definite energy norm, we therefore work with \(-L\) and its diagonal surrogate
\(\widetilde{L}^{-1} = \operatorname{diag}(-L)^{-1}\).
This is purely a matter of convention in how the PDE system is written
and does not affect the residual definition.
With this convention, the velocity energy residual becomes
\begin{equation}
\|v - v^\ast\|_{-L}^{2}
= r_v^{\!\top} (-L)^{-1} r_v
\;\approx\; r_v^{\!\top}\,\widetilde{L}^{-1} r_v
= \|r_v\|_{\widetilde{L}^{-1}}^{2},
\end{equation}
where \(v^\ast\) is the exact velocity solution and \(r_v\) the velocity residual.

\subsubsection*{Pressure Residual}

For the pressure field, the natural energy norm is induced by the Schur complement
\(S = D L^{-1} G\) (see Equation~\eqref{eq:schur_system}).
Direct assembly of \(S\) is infeasible, so we approximate the corresponding energy
via a positive diagonal surrogate \( \widetilde{S}^{-1} \approx S^{-1} \) and define
\begin{equation}
\|p - p^\ast\|_{S}^{2}
= r_p^{\!\top} S^{-1} r_p
\;\approx\;
r_p^{\!\top}\,\widetilde{S}^{-1} r_p
= \|r_p\|_{\widetilde{S}^{-1}}^{2}.
\end{equation}

In Section~\ref{sec:uzawa_iteration} we analyzed the \emph{discrete} scaling of the Schur complement
to design an effective \emph{preconditioner}.
The goal was to approximate the action of a composition of three discrete operators,
and accounting for mesh factors was necessary since they are intrinsically part of these discrete operators.
An energy norm, however, arises from the variational (continuous) formulation of the Stokes problem
and should not encode discretization-specific factors.
For this reason, we first design a surrogate for the continuous equivalent of the Schur complement
and only in a second stage discretize it.

The continuous equivalent of \(L\) is the viscosity-weighted Laplacian operator
\begin{equation}
\mathcal{L} v = \nabla\!\cdot\!\big(\eta(\nabla v + \nabla v^\top)\big).
\end{equation}
The dominant physical effect of \(\mathcal{L}\) is to relate the velocity gradients
to viscous stresses of the same order of magnitude as the pressure.
Accordingly, its inverse has the inverse effect, and we approximate
\(\mathcal{L} \approx -\eta \Rightarrow \mathcal{L}^{-1} \approx -\eta^{-1}\).
The negative sign is added to account for the fact that $\mathcal{L}$ is a negative definite operator under Dirichlet boundary conditions and thus its surrogate must also be negative definite. Notice that this is a continuous approximation and does not consider any discretization-dependent coefficients.
The continuous equivalent of the Schur complement becomes
\begin{equation}
\mathcal{S} = \nabla\!\cdot\,\mathcal{L}^{-1}\nabla
\;\approx\;
\nabla\!\cdot\big(-\eta^{-1}\nabla\big)
\;\approx\;
-\eta^{-1}\,\Delta,
\end{equation}
so that a natural discrete surrogate for \(S^{-1}\) is \(-\eta\,\Delta_h^{-1}\),
where the subscript \(h\) indicates the discrete version of an operator.
Finally, we take a diagonal approximation of its inverse:
\begin{equation}
\widetilde{S}^{-1}
\;\approx\;
\eta\,\operatorname{diag}(-\Delta_h)^{-1},
\quad
\text{that is,}\quad
\widetilde{S}^{-1}_{jj}
\;\approx\;
\eta_{jj}\,(\operatorname{diag}(-\Delta_h))^{-1}_{jj},
\end{equation}
which on a uniform 2D grid reduces to
\begin{equation}
\widetilde{S}^{-1}_{jj}
\;\approx\;
\frac{\eta_{jj}}{\,2/\Delta x^{2} + 2/\Delta y^{2}}.
\end{equation}
The result is a simple and elegant surrogate that preserves the physics (through \(\eta\))
while introducing only the minimal mesh dependence required
for a resolution-independent pressure energy residual.

\subsubsection*{Global Relative Energy Residual}

Since the only nonzero source term in the Stokes system~\eqref{eq:stokes_saddle} is the body force \(f\),
the global residual is normalized by the corresponding energy norm \(\|f\|_{\widetilde{L}^{-1}}\),
yielding the surrogate relative energy residual
\begin{equation}
\frac{\|x - x^\ast\|_{-L \oplus S}}{\|x^\ast\|_{-L \oplus S}}
\;\approx\;
\sqrt{\frac{\|r_v\|_{\widetilde{L}^{-1}}^{2}
+\|r_p\|_{\widetilde{S}^{-1}}^{2}}
{\|f\|_{\widetilde{L}^{-1}}^{2}}},
\end{equation}
where \(x = [v, p]^{\mathsf{T}}\) is the approximate full state solution,
and \(x^\ast = [v^\ast, p^\ast]^{\mathsf{T}}\) the exact one.
The block energy norm is defined as
\begin{equation}
\|[v,p]\|_{-L \oplus S}^{2}
 := v^{\!\top}(-L)v + p^{\!\top} S p,
\end{equation}
with both \(-L\) and \(S\) symmetric positive definite
under the adopted discretization and boundary conditions.

This provides a resolution-independent and physically meaningful convergence criterion:
the velocity term measures unresolved viscous dissipation,
while the pressure term measures residual divergence errors weighted by viscosity.
A residual value of \(10^{-4}\) indicates that only \(0.01\%\) of the mechanical work remains unbalanced,
giving a robust, energy-consistent measure of solver accuracy.

\section{Numerical Stability and Solver Robustness}
\label{sec:stability}

This section evaluates the numerical stability and robustness of the Stokes solver
implemented in Pyroclast under conditions of strong viscosity contrast.
The analysis follows the same test configuration and discussion presented in~\cite{ferrari_3d_blocking},
with the corresponding figures reproduced here for completeness.

\subsection{Benchmark Problem}
\label{sec:benchmark_problem}

\begin{figure}[hb!]
  \centering
  \includegraphics[width=0.75\linewidth]{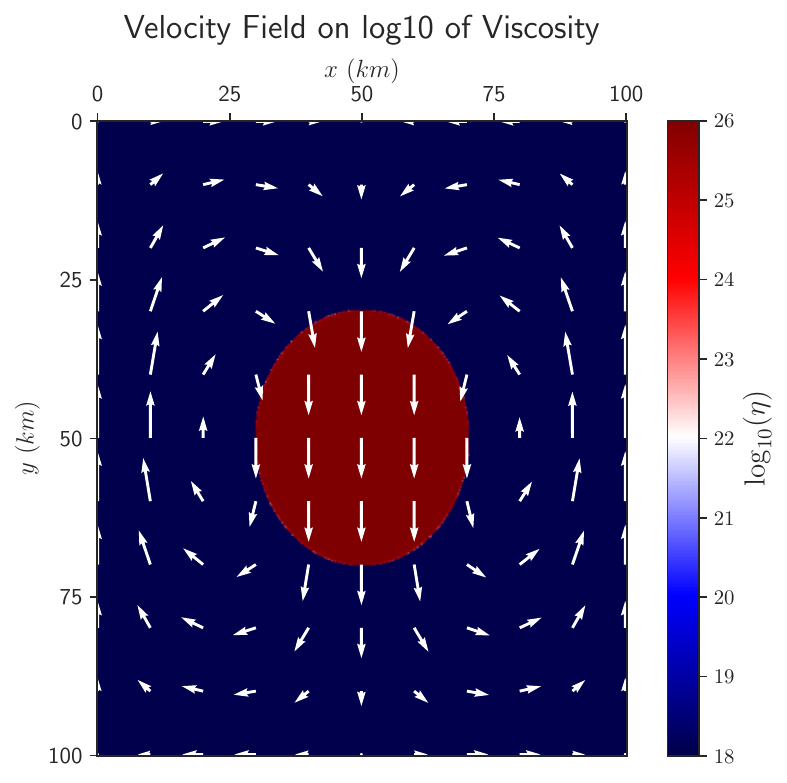}
  \caption{Velocity field (white arrows) superimposed on the logarithm of the viscosity distribution
  for the sinker benchmark, reproduced from~\cite{ferrari_3d_blocking}.
  The spherical inclusion of viscosity \(10^{26}\,\si{\pascal \second}\) and radius \(20\,\si{\kilo\meter}\)
  sinks in a surrounding medium of viscosity \(10^{18}\,\si{\pascal \second}\).
  The motion induces two symmetric vortices in the host material.}
  \label{fig:material_setup}
\end{figure}

To assess the solver’s robustness, we consider a classic benchmark consisting of
a high-viscosity spherical inclusion sinking under gravity within a less viscous medium.
The difficulty of this problem is controlled by the viscosity contrast between inclusion and host,
which strongly affects the stiffness of the resulting Stokes system.
In the present test, the inclusion viscosity is set to
\(10^{26}\,\si{\pascal \second}\) and density to \(3300\,\si{\kilogram \per \meter^3}\),
while the surrounding medium has a viscosity of
\(10^{18}\,\si{\pascal \second}\) and density of \(3200\,\si{\kilogram \per \meter^3}\),
yielding a viscosity contrast of \(10^8\).
A uniform gravitational acceleration \(g_y = 10\,\si{\meter \per \second^2}\)
is applied in the vertical direction.
The computational domain is a square box of size
\(100 \times 100\,\si{\kilo\meter}\),
discretized with \(n_x = 501\) and \(n_y = 601\) nodes.
The spherical inclusion, of radius \(20\,\si{\kilo\meter}\),
is centered in the domain.
Boundary conditions are set to no-slip in \(v_x\) and free-slip in \(v_y\) along the left and right boundaries, while free-slip in \(v_x\) and no-slip in \(v_y\) are imposed along the top and bottom.
The resulting flow field features the expected sinking motion of the dense, viscous inclusion,
producing two symmetric counter-rotating vortices in the surrounding medium
(Figure~\ref{fig:material_setup}).

\subsection{Solver Configuration}
The test configuration follows that described in~\cite{ferrari_3d_blocking}.
A total of \(1000\) Uzawa iterations are executed with acceleration disabled.
Each Uzawa iteration consists of a single multigrid V-cycle,
configured with four grid levels and a geometric scaling factor of \(2.5\).
On the finest grid, each V-cycle applies five pre-smoothing and five post-smoothing iterations,
while on coarser levels the number of smoothing steps increases by a factor of \(2.5\)
per level.

To ensure convergence under such extreme viscosity contrasts, we enable viscosity-rescaling as described in Section~\ref{sec:viscosity_rescaling}, with \(\theta\) gradually increased from \(0\) to \(1\) in increments of \(0.25\) every \(25\) Uzawa iterations (i.e. \(\theta \in \{0,\,0.25,\,0.5,\,0.75,\,1.0\}\)). The initial pressure field is set to the lithostatic distribution, which provides a physically meaningful initial guess.

\subsection{Smoother Configurations}
\label{sec:smoother_setup}

To evaluate solver robustness and convergence stability,
we compare four smoother configurations:

\begin{itemize}
  \item \textbf{RBGS:} classical red–black Gauss–Seidel method, used as a reference baseline.
  \item \textbf{Jacobi:} simple Jacobi iteration, easier to parallelize but typically less robust.
  \item \textbf{RAS:} RAS-type Jacobi smoother with random periodic block shifting
        (\(32\times32\) blocks, \(T_{\text{inner}}=4\)).
  \item \textbf{Mixed:} hybrid setup using Jacobi on the finest level and RAS on coarser levels.
\end{itemize}

All smoothers are run with relaxation factors
\(\omega_v = 0.3\) for velocity and \(\omega_p = 0.6\) for pressure.

\subsection{Results and Discussion}

\begin{figure}[ht!]
  \centering
  \includegraphics[width=0.75\linewidth]{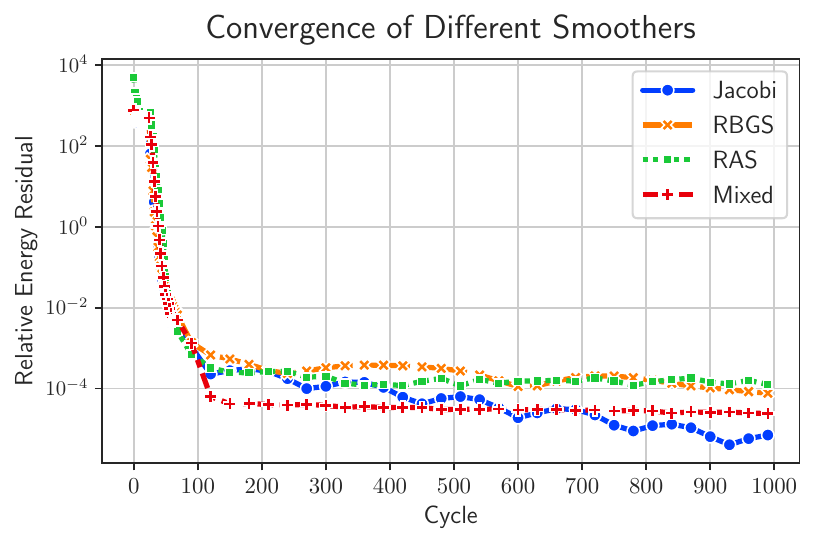}
  \caption{Convergence history of the different smoothers in the sinker benchmark,
  reproduced from~\cite{ferrari_3d_blocking}.
  The plot shows the relative energy residual as a function of Uzawa cycles.
  All smoothers converge robustly by several orders of magnitude.
  The Jacobi smoother attains the lowest residuals, while the mixed configuration
  exhibits the most stable convergence behavior.}
  \label{fig:residuals}
\end{figure}

Figure~\ref{fig:residuals} shows the convergence histories of the tested smoothers,
reported in terms of the relative energy residual defined in Section~\ref{sec:residuals}.
All methods exhibit stable convergence, reducing the residual by several orders of magnitude.
The overall behavior is consistent with that presented in~\cite{ferrari_3d_blocking}:
the Jacobi smoother reaches the lowest final residual values,
while the mixed configuration achieves the most monotone and stable convergence trajectory.
Oscillatory behavior is visible in the pure RBGS and RAS variants,
but remains well controlled throughout the iteration process.

Despite the severe viscosity contrast of \(10^8\),
all smoothers converge to residuals of order \(10^{-4}\) or lower,
demonstrating excellent numerical stability and physical correctness.
Small variations in convergence rate between independent runs are expected,
as the sinker benchmark is inherently sensitive to numerical perturbations
(e.g., thread scheduling, restriction/prolongation order).
To capture this variability, ten independent runs were performed for each smoother,
and the worst case is reported in Figure~\ref{fig:residuals}.

\begin{figure*}[ht!]
  \centering
  \includegraphics[width=\linewidth]{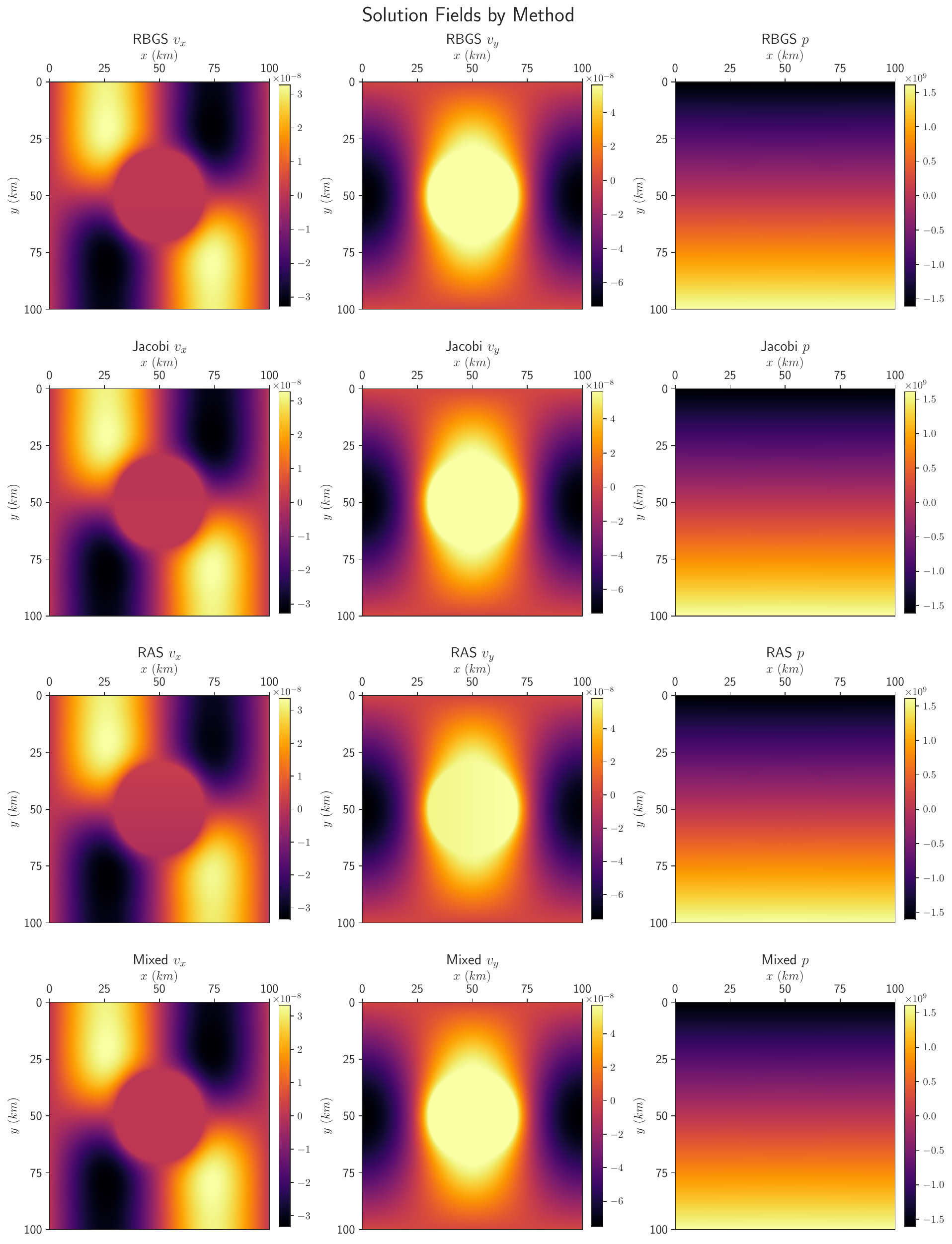}
  \caption{Velocity and pressure fields obtained with the different smoother configurations,
  reproduced from~\cite{ferrari_3d_blocking}.
  Each row corresponds to one smoother (RBGS, Jacobi, RAS, Mixed),
  showing horizontal velocity \(v_x\) (left), vertical velocity \(v_y\) (center),
  and pressure \(p\) (right).
  All methods yield identical physical solutions, with smooth velocity fields
  and the expected pressure gradient driven by density contrast.}
  \label{fig:method_fields}
\end{figure*}

The resulting velocity and pressure fields (Figure~\ref{fig:method_fields})
are consistent across all smoothers.
The inclusion sinks symmetrically, generating two counter-rotating vortices
in the host medium, and the pressure field exhibits the expected vertical gradient
associated with the density contrast.
No spurious oscillations or artefacts are observed.
The magnitude and structure of the solution are identical across methods,
confirming the correctness and robustness of the implementation.

In summary, the benchmark results confirm that the Pyroclast Stokes solver maintains numerical stability and rapid convergence even with extreme viscosity contrast and no Krylov accelerator.
The integration of viscosity rescaling with multigrid-accelerated Uzawa iteration ensures robust performance without compromising accuracy. These findings validate the solver design and its suitability
for large-scale, high-contrast geodynamic applications.

\chapter{Software Architecture and Simulation Workflow}
\label{chap:software_architecture}

In the previous chapter we described the numerical methods that form the basis of Pyroclast,
including the discretization of the governing equations and the coupling between markers and grid.
While these methods define how the physics are solved, the effectiveness of the solver
also depends on how these algorithms are structured within a coherent computational framework.
This chapter presents the software architecture that implements these numerical methods
through a modular and extensible design.

The main goal of the architecture is to maximise modularity, flexibility, and reusability
while maintaining high computational performance on both CPUs and GPUs.
To achieve this, Pyroclast organises the simulation workflow into separate components
that can be selected and combined dynamically at runtime.
The central orchestrator of this workflow is the \emph{Main Solver}, which owns the global
simulation state and delegates operations to three core components:
the \emph{Grid}, the \emph{Marker Pool}, and the \emph{Model}.
Each of these encapsulates a distinct stage of the simulation loop,
as illustrated in Figure~\ref{fig:architecture_overview}.
This separation mirrors the logical flow of the MIC algorithm introduced earlier
and allows users to modify the behaviour of the solver simply by exchanging components.

The same design principles extend to the internal organisation of the physical models.
Rather than implementing each model from scratch, Pyroclast follows a hierarchical structure
based on inheritance and composition.
New models reuse existing functionality by extending simpler ones and adding only the
operations specific to the additional physical processes.
For instance, a visco-elasto-plastic model inherits the thermomechanical model
and implements only the plastic iteration, while the thermomechanical model itself
extends the mechanical model by adding thermal coupling.
This approach, illustrated in Figure~\ref{fig:composition_inheritance},
enables complex multi-physics simulations to be assembled from a small number of reusable components.

The remainder of this chapter expands on these concepts.
It first outlines the guiding principles of composition and inheritance,
then describes how the solver orchestrates data flow between components,
and finally discusses how the high-performance Python software stack
supports this modular design across heterogeneous computing backends.

\section{Design Philosophy}

The architecture of Pyroclast is guided by two central principles: composition and inheritance.
Together they define how functionality is structured, extended, and reused across the solver.

\textbf{Composition} refers to building the solver from a small number of independent components
that interact through well-defined interfaces.
Each component is responsible for a single conceptual task, such as managing grid fields,
advecting markers, or solving the governing equations.
This separation allows components to be developed, tested, and replaced independently,
which simplifies both experimentation and maintenance.
For example, the interpolation or advection scheme can be modified without altering
the rest of the solver, as long as the component interface is respected.

\textbf{Inheritance} governs how new functionality is added.
Rather than implementing new models from scratch, Pyroclast encourages the reuse of existing ones
by extending base classes and overriding only the methods that change.
This allows new physical processes or numerical strategies to be integrated incrementally,
without duplicating code or breaking existing components.
A new visco-elasto-plastic model, for instance, inherits the thermomechanical model
and adds only the logic required for the plastic Picard iteration~\cite{Gerya_viscoelastoplastic}.

These two principles make the architecture particularly suited for large, high-performance scientific codes.
They promote a clear separation of concerns between physics, numerics, and orchestration,
which improves readability and reduces the likelihood of side effects when modifying core routines.
They also enhance backend portability by keeping numerical operators isolated from
the underlying execution environment, enabling the same high-level structure
to run efficiently on both CPU and GPU architectures.
In practice, this design philosophy allows Pyroclast to evolve as a flexible research platform
while maintaining the computational efficiency required for large-scale geodynamic simulations.

\section{Solver Orchestration and Data Ownership}

At the highest level, the simulation pipeline is coordinated by a \emph{Main Solver} class,
which acts as the central controller of the entire workflow.
It is responsible for maintaining the global simulation state, orchestrating the execution
of the main computational stages, and ensuring that each component operates on a consistent view of the system.
All data, parameters, and time-stepping information are owned by the Main Solver,
while the individual components perform specific tasks on this shared state.

The solver delegates the actual numerical work to three core \emph{components}:
the \emph{Grid}, the \emph{Marker Pool}, and the \emph{Model}.
Each encapsulates a distinct step of the MIC method
while adhering to a common interface that allows them to interact seamlessly within the solver loop.
Their conceptual roles are summarized below.

\begin{itemize}
    \item \textbf{Grid.}  
    The Grid object defines the geometry and structure of the Eulerian mesh.
    It stores information about grid coordinates, grid spacing, and mesh refinement,
    and it manages the interpolation of material properties from the marker pool onto the grid nodes.
    Unlike traditional monolithic solvers, the Grid in Pyroclast does not implement
    stencil operators or PDE solvers directly. Its role is purely geometric and organizational,
    providing the spatial framework for all grid-based operations.

    \item \textbf{Marker Pool.}  
    The Marker Pool manages the collection of Lagrangian markers that carry advected material properties.
    It is responsible for interpolating field quantities from the grid to the markers,
    performing marker advection through time, and updating advected fields such as density,
    composition, or viscosity. The Marker Pool thus represents the Lagrangian part of the method.

    \item \textbf{Model.}  
    The Model class implements the governing physics of the system.
    It defines the stencil operators required to solve the partial differential equations
    on the grid and may include internal iterative loops, such as thermomechanical or plastic iterations.
    The Model operates on the grid data provided by the shared simulation state
    and writes its results back to it.
\end{itemize}

\subsection*{The Simulation State Class}

A central element of Pyroclast’s architecture is the \emph{Simulation State} class,
which replaces the tightly coupled data structures typical of monolithic solvers.
In traditional designs, all computational routines operate directly on globally defined arrays.
This approach is efficient from a memory management perspective, since data is allocated once
and shared across all routines. However, it comes at the cost of strong coupling between components:
each routine assumes the existence and structure of global variables, making the codebase rigid
and difficult to extend or modify without affecting other parts of the solver.
Pyroclast addresses this limitation by separating data ownership from execution logic.
Each component operates on a shared state object that contains all arrays and parameters
that need to persist across different stages of the simulation, but access to this data
is mediated through a clear and consistent interface.

At each step of the pipeline, the state is passed to the component responsible
for a particular operation. The component performs its calculations and updates
the relevant quantities in the state before returning control to the main solver object.
In this view, each component acts as an operator on the shared physical state of the system.
This approach offers several advantages:

\begin{enumerate}
    \item \textbf{Dynamic composition at runtime.}  
    Because all components interact with the same abstract state interface,
    different combinations of Grid, Pool, and Model classes can be dynamically imported at runtime.
    The solver can thus switch between physical models or numerical schemes
    based solely on the input file configuration, without requiring a common base implementation or modifying and recompiling the code.

    \item \textbf{Efficient memory management.}  
    Each component may allocate its own local scratch arrays for temporary computations,
    but only the final results are written back to the shared state.
    This makes it possible to reclaim memory when a component completes its task,
    allowing the solver to run significantly larger systems within the same memory budget.
    The benefit is particularly pronounced on GPUs, where available memory is limited and memory allocations are cached~\cite{cupy_memory_management}.
    For example, when solving a thermomechanical system,
    the memory used for the Stokes multigrid hierarchy can be released
    and reused for the thermal solver, effectively doubling the usable problem size
    with negligible overhead.

    \item \textbf{Reduced boilerplate code and simplified extensibility.}  
    Since data ownership is centralized, components can focus solely on their numerical logic.
    Adding a new solver or physical model requires only a few lines of code to define
    how the component reads from and writes to the simulation state,
    without the need to redefine or reallocate shared arrays.
\end{enumerate}

\begin{figure}[h!]
    \centering
    \includegraphics[width=\textwidth]{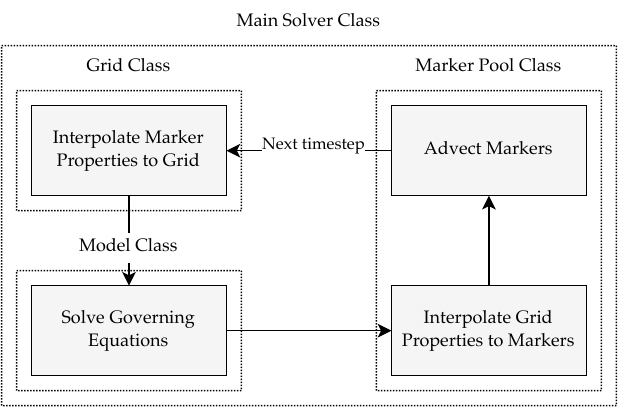}
    \caption{Runtime orchestration and data ownership in Pyroclast.
    The Main Solver manages the simulation state and delegates operations
    to the Grid, Marker Pool, and Model components.
    Each component acts as an operator on the shared state,
    updating relevant quantities before passing it to the next stage in the loop.}
    \label{fig:architecture_overview}
\end{figure}

This architecture transforms the simulation into a sequence of operations
applied to a common evolving state, with each component functioning as a pure operator.
It eliminates the need for deeply coupled procedures,
makes data dependencies explicit, and ensures that memory use remains efficient
throughout the simulation.
The overall orchestration is illustrated in Figure~\ref{fig:architecture_overview}.

\section{Composition and Inheritance of Models}

All major components of Pyroclast, including the \emph{Grid}, the \emph{Marker Pool}, and the \emph{Model},
follow the same architectural principles of composition and inheritance.
Each family of components is organized as a hierarchy of progressively more specialized subclasses,
which makes it possible to introduce new functionality incrementally while reusing existing logic.
This design avoids code duplication and promotes a consistent development workflow across the entire solver.

Each component provides a minimal and well-defined interface that can be extended or combined with others.
For example, a grid implementation may extend a basic uniform grid by adding mesh refinement or
including additional physical quantities on its nodes, while a marker pool may implement a different advection scheme or consider additional material properties to be transported.
These hierarchies remain independent but interoperable, since all components communicate
through the shared simulation state and the interfaces defined by the main solver.

\begin{figure}[h!]
    \centering
    \includegraphics[width=\textwidth]{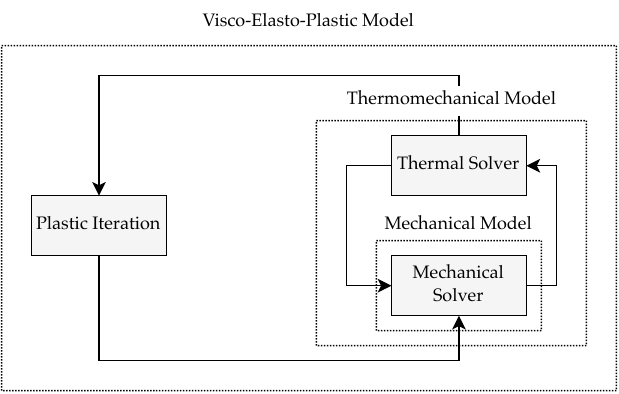}
    \caption{Composition and inheritance between models in Pyroclast.
    All major components follow the same hierarchical design pattern.
    The Model hierarchy shown here illustrates how complex multi-physics systems
    can be built incrementally from simpler components.}
    \label{fig:composition_inheritance}
\end{figure}

Among the three components, inheritance is most intuitively visible in the \emph{Model} hierarchy
because it directly encodes the physics of the simulation.
Figure~\ref{fig:composition_inheritance} illustrates this structure using the mechanical,
thermomechanical, and visco-elasto-plastic models as an example.
The \textbf{Mechanical Model} provides the basic Stokes solver, computing velocity and pressure fields.
The \textbf{Thermomechanical Model} extends it by introducing the energy equation and thermal coupling,
reusing the mechanical solver and adding an iterative coupling between the two systems.
The \textbf{Visco-Elasto-Plastic Model} inherits from the thermomechanical model
and adds a plastic Picard iteration to handle nonlinear yield behavior.
Each successive model introduces only the additional functionality required for the new physics,
while all other routines are inherited from the base classes.

This hierarchical organisation has two important implications.
First, the incremental design makes it straightforward to develop new physical models
by modifying only small and self-contained parts of the code.
Second, the solver remains fully flexible. Once a new model is implemented,
any previously defined model can still be instantiated and executed by selecting it in the input file.
A user can therefore run a mechanical-only simulation using the same codebase that supports
thermomechanical or visco-elasto-plastic coupling.
This is rarely possible in monolithic solvers, where adding new physics often requires
maintaining or recompiling separate versions of the code.

By combining inheritance across all major components with the shared-state interface described earlier,
Pyroclast achieves both extensibility and runtime flexibility.
Complex multi-physics simulations can be assembled from existing building blocks,
while simpler configurations remain accessible without additional overhead.
This approach turns Pyroclast into a unified and highly adaptable framework
rather than a collection of specialized solvers.

\section{High-Performance Python Software Stack}

The modular architecture of Pyroclast is designed to take full advantage of the modern Python high-performance computing ecosystem.
Rather than relying on handwritten low-level kernels or manual memory management, the solver builds on established tools
that provide efficient execution on both CPUs and GPUs while maintaining the clarity and flexibility of Python.
This approach makes it possible to write performance-critical scientific software in a high-level language
without compromising numerical efficiency or scalability.

At the core of the computational stack are NumPy~\cite{numpy} and CuPy~\cite{cupy}, which provide a unified interface
for dense array operations on CPUs and GPUs respectively.
Both libraries share the same API, allowing Pyroclast to execute the same numerical kernels
on different hardware backends without code duplication.
All major data structures in the solver are implemented as array-based objects,
so that switching between CPU and GPU execution requires only the selection of the desired backend.

For just-in-time (JIT) compilation of numerical kernels, Pyroclast relies on Numba~\cite{numba}.
Numba allows Python functions to be compiled to optimized machine code for both CPUs and CUDA-enabled GPUs~\cite{numba_cuda}.
It also supports HIP-based compilation for AMD hardware~\cite{numba_hip}, which ensures portability across different accelerator architectures.
Through Numba, the solver achieves near-native performance for core computational routines such as interpolation,
advection, and matrix-free stencil operations, while retaining the readability and flexibility of Python syntax.

Distributed-memory parallelism is provided through mpi4py~\cite{mpi4py}, which offers a thin and efficient interface
to the standard MPI library.
This allows Pyroclast to run on multiple nodes and exchange data between subdomains
without departing from the Python environment.
The same modular structure that governs single-device execution naturally extends to the distributed setting,
where each component interacts with its local portion of the shared state while communicating halo data with its neighbors.
This MPI-enabled architecture is discussed in more detail in Chapter~\ref{chap:mpi}.

For operations that require very low-level control or integration with existing C++ and CUDA code,
Pyroclast uses nanobind~\cite{nanobind}.
This lightweight binding framework allows high-performance kernels and external solver modules
to be written in C++ and exposed directly to Python with minimal overhead.
Typical use cases include sparse linear algebra routines and low-level optimizations that are difficult to express efficiently in pure Python.

Interoperability with external frameworks is achieved through the DLPack standard~\cite{dlpack_v1.2},
which provides zero-copy data exchange between array libraries.
This mechanism allows Pyroclast to share data with deep learning libraries such as PyTorch,
enabling the integration of machine learning models for parameter inference,
surrogate modeling, or data-driven constitutive laws.
Since all data exchange occurs at the device-memory level,
no intermediate copies are created and performance overhead remains negligible.

Together, these technologies form a cohesive and portable software stack that allows Pyroclast
to operate efficiently across heterogeneous systems while preserving the modular and object-oriented structure
described in this chapter.
The following chapters focus on the implementation of GPU acceleration
and distributed parallelism that build upon this foundation.

\chapter{GPU Support and Device-Agnostic Execution}
\label{chap:gpu_support}

The modular solver architecture introduced in the previous chapter provides a natural foundation for efficient execution on modern GPU hardware. Because the computational logic in Pyroclast is fully decoupled from data ownership and backend-specific details, the same algorithms can run on both CPUs and GPUs without modification. This chapter describes how the solver achieves this level of portability and explains the main design principles that enable GPU acceleration within the existing architecture.

Rather than rewriting the solver in a low-level language or maintaining separate implementations, Pyroclast relies on device-agnostic array programming and just-in-time compilation to target GPU hardware from the same high-level Python code. This design preserves the flexibility of the modular framework while allowing performance-critical operations to execute natively on GPUs. At present, Pyroclast primarily targets NVIDIA GPUs through the CUDA runtime, with support for AMD GPUs via the HIP backend planned for future releases.

This chapter presents the design principles and implementation strategies that enable efficient GPU execution in Pyroclast. It covers the main components of the GPU implementation, including the array-based programming model that provides backend portability, the Lagrangian advection solver designed for GPU execution, the stencil and matrix-free kernels used for grid-based computations, and the hybrid GPU multigrid implementation. A quantitative evaluation of the GPU-enabled components of Pyroclast is presented in Chapter~\ref{chap:benchmarks}. Together, these techniques allow Pyroclast to exploit the computational power of GPUs without compromising modularity or clarity, and they lay the foundation for the distributed and multi-GPU execution model introduced in the following chapter.

\section{GPU Integration Philosophy}
GPU support in Pyroclast is built on two complementary design principles:
device-agnostic array programming and dynamic dispatching of compiled kernels.
Together they ensure that computations execute on the appropriate device
without code duplication or hardware-specific branches.

\subsection{Device-Agnostic Array Programming}

All numerical components in Pyroclast operate on arrays that can reside either in host memory
or on the GPU, depending on the selected backend.
This is achieved through a unified array interface based on the \texttt{xp} abstraction layer~\cite{cupy_gpu_agnostic},
where \texttt{xp} is dynamically assigned to NumPy~\cite{numpy} for CPU execution
or CuPy for GPU execution~\cite{cupy}.
Both libraries share a nearly identical API, allowing the same array operations,
broadcasting rules, and indexing semantics to be used across architectures.

This design makes array-level code fully portable.
Vectorized operations, memory allocation, and element-wise computations
are all handled by the backend, which ensures that data remains on the correct device.
The solver can therefore switch between CPU and GPU execution at runtime
without modifying any numerical logic.
This approach greatly simplifies data management and ensures that
the simulation state is always consistent with the active execution device.

\subsection{Dynamic Dispatching of Compiled Kernels}
While many operations can be expressed as array manipulations,
others require fine-grained parallelization that is not easily implemented
through high-level array syntax.
For these cases, Pyroclast uses dynamic dispatching of just-in-time compiled kernels.
Each backend defines a set of Numba-compiled functions that implement the same numerical routine
for different devices.
At runtime, the framework automatically imports the correct backend
and dispatches calls to the appropriate implementation.

This pattern is used, for example, in the Stokes solver.
The solver defines CPU and GPU backends that expose identical interfaces
for stencil-based operations such as matrix-free kernels.
When GPU support is enabled, the solver initializes device-resident arrays
using CuPy and dynamically dispatches all kernel calls to the GPU implementation.
When run on CPUs, the same solver instance uses the corresponding Numba-compiled CPU kernels.
In both cases, the high-level solver code remains unchanged,
and all device-specific details are isolated within the backend layer.

This combination of device-agnostic array programming and dynamic kernel dispatch provides both portability and performance.
It allows Pyroclast to maintain a single unified solver codebase
while exploiting the capabilities of different hardware targets.

\section{Marker Advection and Array-Based GPU Operations}

Marker advection provides a clear example of how Pyroclast achieves backend-agnostic execution through a combination of array-based programming and dynamic dispatching.
The advection equation is implemented as a sequence of array operations that update marker coordinates
based on interpolated velocity fields.
Since these operations are fully vectorized, they can be executed on either CPUs or GPUs
using the same source code through the unified \texttt{xp} abstraction layer.
When the backend is set to NumPy, the computation runs on the CPU;
when set to CuPy, the same operations are automatically dispatched to GPU libraries such as cuBLAS or cuTensor,
depending on data layout and precision.
This approach eliminates the need for separate CPU and GPU implementations of the advection routine.

Listing~\ref{lst:rk4_agnostic} shows an example of the fourth-order Runge–Kutta (RK4)
integration scheme for marker advection written in this hardware-agnostic style.
The \texttt{get\_xp()} function automatically returns the correct array backend,
and all subsequent array and interpolation operations execute on the appropriate device.
This same mechanism is used throughout the solver to select the backend dynamically at runtime,
ensuring that data remains resident on the correct device during computation.

While the RK4 integration loop itself is written entirely in array operations,
the interpolation functions illustrate the use of \textbf{dynamic dispatching}.
Interpolation cannot be expressed efficiently as simple array expressions,
since it involves evaluating local stencils and gathering data from neighboring grid nodes.
To handle this, Pyroclast implements interpolation using Numba-compiled kernels that exist in both CPU and GPU versions.
At runtime, the framework automatically imports the appropriate backend and dispatches calls
to the correct compiled function.
This allows the same high-level advection loop to run unmodified,
while the interpolation step executes as a low-level, parallel kernel optimized for the active device.

This hybrid approach combines the strengths of array programming and just-in-time kernel compilation.
High-level control flow and vectorized arithmetic remain portable and readable,
while fine-grained parallelism is achieved through dynamic dispatch to compiled kernels.
The result is a fully GPU-resident advection cycle where markers are interpolated and advanced
without redundant host–device data transfers.

\begin{minipage}{\linewidth}
\begin{lstlisting}[language=Python, caption={Hardware-agnostic Runge–Kutta advection of Lagrangian markers in a periodic domain. The same code executes on either CPU or GPU depending on the selected array backend.}, label={lst:rk4_agnostic}]
# ------- RK4 advection of markers -------
xp = get_xp()  # Automatically returns correct array backend

# Dispatch marker coordinates to the correct device
xmA = xp.array(xm)
ymA = xp.array(ym)

# Interpolation routines automatically run on the same device
vxmA = interpolate_vx(xmA, ymA)
vymA = interpolate_vy(xmA, ymA)

# RK4 intermediate stages
xmB = xp.mod(xmA + 0.5*dt*vxmA, xsize)
ymB = xp.mod(ymA + 0.5*dt*vymA, ysize)
vxmB = interpolate_vx(xmB, ymB)
vymB = interpolate_vy(xmB, ymB)

xmC = xp.mod(xmA + 0.5*dt*vxmB, xsize)
ymC = xp.mod(ymA + 0.5*dt*vymB, ysize)
vxmC = interpolate_vx(xmC, ymC)
vymC = interpolate_vy(xmC, ymC)

xmD = xp.mod(xmA + dt*vxmC, xsize)
ymD = xp.mod(ymA + dt*vymC, ysize)
vxmD = interpolate_vx(xmD, ymD)
vymD = interpolate_vy(xmD, ymD)

# Compute effective velocity and update marker positions
vx_eff = (1/6)*(vxmA + 2*vxmB + 2*vxmC + vxmD)
vy_eff = (1/6)*(vymA + 2*vymB + 2*vymC + vymD)
xm = xp.mod(xmA + dt*vx_eff, xsize)
ym = xp.mod(ymA + dt*vy_eff, ysize)
\end{lstlisting}
\end{minipage}

\subsection{Memory Constraints and Mixed Precision Advection}

One of the main challenges of GPU-based advection lies in the memory footprint of marker data.
Each Eulerian cell typically contains between 8 and 16 markers,
which means that marker-based quantities consume roughly one order of magnitude more memory
than grid-based fields.
On large 2D or 3D domains, the total number of markers can easily exceed the available GPU memory,
making it difficult to perform advection for all markers simultaneously.

Two complementary strategies are under investigation to address this issue.
The first is the use of mixed-precision simulations,
where marker quantities are stored in single precision (FP32)
while grid quantities remain in double precision (FP64).
This significantly reduces memory usage with minimal impact on accuracy for most flow regimes.
The second approach is marker batching,
in which markers are processed in smaller groups that fit within the available GPU memory.
Each batch is advected and interpolated independently, and the updated values are written back to the shared simulation state.
Batching introduces some additional kernel launch overhead
but allows simulations of much larger systems to be executed without exceeding memory limits. 

\section{Stencil Operations and Numba Kernels}

Stencil-based operators form the computational core of the model component in Pyroclast.
They are used to evaluate spatial derivatives and apply the discretized governing equations
on the staggered grid, including the momentum and continuity equations of the Stokes system.
These operators are implemented in a matrix-free form, where local finite-difference updates
are computed directly at each grid node rather than through global sparse matrix assembly.
This approach minimizes memory overhead and allows efficient execution on both CPUs and GPUs.

Each stencil is expressed as a simple point-wise function that applies the discrete operator
to a single grid node based on its local neighborhood.
This function is then JIT-compiled for both CPU and GPU targets using Numba,
producing optimized machine code for the respective architecture.
The same operator definition is therefore shared across backends,
and the appropriate implementation is imported at runtime based on the selected execution mode.

Listing~\ref{lst:vx_op_point} shows an example of the \textit{x}-momentum operator.
The function \texttt{vx\_op\_point} computes the value obtained at node $(i, j)$
when applying the discrete operator corresponding to the discretized \textit{x}-momentum equation.
It uses the local viscosities, velocities, and pressure values to compute the result implicitly.

\begin{minipage}{\linewidth}
\begin{lstlisting}[language=Python, caption={Point-wise definition of the $x$-momentum operator on a staggered grid. The function computes the local update at node $(i, j)$ using neighboring values of viscosity, velocity, and pressure.}, label={lst:vx_op_point}]
def vx_op_point(i, j, dx, dy, etap, etab, vx, vy, p):
    # viscosities around (i, j) on staggered grid
    etaA = etap[i,   j]
    etaB = etap[i,   j+1]
    eta1 = etab[i-1, j]
    eta2 = etab[i,   j]

    # coefficients
    vx1 = 2.0 * etaA / (dx * dx)
    vx2 = eta1 / (dy * dy)
    vx3 = -(eta1 + eta2)/(dy * dy) - 2.0*(etaA + etaB)/(dx * dx)
    vx4 = eta2 / (dy * dy)
    vx5 = 2.0 * etaB / (dx * dx)

    # cross terms with vy
    vy1 =  eta1 / (dx * dy)
    vy2 = -eta2 / (dx * dy)
    vy3 = -eta1 / (dx * dy)
    vy4 =  eta2 / (dx * dy)

    # pressure gradient
    dp_right = -p[i, j+1] / dx
    dp_left  = +p[i, j]   / dx

    return (
        vx1 * vx[i,   j-1] +
        vx2 * vx[i-1, j  ] +
        vx3 * vx[i,   j  ] +
        vx4 * vx[i+1, j  ] +
        vx5 * vx[i,   j+1] +
        vy1 * vy[i-1, j  ] +
        vy2 * vy[i,   j  ] +
        vy3 * vy[i-1, j+1] +
        vy4 * vy[i,   j+1] +
        dp_right + dp_left
    )
\end{lstlisting}
\end{minipage}

For CPU execution, this function is compiled with Numba’s \texttt{@njit} decorator
and called inside a standard nested loop that iterates over the interior nodes of the grid.
This produces a vectorized and multithreaded implementation.

\begin{minipage}{\linewidth}
\begin{lstlisting}[language=Python, caption={CPU implementation of the $x$-momentum operator using Numba parallel loops.}, label={lst:vx_operator_cpu}]
vx_op_point = nb.njit(vx_op_point, inline='always', cache=True)

@nb.njit(cache=True, parallel=True)
def vx_operator(nx1, ny1, dx, dy, etap, etab, vx, vy, p, out):
    for i in nb.prange(1, ny1 - 1):
        for j in range(1, nx1 - 2):
            out[i, j] = vx_op_point(i, j, dx, dy, etap,
                                     etab, vx, vy, p)
    return out
\end{lstlisting}
\end{minipage}

For GPU execution, the same function is compiled with \texttt{@cuda.jit}
and executed within CUDA’s grid–block parallel model.
Each CUDA thread updates one grid node,
while thread blocks form a two-dimensional grid that covers the entire computational domain. 

\begin{minipage}{\linewidth}
\begin{lstlisting}[language=Python, caption={GPU implementation of the same operator using Numba’s CUDA backend. Each thread updates one grid node according to the point-wise operator.}, label={lst:vx_operator_gpu}]
vx_op_point_device = cuda.jit(inline="always")(vx_op_point)

@cuda.jit
def _vx_operator_kernel(dx, dy, etap, etab,
                            vx, vy, p, out, nx1, ny1):
    i = cuda.blockIdx.y * cuda.blockDim.y + cuda.threadIdx.y
    j = cuda.blockIdx.x * cuda.blockDim.x + cuda.threadIdx.x
    if 1 <= i < ny1 - 1 and 1 <= j < nx1 - 2:
        out[i, j] = vx_op_point_device(i, j, dx, dy, etap,
                                        etab, vx, vy, p)

def vx_operator(nx1, ny1, dx, dy, etap,
                    etab, vx, vy, p, out):
    grid, block = launch_2D(vx.shape)
    _vx_operator_kernel[grid, block](dx, dy, etap, etab,
                                       vx, vy, p, out, nx1, ny1)
    return out
\end{lstlisting}
\end{minipage}

This pattern provides a unified way to implement differential operators on different backends.
The same operator definition and interface are used for both CPU and GPU targets;
only the execution layer differs.
At runtime, Pyroclast imports the correct backend based on the configuration file, ensuring that the high-level solver code remains identical.

\section{GPU Multigrid and Hybrid Execution}

Implementing an efficient multigrid solver with GPU support requires particular care.
Stencil operations are memory-bound, and their performance depends critically on locality and data movement.
To fully exploit GPU bandwidth, data must remain resident on the correct device,
and host–device transfers must be minimized.
This is challenging in multigrid methods, where each level represents a problem
of exponentially smaller size.
Fine levels contain most of the computational work and benefit from GPU parallelism,
whereas coarser levels quickly become too small to utilize the GPU efficiently.
It is therefore essential that each level executes on the most suitable backend:
GPU for fine grids, multithreaded CPU for intermediate ones, and serial CPU for the coarsest grids.

A naive implementation typically treats the GPU as an auxiliary accelerator.
All multigrid levels and data reside in CPU memory, and when a compute-intensive kernel
such as a smoother is called, the relevant arrays are copied to the GPU.
After computation, results are written back to the CPU.
This offloading approach is simple but inefficient, as every smoother invocation incurs
host–device transfer overhead.
Because stencil-based smoothers are limited by memory bandwidth,
the solver becomes bottlenecked by PCIe or NVLink throughput instead of GPU memory bandwidth,
significantly reducing the potential performance gains.

To mitigate this, naive implementations often use a \emph{threshold} to offload
only grid levels fine enough to benefit from GPU execution.
While this avoids underutilization on coarse grids, it still involves repeated data transfers
for every smoothing step, which severely limits performance.

Pyroclast instead adopts a GPU-resident multigrid design.
The hierarchy is split between CPU and GPU memory:
fine levels above a chosen threshold are kept permanently on the GPU,
while coarse levels remain on the CPU.
All GPU-resident levels perform smoothing, residual calculation, restriction, prolongation, and correction without any host–device communication.
Only cross-device restriction and prolongation exchange data between CPU and GPU,
and these operate on exponentially smaller grids, significantly reducing the cost of data transfer.

The threshold determines where the hierarchy is divided.
All levels finer than the threshold remain on the GPU,
and all coarser levels stay on the CPU.
This minimizes communication while allowing the split to be adjusted
to balance memory usage and computational load.

\begin{figure}[h!]
    \centering
    \includegraphics[width=\textwidth]{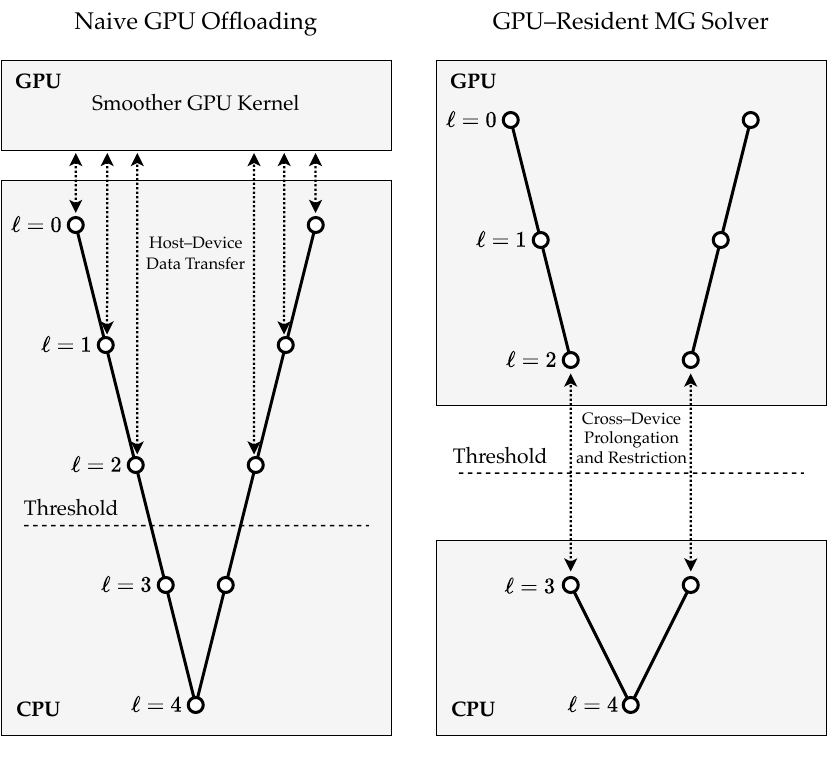}
    \caption{Comparison between naive GPU offloading (left) and the GPU-resident multigrid design (right). 
    In the naive approach, all levels reside on the CPU and each smoother invocation requires host–device data transfer.
    In Pyroclast’s GPU-resident implementation, fine grid levels remain on the GPU,
    while coarse levels are handled by the CPU.
    The dashed line marks the threshold separating GPU- and CPU-resident levels.
    Cross-device data movement occurs only during restriction and prolongation across this boundary.}
    \label{fig:gpu_mg_design}
\end{figure}

Figure~\ref{fig:gpu_mg_design} compares the two strategies.
In the naive offloading model (left), each smoother call triggers data transfers
between CPU and GPU.
In the GPU-resident design (right), fine levels remain entirely on the GPU,
and only cross-device restriction and prolongation occur across the threshold level,
drastically reducing communication volume.

A GPU-resident multigrid solver is more complex to implement,
as all multigrid operations, including smoothers, residual evaluation, restriction, and prolongation,
must exist on the GPU.
It also introduces a modest memory overhead,
since parts of the hierarchy are stored simultaneously on both devices.
This typically amounts to around 30\% additional GPU memory in 2D
and about 15\% in 3D, assuming a grid refinement ratio of two.
Despite this cost, the benefits are substantial,
as the majority of computation occurs on fine levels that fully exploit GPU bandwidth.

Pyroclast’s modular architecture makes this design straightforward to integrate.
The multigrid cycle is implemented once in a backend-agnostic form,
and each level is assigned to a device during initialization.
At runtime, the solver imports the correct backend and allocates data
on the corresponding device automatically.
Only the device-specific kernels for smoothing, restriction, and prolongation
must be implemented separately, using the same stencil-based techniques
described in the previous section.

This hybrid execution model enables Pyroclast to maintain a simple and modular design
while achieving high GPU efficiency.
Fine grids operate entirely on the GPU, maximizing bandwidth utilization,
whereas coarse levels remain on the CPU to save memory and avoid idle GPU resources.
The result is a flexible, scalable multigrid implementation
that adapts efficiently to heterogeneous computing environments.

\chapter{Distributed Execution and MPI Parallelism}
\label{chap:mpi}

The GPU implementation discussed in the previous chapter allows Pyroclast to exploit the
fine-grained parallelism of a single GPU on a single node.
To scale across multiple nodes and leverage larger compute clusters comprising multiple GPUs,
Pyroclast extends this model to distributed-memory execution using
the Message Passing Interface (MPI) standard~\cite{MPI-5.0}.

MPI provides a process-based parallel programming model in which each process, or \emph{rank}, 
operates as an independent solver instance with its own memory space and local subset of the computational domain. 
Parallelism is achieved through explicit communication between ranks using message passing, 
allowing data to be exchanged between subdomains. 
This model is particularly well suited for large-scale scientific simulations, 
as it offers explicit control over synchronization and data movement, 
enabling scalability across many CPUs and GPUs.

In Pyroclast, MPI functionality is implemented through the mpi4py package~\cite{mpi4py},
which offers Python bindings to the MPI standard and exposes nearly all of its features,
including point-to-point communication, collective operations, as well as different communicator topologies.
This enables the solver to maintain a pure Python interface
while integrating seamlessly with high-performance distributed systems.

This chapter describes how Pyroclast extends its hybrid CPU–GPU architecture
to distributed-memory parallelism through MPI.
The discussion focuses on the design of domain decomposition,
halo exchange, and distributed marker advection.
Although full MPI support is currently implemented only for the Lagrangian advection solver,
the same algorithms and design principles are being used to extend distributed execution
to the Eulerian components of Pyroclast, including interpolation, smoothing,
and multigrid preconditioning.

\section{Overview of the MPI Programming Model}

The Message Passing Interface (MPI) defines a standardized model for distributed-memory parallel computing.
In MPI, each process, referred to as a \emph{rank}, operates as an independent instance of the program
with its own local memory space.
Parallelism is achieved through explicit communication between ranks using
message-passing primitives instead of shared-memory synchronization.
This model offers full control over data locality and communication,
which makes it highly suitable for large-scale scientific simulations.
A full technical specification of the latest MPI 5.0 standard is available in~\cite{MPI-5.0};
here we limit ourselves to a high-level overview of the most relevant concepts.

MPI supports two main communication paradigms.
The first consists of \textbf{point-to-point operations}, where ranks exchange data directly
using explicit send and receive calls.
The second involves \textbf{collective operations}, which coordinate data exchange
among all ranks in a communicator, for example broadcasting a variable from one rank to all other peers,
or reducing values across the entire process group.
A communicator defines the set of ranks that participate in a given communication context,
and Pyroclast uses this mechanism to organize processes into logical topologies
that mirror the structure of the computational domain.

Both \textbf{blocking} and \textbf{non-blocking} communication modes are used within Pyroclast.
Blocking communication ensures that data transfer is complete before execution continues,
providing strict synchronization between ranks.
Non-blocking operations, on the other hand, return control immediately
and allow computation to proceed while communication is still in progress.
This enables \textbf{asynchronous communication}, where data transfer and computation
can overlap in time to improve scalability and resource utilization.
Non-blocking communication enables critical optimizations like pipelining, where an operation is split into stages so that communication is overlapped and interleaved with computation. This pattern will be particularly important in the following sections for scaling the advection solver efficiently at large process counts.

For GPU execution, Pyroclast relies on \textbf{GPU-enabled MPI} implementations
that support direct communication from device memory.
These make use of technologies such as NVIDIA GPUDirect~\cite{nvidia_gpudirect}
and Remote Direct Memory Access (RDMA)~\cite{rdma},
which enable data transfer between GPUs or network interfaces
without staging through host memory.
The same mechanisms are used for CPU-based RDMA communication,
allowing both CPU and GPU ranks to exchange halo or marker data asynchronously
while other computations continue.
This design enables Pyroclast to achieve low-latency communication
and efficient overlap of computation and data exchange
across multiple nodes.

\subsection{Rank-to-Device Mapping}

In GPU-enabled distributed configurations, Pyroclast employs a one-rank-per-GPU execution model. 
Each MPI rank is bound to a specific GPU device and manages all associated data and computation. 
This avoids resource contention and allows the solver to leverage CUDA-aware MPI communication for direct 
device-to-device data transfers. 
Although multi-GPU per rank configurations may be theoretically possible, they are not currently supported.

\section{Domain Decomposition of Structured Grids}
\label{sec:domain-decomposition}

Distributed-memory parallelism in Pyroclast is based on a structured domain decomposition,
where the global computational grid is divided into subdomains,
each managed by a separate MPI rank.
Every rank is responsible for updating its portion of the solution
and exchanging boundary data with neighboring ranks
to maintain global consistency.

\begin{figure}[h!]
    \centering
    \includegraphics[width=\textwidth]{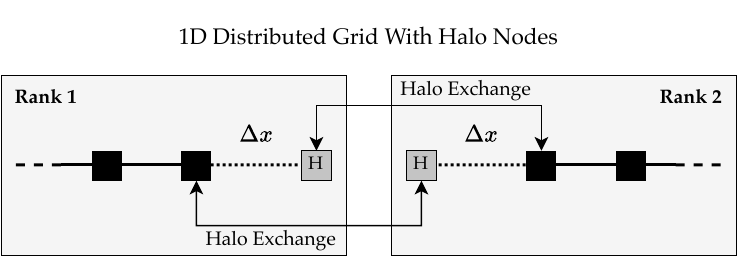}
    \caption{One-dimensional domain decomposition of a structured grid into two MPI ranks.
    Each subdomain owns its interior grid points and a set of halo nodes (H)
    that store copies of boundary values from neighboring ranks.
    After each iteration, data are exchanged across the interface through a halo exchange.}
    \label{fig:1D-domain-decomposition}
\end{figure}

To illustrate the basic concept, Figure~\ref{fig:1D-domain-decomposition}
shows a simple 1D decomposition of a uniform grid.
Each rank owns a contiguous segment of the domain,
including a number of additional grid points at its boundaries known as \emph{halo nodes}.
Halos store copies of data from neighboring subdomains
and are used to evaluate stencil operations that extend beyond local boundaries.
After each computational step, the updated boundary data are exchanged
between neighboring ranks in a process known as a \emph{halo exchange}.
This ensures that each subdomain has up-to-date values in its halo region
before the next stencil operation is applied.

\begin{figure}[h]
\centering
\includegraphics[trim={2cm 2cm 2cm 3cm}, clip, width=0.9\linewidth]{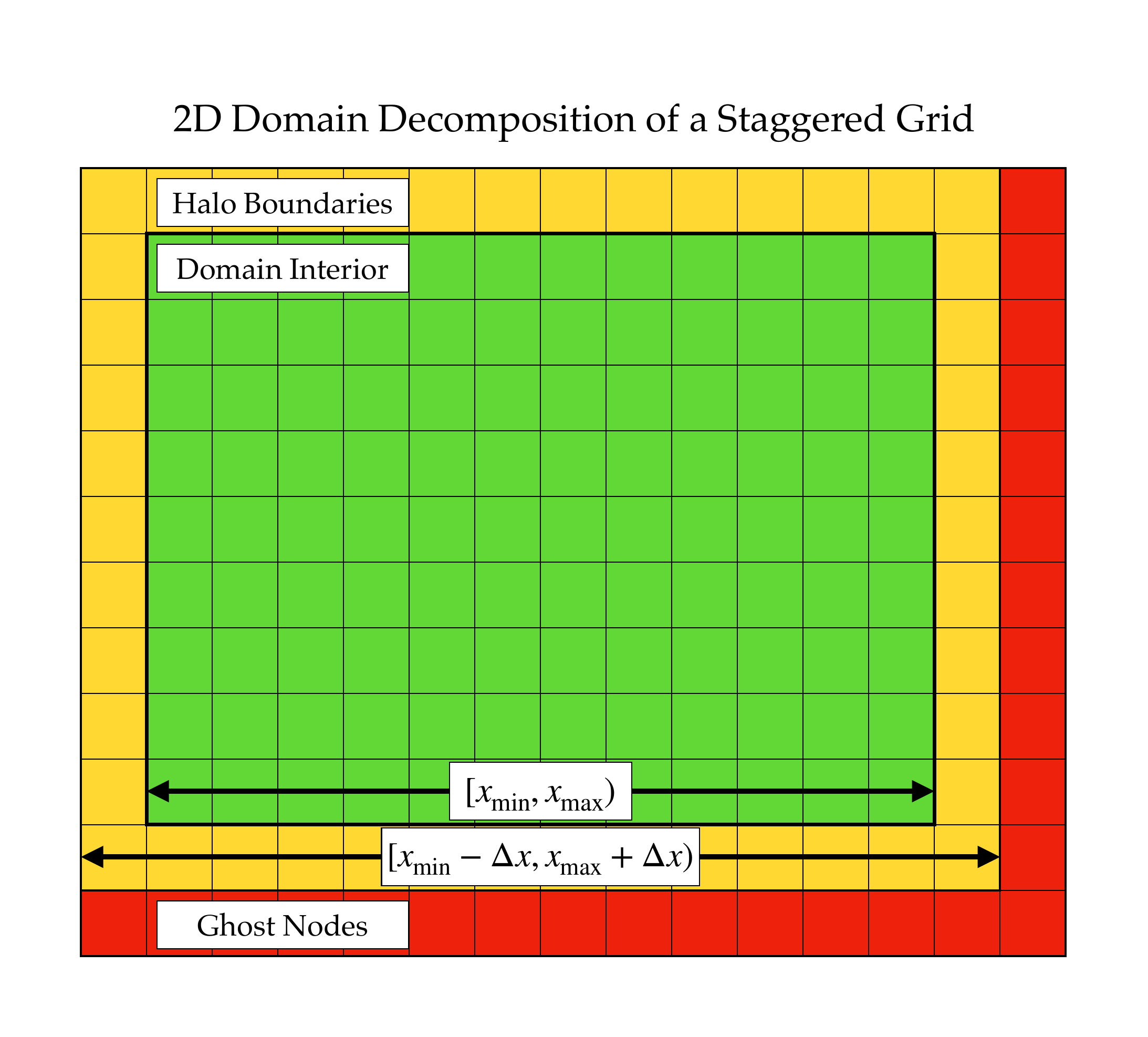}
\caption{Two-dimensional domain decomposition of a staggered grid.
Each subdomain includes an interior region (green), halo boundaries (yellow),
and ghost nodes (red).
Halo nodes store boundary data exchanged with neighboring ranks,
while ghost nodes ensure consistent array shapes for all staggered fields
and simplify grid operations.}
\label{fig:2d-staggered-domain-decomposition}
\end{figure}

For staggered grids, the concept extends naturally to multiple dimensions.
Figure~\ref{fig:2d-staggered-domain-decomposition} illustrates a 2D domain decomposition
showing the different regions of a subdomain.
The interior (green) represents the local computational domain,
the halo region (yellow) holds data referencing neighboring ranks,
and the outermost layer (red) contains \emph{ghost nodes}.
While halos and ghost nodes are both additional cells beyond the physical domain,
their purpose is very different.

Halo nodes represent quantities that reside on other ranks but are physically adjacent to the domain interior and must remain synchronized during execution.
Reads from halo nodes access values that belong to neighboring subdomains,
while writes to halo nodes represent updates to remote data.
Depending on context, these updates may correspond to replacing a value,
adding an increment, or performing a reduction such as accumulation or subtraction.
In contrast, ghost nodes exist only to maintain consistent array shapes
across all staggered quantities.
They serve the same purpose as described in Section~\ref{sec:staggered-grid-layout}
for the single-node staggered grid setup.
Ghost nodes allow all fields to share the same array dimensions
without introducing conditional branching in stencil or interpolation kernels.
They may be written to for convenience,
but they never hold meaningful physical data and are not included in halo exchanges.

Boundary conditions are handled consistently within this decomposition framework.
Periodic boundaries are straightforward to implement:
when a subdomain at the edge of the domain performs a halo exchange
with an out-of-bounds neighbor, the communication is redirected
to the corresponding subdomain on the opposite edge,
creating a periodic wrap-around.
This ensures periodic continuity of all quantities across boundaries
without requiring special treatment in the solver.

No-slip and free-slip boundary conditions require additional care.
At initialization, each rank checks whether its subdomain lies on a physical boundary.
If so, the subdomain interior is extended by one row and/or column of grid points
beyond the physical domain.
This repurposes the halo nodes as physical boundary nodes,
allowing boundary conditions to be applied directly on the same locations
used for halo storage.
This approach ensures that boundary subdomains require no special treatment beyond repurposing their halo regions
and allows the same stencil operations to work seamlessly with non-periodic boundary con

This decomposition strategy provides a clear mapping between local and global data,
facilitates efficient stencil evaluation near boundaries,
and forms the foundation for all distributed operations in Pyroclast,
including halo exchanges, marker advection, and distributed multigrid.

\section{MPI Communicators and Cartesian Topology}

In MPI, all communication occurs within the context of a \emph{communicator}.
A communicator defines a group of ranks that can exchange messages with one another
and provides the communication scope for collective and point-to-point operations.
Every MPI program starts with the global communicator \texttt{MPI\_COMM\_WORLD},
which includes all ranks in the simulation,
but additional communicators can be created to represent specific subgroups
or topologies that match the structure of the problem.
By organizing ranks into communicators that mirror the data layout,
MPI allows communication patterns to be expressed in a way that directly corresponds
to the underlying virtual logical domain.

The MPI standard provides several types of communicators designed to organize ranks
in a way that reflects the topology of the workload.
In addition to general graph topologies, MPI offers a specialized communicator
for Cartesian domains.
Since Pyroclast operates on rectangular or cuboidal grids,
the natural choice is to use a Cartesian communicator.

Using a Cartesian communicator offers two main advantages.
First, it maps ranks directly onto a logical 2D or 3D process grid,
allowing the solver to reason in terms of subdomain indices
\((p_x, p_y, p_z)\) rather than hardware-specific rank identifiers.
This abstraction removes the need for manual mapping
between domain coordinates and rank IDs
and decouples the virtual layout of the computational domain
from the actual process pinning on the machine.
As a result, the communication pattern is defined purely in terms of the simulation domain,
independent of the underlying hardware configuration.

Second, because the communicator is aware of the logical adjacency of ranks,
MPI can automatically reorder or “shuffle” ranks at initialization
to minimize communication distance on the hardware interconnect~\cite{MPI-5.0}.
Neighboring subdomains that exchange halo data are thus placed as close as possible
in the network topology, significantly reducing latency and data transfer costs.
The communicator also supports periodicity along individual axes,
allowing subdomains on opposite boundaries to be linked automatically
for periodic boundary conditions.

In Pyroclast, this Cartesian topology forms the backbone of all halo and marker exchange operations.
It provides a clean, geometry-aware layer for communication
that scales naturally for both two-dimensional and three-dimensional problems.

\section{Markers in Distributed Domains}

Extending the domain decomposition model to include Lagrangian markers
requires careful definition of ownership, reference frames, and communication rules.
Unlike grid-based quantities, which are defined on fixed Eulerian nodes,
markers move freely across the domain and may cross subdomain boundaries during advection.
Each MPI rank must therefore determine which markers belong to its local domain
and be able to transfer those that leave to neighboring ranks.

\begin{figure}[h!]
    \centering
    \includegraphics[trim={2cm 2cm 2cm 3cm}, clip, width=0.9\linewidth]{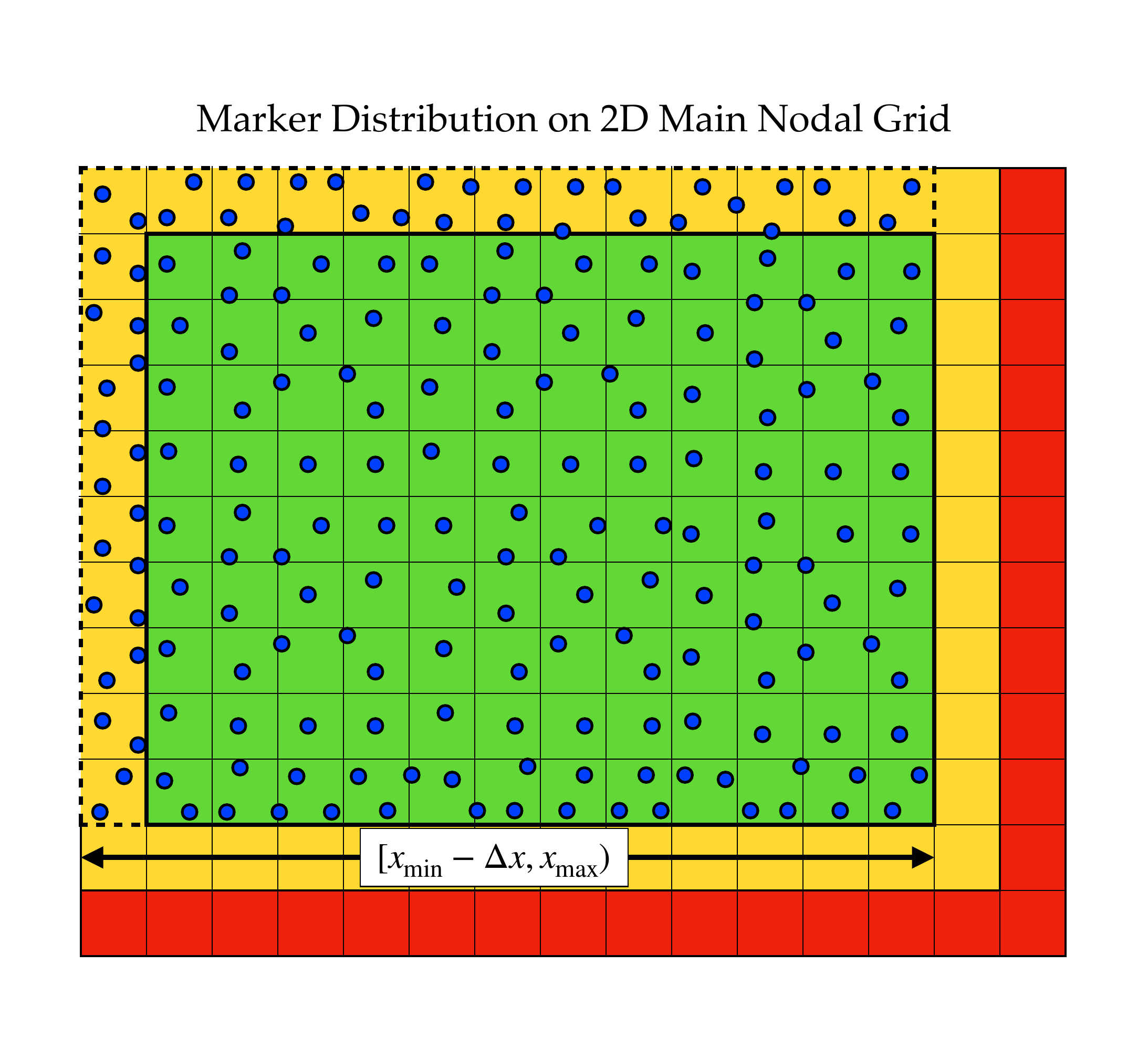}
    \caption{Distribution of Lagrangian markers within a 2D subdomain of the main nodal grid.
    The interior region (green) contains local markers,
    while additional markers occupy the north and west halo gaps (yellow),
    which are managed by this rank as indicated by the dashed boundary.
    Ghost cells (red) do not hold markers and serve only as padding.}
    \label{fig:marker-main-grid}
\end{figure}

To understand the issue, it is useful to recall the 1D distributed example
shown previously in Figure~\ref{fig:1D-domain-decomposition}.
At the interface between two subdomains,
there exists a small gap corresponding to the space between the last node of one rank
and the first node of the next.
This region does not belong to the interior of either rank in the Eulerian sense,
but it must still contain markers, since markers are located between grid nodes.
To avoid ambiguity, Pyroclast adopts the convention that each rank
is responsible for managing all markers in its interior domain
and in the north and west interface gaps shared with neighboring subdomains.
This ensures that every point in the global domain is uniquely assigned to a rank.

Figure~\ref{fig:marker-main-grid} illustrates how markers are distributed
with respect to the main nodal grid of a subdomain.
Markers populate both the interior and the north and west halo gaps,
ensuring complete spatial coverage of the physical domain
without overlap or redundancy between neighboring ranks.
When boundary conditions require extending the domain,
as described in the previous section,
the additional grid cells become part of the interior domain,
and markers in those regions are handled accordingly,
ensuring consistency between Eulerian and Lagrangian reference frames.

Each rank stores marker coordinates in the global reference frame,
consistent with the coordinate system used by the grid.
While it would be possible to store positions in a local coordinate frame
relative to the subdomain origin, this would provide no computational benefit
as long as all operations use a consistent reference.
The parameters \(x_{\min}\) and \(x_{\max}\) (and analogously \(y_{\min}\), \(y_{\max}\))
denote the coordinates of the extremities of the interior domain
handled by a given rank.
During advection, a marker may move outside the extended region
\([x_{\min} - \Delta x, x_{\max}) \times [y_{\min} - \Delta y, y_{\max})\),
which includes the halo gaps assigned to the rank.
When this occurs, ownership of the marker is transferred to the neighboring rank
whose interior domain covers the marker’s new position.
This migration step ensures that markers remain correctly assigned within the process grid
and that material properties are consistently advected across subdomain boundaries.

\subsection{Distributed Marker Interpolation}

Extending marker-grid interpolation to a distributed-memory setting
requires particular care to ensure that all contributions are correctly accumulated
across subdomain boundaries.
In Pyroclast, the same interpolation routines used in the single-node solver and discussed in Section~\ref{sec:marker-interpolation} are also used in the distributed setting,
with the only difference being that halo exchanges are required
whenever data dependencies cross subdomain boundaries.

Following the principle introduced in Section~\ref{sec:domain-decomposition},
reads from halo nodes represent remote reads,
and writes to halo nodes correspond to remote writes.
This means that every interpolation operation that accesses data in a halo region
implicitly requires a halo exchange to propagate information
between neighboring ranks.

\begin{figure}[h!]
    \centering
    \includegraphics[trim={2cm 2cm 2cm 3cm}, clip, width=0.9\linewidth]{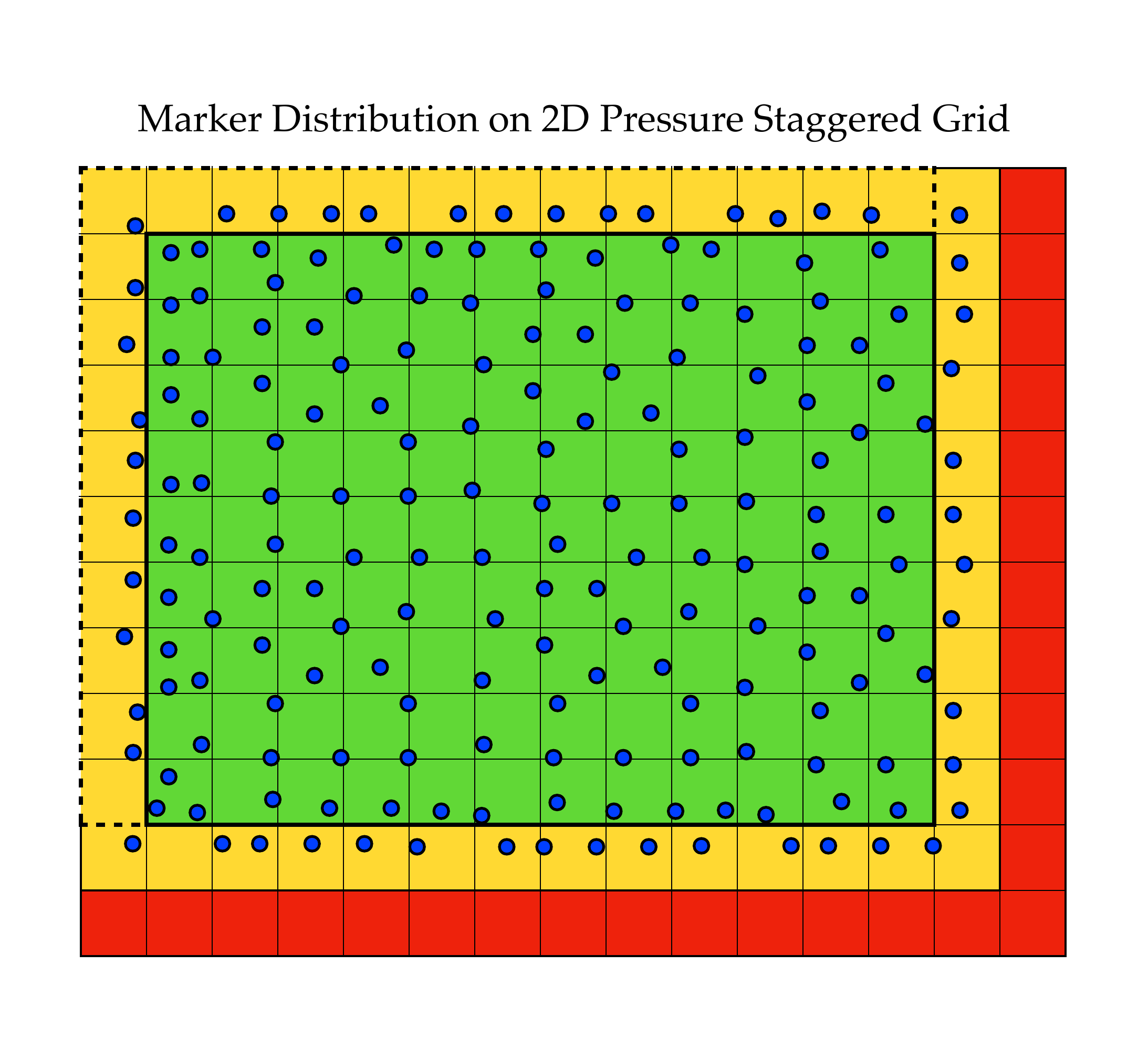}
    \caption{Distribution of markers on the 2D pressure-staggered grid.
    The pressure grid is shifted by \(-\Delta x\) and \(-\Delta y\)
    relative to the main nodal grid, which causes the marker positions
    to appear offset by \(+\Delta x\) and \(+\Delta y\).
    As a result, markers near the subdomain boundaries
    contribute to grid nodes belonging to all four neighboring ranks.
    All halo cells (yellow) containing markers require halo exchanges
    to correctly propagate interpolation data between subdomains.}
    \label{fig:marker-pressure-grid}
\end{figure}

At first glance, the marker distribution shown in
Figure~\ref{fig:marker-main-grid} might suggest
that only the north and west halos are modified during marker-to-grid interpolation.
One might therefore assume that communicating updates only to the north and west neighbors
would be sufficient to complete the operation.
However, this reasoning overlooks the fact that Pyroclast employs
a staggered grid layout, where each physical quantity is defined
on a different reference grid that may be shifted by
\(-\Delta x\), \(-\Delta y\), or both.
As a result, each subdomain actually hosts multiple overlapping grids,
and markers associated with one quantity may project their contributions
onto grid points belonging to up to four neighboring subdomains.

This situation can be visualized by considering the distribution of markers
on the pressure grid, shown in Figure~\ref{fig:marker-pressure-grid}.
Since the pressure grid is staggered by \(-\Delta x\) and \(-\Delta y\)
relative to the main nodal grid,
its associated markers appear shifted by \(+\Delta x\) and \(+\Delta y\).
Consequently, some of these markers contribute to nodes located
beyond the local subdomain boundaries,
requiring interpolation results to be exchanged with all four adjacent ranks.

\FloatBarrier

\subsubsection*{Markers-to-Grid Distributed Interpolation}

As usual, interpolating marker values to the grid is considerably more complex than interpolating grid values back to markers.
Each marker contributes to the grid nodes surrounding its position,
and, as shown in Figure~\ref{fig:marker-pressure-grid}, some of these nodes may belong to neighboring subdomains.
Ensuring that all contributions are correctly accumulated across process boundaries
therefore requires a distributed reduction procedure.

In Pyroclast, this operation is implemented as a two-stage process
that guarantees consistency between subdomains
while maintaining full compatibility with the single-node interpolation kernels.

\textbf{Stage 1: Local accumulation and first halo exchange.}
Each rank begins by computing the local contributions of its markers
to the weighted values and weights arrays defined on the local grid.
Interior grid nodes already contain complete data after this stage,
but the nodes along subdomain boundaries contain only partial sums,
since they are missing the contributions of markers residing in neighboring ranks.
Simultaneously, the halo regions contain contributions from local markers
to grid nodes that belong to adjacent subdomains.
To reconcile these partial values, a first halo exchange is performed:
each rank sends its halo values to its neighbors
and receives their corresponding contributions.
The received data is added to the local grid’s interior edge nodes, ensuring that the effective contribution of remote markers is accounted for correctly.

\textbf{Stage 2: Normalization and second halo exchange.}
Once all contributions have been received and accumulated,
each rank normalizes its weighted values by the accumulated weights
to obtain the final interpolated quantities.
At this point, the halo regions still contain only local contributions
to remote grid nodes.
A second halo exchange is therefore performed to synchronize the normalized data,
ensuring that all halo regions reflect the correct boundary values.
After this step, every rank holds a consistent and fully updated copy
of the distributed grid field.

The overall algorithm is summarized in Algorithm~\ref{alg:distributed_marker2grid}.
This design ensures that distributed interpolation behaves identically
to the single-node implementation, while minimizing redundant communication.

\begin{algorithm}[h!]
\caption{Distributed marker-to-grid interpolation\\using two-stage reduction}
\label{alg:distributed_marker2grid}
\KwIn{Marker positions $(x_m, y_m)$, marker values $v_m$,\\grid coordinates $(x, y)$.}
\KwOut{Distributed grid arrays $\texttt{grid\_values}$ and $\texttt{grid\_weights}$.}

\BlankLine
\textbf{Stage 1: Local accumulation}\\
\ForEach{marker $m$ in local subdomain}{
    \ForEach{node $(i,j)$ surrounding marker $m$}{
        Compute interpolation weight $w_m^{(i,j)}$\;
        $\texttt{grid\_values}[i,j] \mathrel{+}= v_m w_m^{(i,j)}$\;
        $\texttt{grid\_weights}[i,j] \mathrel{+}= w_m^{(i,j)}$\;
    }
}
\BlankLine
\textbf{First halo exchange and reduction}\\
Each rank sends its halo regions (local contributions to remote grids) to neighboring ranks\;
Each rank receives the corresponding boundary regions (remote contributions to its own interior grid nodes)\;
Accumulate received data into the interior edge nodes of $\texttt{grid\_values}$ and $\texttt{grid\_weights}$\;
\BlankLine
\textbf{Stage 2: Normalization and synchronization}\\
Normalize interior nodes: $\texttt{grid\_values} := \texttt{grid\_values} / \texttt{grid\_weights}$ (element-wise)\;
\BlankLine
\textbf{Second halo exchange}\\
Each rank sends the normalized boundary layers of its interior domain
to neighboring ranks so they can update their halo regions\;
Each rank receives normalized values from neighbors
and writes them into its local halo regions\;
\Return{$\texttt{grid\_values}$}\;
\end{algorithm}

\subsubsection*{Grid-to-Markers Distributed Interpolation}

The grid-to-marker interpolation is considerably simpler than the marker-to-grid case.
Each marker computes its value from the four surrounding grid nodes,
which are guaranteed to be available locally as long as the halo regions
of the grid fields are up to date.
If the field halo values are stale, before performing interpolation, a single halo exchange of the relevant grid quantities
is sufficient to fetch the most recent data from neighboring ranks.
Once the halos are synchronized, every rank can interpolate marker values
independently using the same kernels as in the single-domain implementation.
No further communication is required.

\section{Marker Advection Strategies}

With distributed interpolation established, we can now address the problem
of advecting markers across subdomain boundaries.
Each MPI rank is responsible for tracking the markers that reside
within its local portion of the domain.
When a marker leaves this region during advection,
it must be transferred to the rank that owns the subdomain it has entered.
This process is handled through a \emph{compaction} and \emph{migration} operation
performed at every advection step.

After advection, all markers are classified as either active (remaining in the local domain)
or outbound (leaving the domain).
Outbound markers are packed into send buffers and communicated to the appropriate ranks.
Incoming markers from neighboring ranks are then appended to the local arrays,
and all active markers are compacted into contiguous memory.
This guarantees that no inactive markers are left behind
and that subsequent advection and interpolation steps operate on a dense data layout.

Currently, Pyroclast reallocates memory at each advection step
to match the exact number of active plus incoming markers.
In principle, this overhead could be reduced by adopting an exponential allocation strategy.
In such a scheme, the marker arrays are expanded only when necessary,
for example by a factor of $1.3\times$--$1.5\times$, and the compaction step ensures that
all active markers are grouped at the start of the array.
This amortizes memory reallocation costs and improves cache efficiency.

Based on this data management model,
two different advection strategies can be implemented,
each with its own trade-offs in performance and scalability.

\subsection*{Halo-Region Constrained Advection}

In this approach, the timestep is restricted such that
markers cannot move beyond the halo regions of the neighboring subdomains
within a single advection step.
In two dimensions, this means that markers can travel only to one
of the eight adjacent locations (N, S, E, W, NE, NW, SE, SW),
and in three dimensions to one of the twenty-six adjacent subdomains.
The destination is determined by checking which boundaries the marker crosses:
for instance, if $x > x_{\max}$ but $y < y_{\max}$, the marker moves east,
while crossing both conditions implies a northeast transfer.

This strategy has the advantage of keeping communication strictly local,
as each rank exchanges markers only with its immediate neighbors.
However, it also imposes a non-physical constraint on the timestep.
As resolution increases, the maximum allowable timestep decreases,
even if stability or physical considerations would permit a larger step.
A possible remedy is to employ \emph{subcycling},
where several smaller advection updates are performed per global timestep.
At present, Pyroclast implements this halo-constrained advection model.

\subsection*{Free Marker Advection}

The free-advection strategy removes the constraint on marker displacement.
Each rank stores the spatial bounds of all subdomains,
which are exchanged once at initialization.
After advection, every marker determines its new owner rank
through a bisection search over the domain bounds.
Ranks then perform an \texttt{MPI\_Alltoall} operation
to exchange the number of markers to send and receive.
This step involves communicating a single integer per rank
and is therefore inexpensive.

Once the communication pattern is known,
each rank posts non-blocking point-to-point transfers
to send and receive marker data only from the necessary ranks.
This allows markers to migrate across multiple subdomains in a single step
without timestep restrictions.
The drawback is that communication can become non-local,
depending on the spatial distribution and hardware mapping of ranks.
This may increase latency and data transfer cost,
and requires additional synchronization and bookkeeping.
While more flexible, this approach is also more complex to implement
and may introduce communication bottlenecks in large-scale runs.

\subsection*{A Note on Distributed Runge-Kutta Advection}

Implementing multi-stage Runge-Kutta (RK) advection schemes in a distributed environment
introduces several challenges that do not arise in single-node execution.
As described in Chapter~\ref{chap:numerical_methods},
RK schemes improve trajectory accuracy by evaluating the velocity field
at multiple intermediate positions within each timestep.
In a distributed setting, however, these intermediate evaluations complicate
the advection process, since markers may temporarily leave the local subdomain
between successive stages.

At each RK substep, the velocity must be interpolated at the marker’s predicted position.
If that position lies outside the local domain, the required grid values reside
on a different MPI rank, necessitating additional halo exchanges or temporary marker migration.
Performing such data transfers for every substage would introduce substantial communication overhead
and synchronization.

For this reason, Pyroclast currently employs the simple forward Euler scheme
for distributed advection, as it requires only one velocity evaluation per timestep.
An alternative scheme that is under investigation is the locally polynomial integration (LPI) scheme
introduced in Chapter~\ref{chap:numerical_methods}.
Because LPI analytically integrates a local Taylor expansion of the velocity field,
it approximates the effect of higher-order RK schemes without requiring
multiple velocity interpolations.
This makes it particularly well suited for distributed advection,
as markers can be advanced correctly even if their trajectories
temporarily cross subdomain boundaries.

\section{Distributed Multigrid and RAS Preconditioning}

Extending multigrid solvers to distributed-memory architectures requires balancing
local computation with global communication.
In Pyroclast, this is achieved through a domain decomposition approach
where each MPI rank performs a local multigrid solve on its subdomain,
using boundary data exchanged with neighboring ranks.
This strategy follows the principles of the Restricted Additive Schwarz (RAS)
preconditioning method~\cite{hackbusch_book, yousef_saad_iterative_methods_linear_systems}.

In the RAS framework, the global domain is divided into overlapping subdomains,
each of which includes an extended halo region that acts as a buffer zone.
Within each subdomain, a local multigrid cycle is executed independently,
treating the values at the edge of the overlap as fixed Dirichlet boundaries.
After a few local smoothing and restriction–prolongation steps,
the subdomain solutions are combined across ranks through halo exchanges
to maintain global consistency.
The restricted overlap ensures that the Dirichlet boundaries do not introduce
artificial discontinuities, while keeping communication localized and asynchronous.

In Pyroclast, the distributed multigrid implementation follows a hybrid model:
each rank maintains a complete multigrid hierarchy for its local domain,
with fine levels optionally offloaded to GPUs and coarse levels executed on CPUs.
Communication between ranks occurs only at the top level of the hierarchy,
or periodically every few V-cycles, depending on the convergence rate.
This minimizes synchronization frequency while allowing the multigrid preconditioner
to capture long-range coupling effectively.

To further reduce boundary artifacts, the overlap between subdomains can be increased
beyond a single halo width.
These redundant updates extend the smoothing region across neighboring domains,
providing a buffer that mitigates the effect of artificial Dirichlet constraints.
The result is an efficient, scalable preconditioner that combines the advantages
of geometric multigrid within subdomains with additive Schwarz coupling across ranks.

The distributed multigrid and RAS preconditioning framework in Pyroclast
is currently under active development.
While the design and communication model have been developed,
quantitative performance and scalability tests have not yet been conducted.
These experiments will form part of future work
as the distributed solver components mature.

\chapter{Benchmarks and Validation}
\label{chap:benchmarks}

This chapter presents a series of numerical benchmarks designed to validate
the correctness, stability, and parallel performance of Pyroclast.
The benchmarks cover both shared-memory execution of the Stokes solver
on CPU and GPU architectures, and distributed-memory execution of the
Lagrangian advection solver using MPI.
All tests were performed under the same hardware and software environment
to ensure consistent conditions, and unless otherwise noted,
all runs were carried out in double precision.

The results shown here represent preliminary performance assessments
of the current Pyroclast implementation.
Although they are early in the development cycle,
the benchmarks already demonstrate excellent numerical robustness
and very promising scalability across both CPU and GPU architectures.
Significant optimization and tuning remain to be done,
particularly for GPU kernels, multigrid hierarchy management,
and distributed-memory components.
Future work will focus on improving performance portability
and extending the benchmark suite to fully coupled thermo-mechanical simulations.

\section{Hardware and Software Setup}
\label{sec:hardware_setup}

All benchmarks were executed on the \emph{Piora} cluster at the Swiss National Supercomputing Centre (CSCS).
The cluster consists of seven compute nodes connected via an InfiniBand network.
The system is heterogeneous, comprising two new-generation GPU-equipped nodes
and five earlier-generation CPU nodes.

The two new nodes are equipped with dual-socket AMD EPYC~7773X processors
(64~cores per socket, 128~cores total),
each paired with four NVIDIA~A100 GPUs and 1~\si{\tera\byte} of system memory.
The remaining five nodes use dual-socket AMD EPYC~7742 CPUs
with 128 total cores and 512~\si{\giga\byte} of memory.

InfiniBand interconnect speeds vary between
100~\si{\giga\bit\per\second} on the older nodes
and 200~\si{\giga\bit\per\second} on the newer ones,
corresponding respectively to Mellanox MT4115 and MT4123 host adapters.

All computations were performed in a uniform software environment
based on the NVIDIA PyTorch container~\cite{nvidia_pytorch_container},
executed via Apptainer~\cite{apptainer_1, apptainer_2}.
The container includes CUDA~12.9 and a modern Python HPC software stack.
Key packages used by Pyroclast include
\texttt{NumPy~1.26.4}, \texttt{CuPy~13.2}, \texttt{Numba~0.62}, \texttt{Numba-Cuda~0.20.0},
\texttt{mpi4py~4.0.3}, and \texttt{nanobind~2.2.0}.
MPI communication was provided by Open~MPI~v4.1.7,
initialized via PMIx and managed by Slurm.
Multithreading in Numba was handled through the OpenMP backend.

All runs were executed with explicit process pinning to ensure full control
over process placement and memory locality.
CPU affinity and NUMA bindings were enforced via Slurm as well as the appropriate OpenMP environment variables.
Each benchmark included a dedicated warm-up phase preceding the timed execution.
During the benchmark run, multiple iterations of the solver were performed,
and the total runtime over all iterations was recorded.
This approach minimizes the influence of transient effects
and provides a stable measure of sustained performance.
All tests produced deterministic and reproducible results
across different nodes with identical hardware configurations.

This environment provides a representative hybrid CPU–GPU platform
with modern compilers, communication libraries, and runtime management,
ensuring that the benchmarks accurately reflect the performance characteristics
of Pyroclast on current high-performance computing hardware.

\section{Shared-Memory Stokes Solver Benchmarks}
This section evaluates the performance of the Stokes solver in a shared-memory environment,
focusing on CPU and GPU execution within a single compute node.
The goal is to assess how efficiently Pyroclast exploits thread-level parallelism
and hardware acceleration for large two-dimensional systems.
All tests were performed on the same setup used in the numerical stability study
presented in Section~\ref{sec:stability}, using the sinker benchmark
as a representative visco-mechanical problem.

The Stokes solver benchmarks primarily target three aspects:
the convergence behavior at increasing grid resolutions,
the strong scaling performance across CPU cores,
and the achievable speedup when executing on a GPU compared to multicore CPUs.
Together, these experiments provide an initial assessment of the computational
efficiency of Pyroclast’s matrix-free multigrid implementation
and its suitability for high-resolution geodynamic simulations.

\subsection{Benchmark Setup}

The benchmark configuration follows the \emph{sinker problem}
described in Section~\ref{sec:stability}.
All material parameters, boundary conditions, and solver settings
are identical to those used in the numerical stability tests,
ensuring consistency between validation and performance analyses.

Four grid resolutions were tested:
\(2500\times2500\), \(5000\times5000\), \(10000\times10000\), and \(15000\times15000\).
For each resolution, the multigrid hierarchy was configured with six grid levels,
using a geometric scaling factor determined at runtime
so that the coarsest grid contained approximately \(30\times30\) nodes.
A standard Jacobi smoother was used throughout,
with five pre-smoothing and five post-smoothing iterations on the finest grid.
The number of smoothing iterations on coarser levels followed the same scaling strategy
as in Section~\ref{sec:stability}.
The solver was run for 500 Uzawa iterations,
each consisting of a single multigrid V-cycle and pressure correction,
and the total runtime was recorded.

All tests employed the standard Jacobi smoother.
The GPU benchmarks were executed on the newer AMD~EPYC~7773X nodes,
using 32~CPU threads.
CPU-only benchmarks were performed on the earlier AMD~EPYC~7742 nodes.
This hardware difference has negligible influence on performance
since the Stokes solver is primarily memory-bound
and exhibits similar scaling characteristics across both processors. The GPU threshold was set to \(2000\times2000\).

\subsection{Convergence}

Before analyzing performance, we first verify that the Stokes solver exhibits
resolution-independent convergence across all benchmark configurations.
The observed behavior closely matches the results presented in
Section~\ref{sec:stability}, where robustness under extreme viscosity contrast
was demonstrated.

\begin{figure}[h!]
    \centering
    \includegraphics[width=0.95\textwidth]{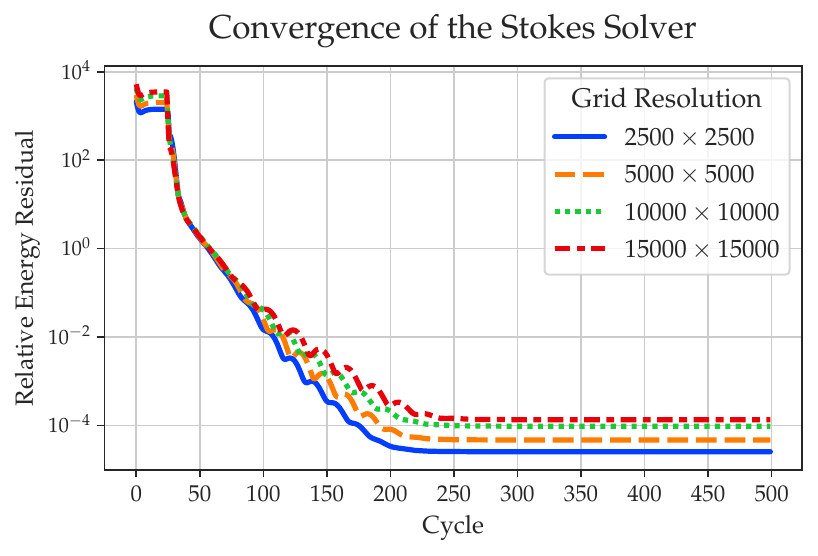}
   \caption{Relative energy residual over 500 Uzawa cycles for four grid
resolutions. All runs converge to \(\sim10^{-4}\) with nearly identical
trajectories, indicating resolution-independent solver performance.}
    \label{fig:convergence}
\end{figure}

Figure~\ref{fig:convergence} reports the relative energy residual over
500 Uzawa iterations for all tested grid resolutions.
In every case, the solver reaches a final residual on the order of
$10^{-4}$, corresponding to a relative energy error of roughly $0.01\%$.
The convergence trajectories for the four resolutions are nearly identical:
both the decay rate and the final error level are effectively independent of
grid spacing.
This confirms that the surrogate residual metric used by Pyroclast behaves
as intended and that the multigrid hierarchy and smoother configuration
produce consistent, resolution-independent convergence.

These results show that all test cases converge to the same physical solution
with no degradation in numerical stability.

\FloatBarrier
\subsection{Strong Scaling Performance}

\begin{figure}[h!]
    \centering
    \includegraphics[width=0.95\textwidth]{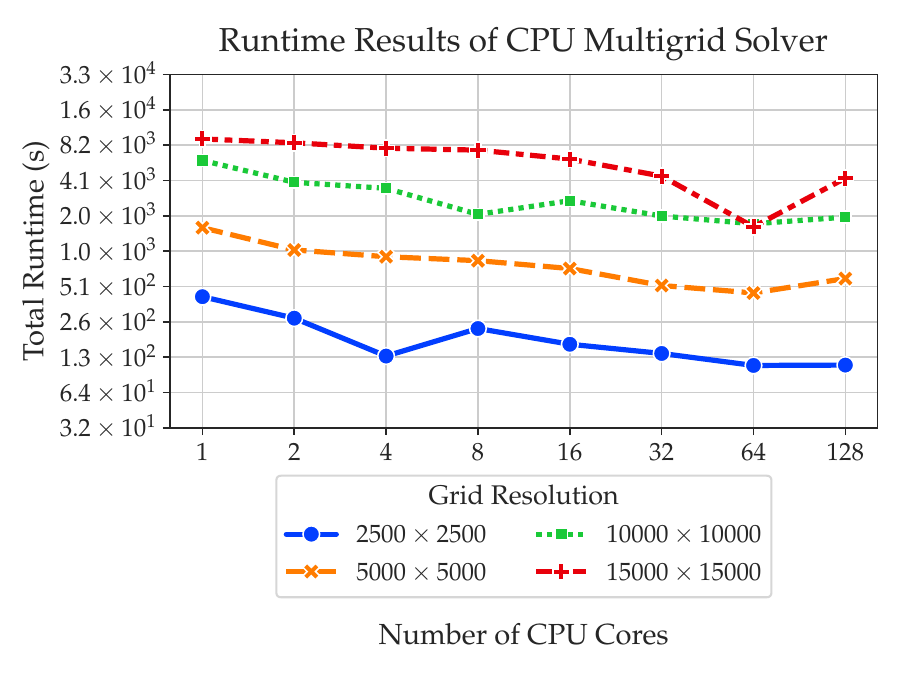}
   \caption{Total runtime of the shared-memory Stokes solver for
500 Uzawa iterations across four grid resolutions.
Performance improves with thread count, but scaling is limited
by the memory-bound nature of the multigrid kernels.}

    \label{fig:stokes_runtime_results}

    \centering
    \includegraphics[width=0.95\textwidth]{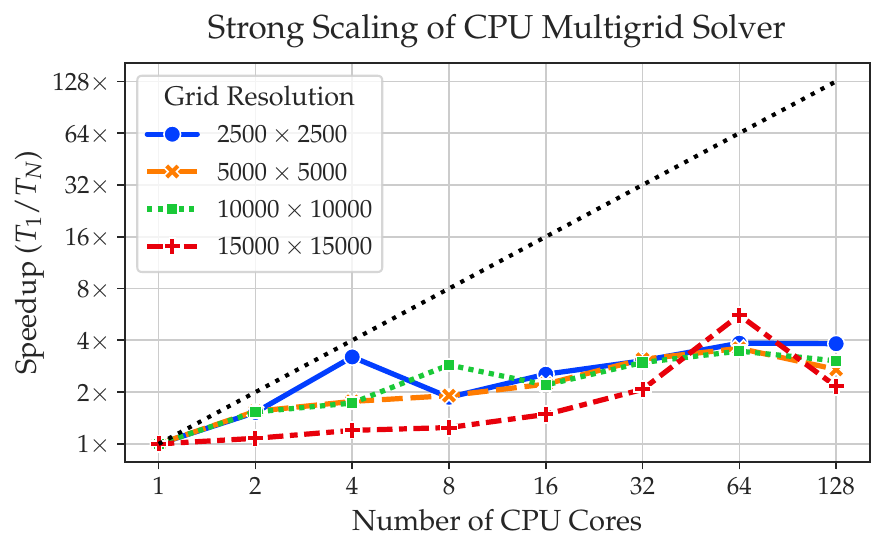}
    \caption{Strong scaling of the shared-memory Stokes solver.
Speedup improves up to 32--64 threads before saturating due to
memory-bandwidth constraints.
The dashed line indicates ideal linear speedup.}

    \label{fig:strong_scaling}
\end{figure}

To evaluate the shared-memory parallel efficiency of the Stokes solver,
we measure strong scaling on four grid resolutions
(\(2500\times2500\), \(5000\times5000\), \(10000\times10000\),
and \(15000\times15000\))
using between 1 and 128 OpenMP threads.
Figure~\ref{fig:stokes_runtime_results} reports the total runtime of
500 Uzawa iterations for each configuration, while
Figure~\ref{fig:strong_scaling} shows the corresponding parallel speedup
defined as \(S_p = T_1 / T_N\), where $T_1$ is the runtime with 1 thread and $T_N$ with $N$ threads.

The results indicate that OpenMP scaling is modest across all resolutions.
This behaviour is expected: the multigrid V-cycle is dominated by
memory-bound stencil operations whose arithmetic intensity is too low
to take advantage of many CPU cores.
As the number of threads increases, performance quickly becomes limited
by memory bandwidth and by cache- and NUMA-related effects.
Consequently, speedup improves up to roughly 32--64 threads but
stagnates or even slightly degrades at 128 threads,
consistent with the characteristics of dual-socket systems~\cite{ferrari_3d_blocking}.

Despite these limitations, the absolute runtimes remain competitive,
and the solver maintains stable performance up to the highest tested core counts.
We expect significantly better strong-scaling efficiency in three-dimensional
problems, where the increased computational work per grid point
leads to higher arithmetic intensity and better utilisation of
shared-memory parallelism.

\FloatBarrier

An interesting outlier appears for the \(15000\times15000\) configuration at
64~CPU cores, which exhibits significantly better speedup than neighbouring
thread counts or smaller problem sizes.
The underlying cause of this behaviour is not yet fully understood.
However, the result was reproduced consistently across multiple runs
and on different nodes with identical hardware, suggesting that it is
a genuine performance characteristic rather than a measurement artefact.
Further profiling will be required to determine whether this peak is due
to favourable NUMA effects, cache behaviour, or other architectural interactions.

\subsection{GPU Acceleration and CPU Comparison}
\begin{figure}[h!]
    \centering
    \includegraphics[width=\textwidth]{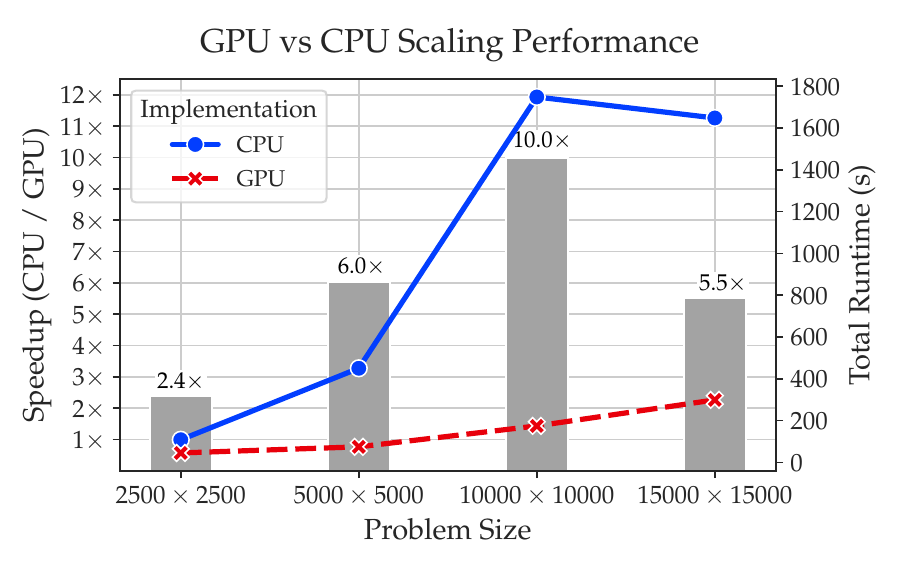}
    \caption{
    GPU acceleration compared to the fastest CPU configuration for each grid size.
    Lines show absolute runtimes (right axis), while bars show the speedup of
    the A100 over the best CPU run (left axis).
    GPU efficiency increases markedly with problem size, peaking at a \(10\times\)
    speedup for the \(10000\times10000\) grid, reflecting the advantages of
    device-resident stencil operations and high memory bandwidth.}
    \label{fig:gpu_speedup}
\end{figure}

To assess the benefit of GPU acceleration, we compare the runtime of the
single--GPU implementation (one NVIDIA~A100) against the best
multi-threaded CPU configuration for each grid resolution.
Figure~\ref{fig:gpu_speedup} reports both the absolute GPU runtimes
(right axis) and the speedup of the GPU over the fastest CPU run
(left axis).

For small to moderate grid sizes, the GPU is underutilized:
the \(2500\times2500\) case achieves only a modest speedup of
approximately \(2.4\times\). As problem size increases, GPU performance improves substantially.
The speedup peaks at \(10\times\) for the \(10000\times10000\) grid,
reflecting excellent utilisation of the A100's high memory bandwidth and
device-resident multigrid execution.
The \(15000\times15000\) case still shows a solid speedup of \(5.5\times\),
although this value is artificially reduced by the unusually fast
CPU runtime observed for this specific configuration.
The GPU runtime continues to scale smoothly with problem size,
indicating that the drop in speedup is caused by the CPU outlier rather
than any limitation on the GPU side.

Overall, the results confirm that the GPU implementation is highly
efficient for large-scale Stokes problems.
The full GPU-resident multigrid hierarchy avoids PCIe/NVLink transfers,
and is able to efficiently exploit the GPU's memory bandwidth to accelerate the stencil operations that dominate the runtime cost.
These findings highlight the importance of problem size:
while small grids may not saturate the GPU, large grids
fall squarely in the regime where GPU acceleration provides
significant performance gains.

\section{Distributed Advection Benchmarks}

This section evaluates the performance and scalability of Pyroclast’s
distributed Lagrangian advection solver implemented with MPI.  
While full MPI support for the Eulerian Stokes components is still under
development, the advection solver already provides a complete and
fully distributed implementation, making it an ideal test case for
assessing parallel efficiency across multiple nodes.

Our benchmarks focus on weak-scaling behaviour, where the global problem
size increases proportionally to the number of MPI ranks so that the
computational load per rank remains constant.  
This choice reflects the intended usage of marker-based advection in
large-scale geodynamic simulations, where the total number of markers
grows with domain size.

The tests examine how halo exchange, marker migration, and stencil-based
interpolation perform under increasing levels of parallelism and heterogeneous
cluster hardware.  
All measurements were performed with the same warm-up procedure and pinning
strategy described in Section~\ref{sec:hardware_setup}, ensuring consistent
and reproducible results.

\subsection{Benchmark Setup}

To assess the performance of Pyroclast’s distributed marker advection
capabilities, we conducted a weak-scaling benchmark in which each compute
node is assigned a fixed local problem size of \(5000\times 5000\) grid cells.
The global domain size therefore increases proportionally with the number
of nodes:  
\[
5000\times 5000,\quad
7100\times 7100,\quad
10000\times 10000,\quad
13300\times 13300
\]
for runs on 1, 2, 4, and 7 nodes respectively.
This ensures that the computational load per rank remains constant,
allowing us to measure the scalability of communication, halo exchange,
and marker migration.

\begin{figure}[hb!]
  \centering
  \includegraphics[width=0.75\linewidth]{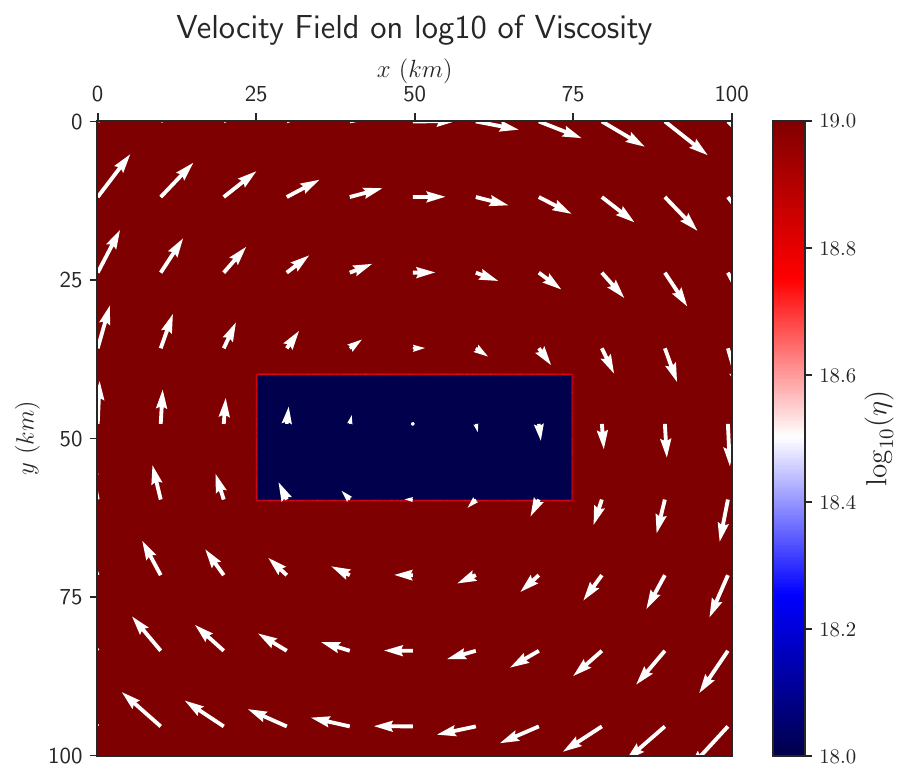}
  \caption{
  Benchmark configuration for the distributed advection test:
  a rotating velocity field superimposed on a viscosity map containing
  a high-viscosity rectangular inclusion.  
  The smooth rotational flow causes markers to migrate across subdomain
  boundaries, providing a controlled test of distributed advection,
  halo exchange, and marker migration.}
  \label{fig:advection_material_setup}
\end{figure}

The benchmark problem consists of a rotating slab of viscous material
advected by a prescribed, divergence-free velocity field.
The velocity field induces a smooth rotational motion around the centre
of the domain, with a high-viscosity rectangular inclusion embedded
in the background material (Figure~\ref{fig:advection_material_setup}).
Although marker-to-grid distributed interpolation is not necessary when dealing with a constant velocity field, we still perform marker interpolation exactly as if we were solving the Stokes system between advection steps.
This configuration provides a controlled test in which marker trajectories
continuously cross subdomain boundaries, ensuring that the benchmark
exercises both halo exchange and marker compaction.

Simulations were performed on up to seven nodes
(\(896\) CPU cores in total), using hybrid MPI+OpenMP parallelism.
For each node count, we tested three different decomposition strategies:
\[
8\times 16,\qquad 16\times 8,\qquad 32\times 4,
\]
where the notation denotes
\(\text{MPI ranks per node} \times \text{OpenMP threads per rank}\).
Each MPI rank owns exactly one subdomain with non-periodic boundaries,
and marker advection was performed using the explicit Euler method
on the CPU.

All runs used the same warm-up and pinning strategy described in
Section~\ref{sec:hardware_setup}.
Each benchmark consists of 20 consecutive advection iterations,
and we report the total runtime over these 20 steps.
All tests were deterministic and reproducible.

\subsection{Weak Scaling Results}

\begin{figure}[h!]
    \centering
    \includegraphics[width=\textwidth]{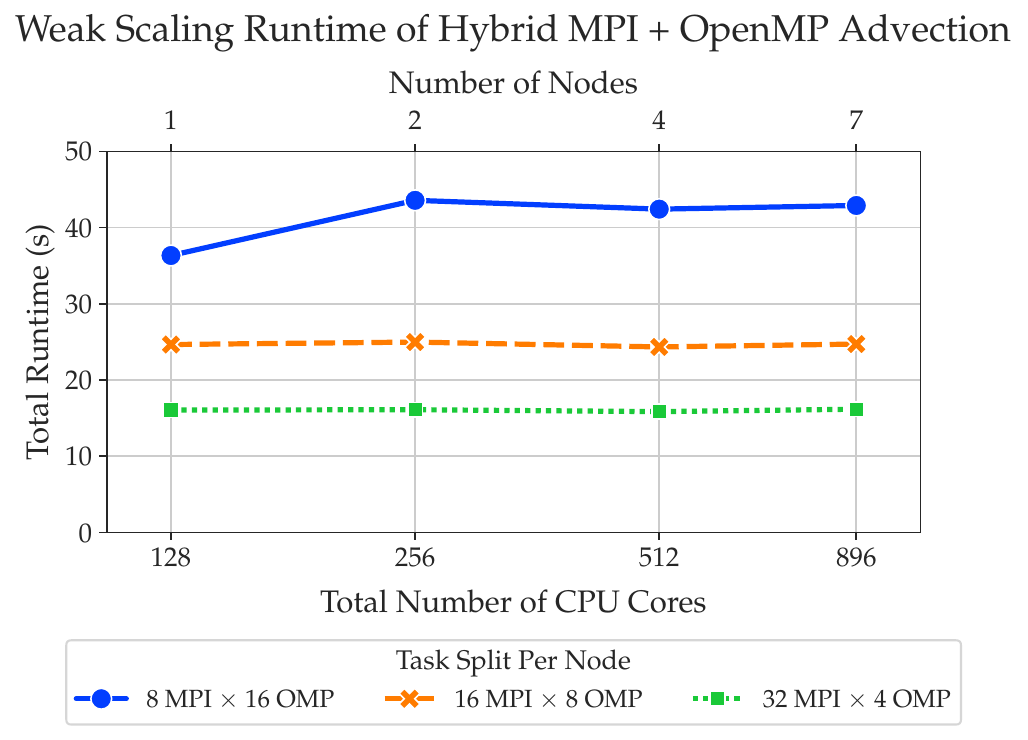}
    \caption{
    Weak-scaling runtime for the distributed advection benchmark.
    Each node processes a fixed \(5000\times5000\) subdomain while the global
    domain grows with the number of nodes.
    Configurations with fewer OpenMP threads per MPI rank show the best
    performance, with nearly constant runtime up to seven nodes.}
    \label{fig:weak_scaling_runtime}
\end{figure}

Figure~\ref{fig:weak_scaling_runtime} reports the total runtime of the
distributed advection benchmark for increasing node counts and for all
three MPI--OpenMP configurations.
Overall, the weak-scaling behaviour of the advection solver is excellent:
for the best-performing configuration (\(32\times 4\)),
the runtime remains nearly constant as the simulation expands from
\(5000\times5000\) on a single node to \(13300\times13300\) on seven nodes.
This demonstrates that the MPI-based halo exchanges and marker-migration
mechanisms scale efficiently across the cluster.

A notable feature of the results is the performance sensitivity to
OpenMP parallelism.
Similarly to the shared-memory Stokes benchmarks, the advection solver
does not benefit significantly from additional OpenMP threads.
Instead, increasing the number of threads tends to slow down the local work
by a nearly constant factor.
The primary cause is the marker-to-grid interpolation routine, which relies
on atomic reductions for accumulating nodal weights.
These reductions introduce contention, especially when increasing the number of threads, leading to limited OpenMP scalability.
In contrast, the MPI-level distributed interpolation does not suffer from this
contention, resulting in excellent scalability across nodes.

\begin{figure}[h!]
    \centering
    \includegraphics[width=\textwidth]{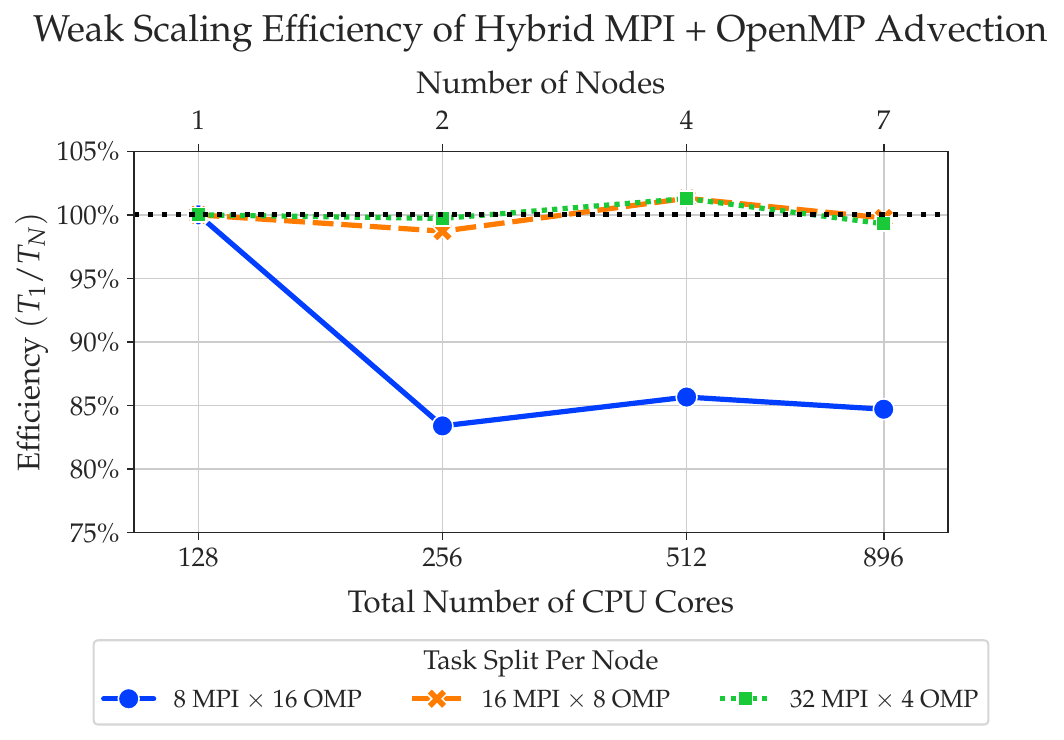}
    \caption{
    Weak-scaling efficiency of the distributed advection benchmark.
    The \(16\times 8\) and \(32\times 4\) configurations maintain efficiency
    close to \(100\%\) across all node counts.
    Minor superlinear scaling at intermediate sizes likely reflects cache
    effects or small runtime fluctuations.}
    \label{fig:weak_scaling_efficiency}
\end{figure}

Figure~\ref{fig:weak_scaling_efficiency} shows the normalized weak-scaling
efficiency, computed as \(T_1/T_N\) where \(T_1\) is the single-node runtime and $T_N$ is the runtime on $N$ nodes.
Efficiencies remain close to \(100\%\) for the \(16\times 8\) and \(32\times 4\)
configurations, with minor superlinear behaviour at four nodes.
This small improvement may arise from cache effects or statistical variation
in runtime measurement, both common in distributed-memory scaling tests.
The \(8\times 16\) configuration shows lower efficiency, reflecting the same
OpenMP-related limitations noted above, yet still maintains an overall scaling efficiency over $80\%$.

Taken together, these results demonstrate that the distributed advection
framework in Pyroclast scales efficiently on modern HPC clusters, provided
that the MPI/OpenMP balance avoids excessive use of OpenMP threads.
Future improvements to the interpolation kernels may further enhance
shared-memory scaling and reduce the performance gap between the
different hybrid configurations.

\section{Discussion and Summary}

The benchmarks presented in this chapter provide an initial assessment of the
numerical robustness and computational performance of Pyroclast across a range
of architectures and problem sizes.  
Despite being early in the development cycle, the results already reveal several
important strengths of the framework, as well as clear opportunities for future
optimisation.

\subsubsection*{Stokes Solver Performance}

The shared-memory benchmarks demonstrate that Pyroclast’s matrix-free
multigrid Stokes solver delivers stable, resolution-independent convergence
across all tested grid sizes.
The solver consistently reaches a relative energy residual of
$\mathcal{O}(10^{-4})$ and maintains identical convergence trajectories across
resolutions, confirming the correctness and robustness of both the multigrid
hierarchy and the Uzawa iteration.

Strong-scaling results on CPUs reflect the memory-bound nature of
stencil-dominated multigrid kernels: speedup improves up to 32--64 threads before
saturating due to bandwidth limitations and NUMA effects.
This behaviour is consistent with expectations for two-dimensional structured
grid solvers and aligns with findings in the literature~\cite{ferrari_3d_blocking}.
An interesting performance anomaly was observed for the
$15000\times15000$ case at 64 threads, which repeatedly produced unusually good
speedup.
Further profiling will be required to determine whether this behaviour results
from favourable memory placement, cache effects, or other architectural factors.

GPU acceleration proves highly effective for large systems.
Because the entire multigrid hierarchy above the threshold resides on the device, the A100 GPU
can exploit high memory bandwidth and avoid host–device transfers.
As problem size increases, GPU utilisation improves markedly, reaching a peak
speedup of $10\times$ over the fastest CPU configuration.
Even the largest test case shows substantial acceleration, with the observed dip
in speedup attributable to an anomalously fast CPU measurement rather than a GPU
limitation.  
These results confirm that Pyroclast’s GPU backend already provides strong
computational benefits for large Stokes systems.

\subsubsection*{Distributed Advection Performance}

The distributed-memory benchmarks show that Pyroclast’s marker advection
infrastructure scales efficiently across multiple nodes in a hybrid MPI+OpenMP
configuration.
Weak scaling is excellent for the best-performing decompositions, with the
runtime remaining nearly constant up to seven nodes and $896$~CPU cores.
The MPI-level halo exchange and marker migration mechanisms demonstrate
consistent, predictable performance even as the global domain size increases.

OpenMP scalability is more limited, reflecting bottlenecks in the shared-memory
marker-to-grid interpolation routine.
Atomic reductions introduce contention when many threads operate on nearby grid
nodes, causing a near-constant slowdown as the thread count increases.
Despite this, the MPI component of the solver scales almost ideally, and overall
efficiency remains close to $100\%$ for decompositions with fewer OpenMP threads
per rank.

\subsubsection*{Outlook}

Taken together, these benchmarks highlight the strong numerical robustness and
promising parallel performance of Pyroclast’s initial solver infrastructure.
The GPU backend already delivers substantial acceleration for large-scale Stokes
problems, while the distributed advection solver scales effectively across
multi-node CPU configurations.

It is important to emphasize that these results are preliminary.
Many components, including GPU kernels, the multigrid hierarchy,
interpolation routines, and communication patterns, have not yet undergone
systematic optimisation.
Significant improvements in both absolute performance and scalability are
therefore expected as development progresses.

Overall, the benchmarks presented here demonstrate a solid foundation for
Pyroclast’s long-term goals.
With continued optimisation and extension toward fully coupled
thermo-mechanical simulations, Pyroclast is well positioned to become a
high-performance, flexible, and scalable framework for large-scale geodynamic
modelling.

\chapter{Conclusions and Future Work}
\label{chap:conclusions}

The goal of this work was to develop a modern, modular, and extensible framework for high-performance geodynamic simulation. The result is \textit{Pyroclast}, a Python-based solver that demonstrates that flexibility, modularity, and high performance can be combined effectively. Pyroclast addresses a long-standing gap in geodynamics computing, where many widely used solvers rely on monolithic and CPU-oriented codebases that are increasingly difficult to maintain, extend, or adapt to emerging computational paradigms such as heterogeneous hardware and machine learning.

The work presented here establishes the mechanical core of a next-generation geodynamic modeling framework. Through a matrix-free and GPU accelerated solver for the incompressible Stokes system, Pyroclast shows that Python, together with the modern scientific computing ecosystem, can serve as a practical foundation for large-scale and research-grade simulation software. The framework brings together numerical robustness, architectural flexibility, and high performance while significantly lowering technical barriers for researchers.

\section{Core Contributions}

\subsection{Modern Software Architecture for Geodynamics}

Pyroclast introduces a clean and modular architecture in which each major component of the MIC pipeline is self-contained, testable, and extendable. The design is based on strict separation of concerns and on device-agnostic programming, which ensures that the same code can run efficiently on CPUs or GPUs. This architectural shift represents an important contribution. By avoiding a monolithic structure, very common in legacy codes, Pyroclast enables rapid experimentation with new physical models and numerical methods. Researchers can modify high-level Python components without navigating complex build systems or rewriting low-level kernels. This level of accessibility is uncommon among HPC-oriented geodynamics codes and positions Pyroclast as both a practical tool and a modern reference implementation.

\subsection{Matrix-Free Multigrid Solver}

The central technical result of this project is the implementation of a matrix-free geometric multigrid solver for incompressible viscous flow. All operators, including stencil kernels, restriction, and prolongation, are applied without forming global sparse matrices. This approach minimizes memory usage and maximizes throughput, especially on GPUs. The solver maintains stable and resolution-independent convergence even in the presence of strong viscosity contrasts. It performs reliably in the classical sinker benchmark, including contrasts up to $10^8$, and consistently reaches relative energy residuals of $\mathcal{O}(10^{-4})$ or lower without manual tuning of parameters. The robustness of the method is notable given how severely ill-conditioned realistic geodynamic problems can be. These results show that Pyroclast's solver is not only correct but also competitive with established low-level codes, despite being written in Python and relying on just-in-time compilation.

\subsection{Performance and Scalability on Modern Hardware}

A major objective of this work was to demonstrate that Python-based high-performance computing does not require compromising performance and efficiency. The results confirm this objective and highlight the effectiveness of the approach adopted in Pyroclast. The GPU-resident multigrid solver achieves speedups of an entire order of magnitude compared to optimized multicore CPU configurations. All two-dimensional benchmarks tested, ranging up to $15000 \times 15000$ grid points, run efficiently and converge stably even at this scale. 

At the same time, the marker advection solver exhibits nearly ideal weak scaling in a distributed setting, supported by efficient MPI-based halo exchange and ghost layer management. These findings validate the use of CuPy, Numba, and mpi4py as a reliable foundation for large-scale scientific computing in Python. Pyroclast stands among the first geodynamic solvers to combine GPU acceleration and distributed memory parallelism within a single and fully Python-based interface.

\section{Impact and Scientific Significance}

Pyroclast's significance extends beyond the numerical results and architectural innovations presented in this manuscript. It offers a path forward for modern and flexible geodynamic modeling that aligns with major trends in scientific computing. These include the increasing use of heterogeneous hardware, the growing emphasis on reproducible research, and the rapid integration of machine learning into scientific workflows.

\subsection{Lowering Barriers to Large-Scale Simulation}

A defining feature of Pyroclast is its accessibility. Traditional geodynamic solvers often require deep expertise in languages such as C or Fortran, in addition to familiarity with MPI, GPU programming, and complex build systems. Pyroclast removes many of these barriers through its high-level Python interface, its modular structure, and its device-agnostic numerical kernels. Researchers can work directly with clear and concise components, incorporate new physical models with minimal changes, and avoid navigating legacy codebases. This accessibility allows a broader community of geoscientists to engage with high-resolution and physics-rich simulations, and it supports scientific creativity by enabling rapid prototyping of new ideas.

\subsection{Integration with Machine Learning}

Pyroclast is naturally compatible with machine learning frameworks such as PyTorch and TensorFlow. Its array-based design and GPU-resident data structures make it straightforward to integrate neural surrogate models, data-driven closure laws, or differentiable operators. These capabilities open the door to a wide range of hybrid approaches that combine physical insight with data-driven methods. Examples include accelerated solvers based on neural networks, inverse modeling through differentiable simulation, or adaptive strategies guided by learned models. Very few geodynamic codes can interface with such methods without extensive modification. Pyroclast fills this gap and positions itself at the intersection of high-performance computing and artificial intelligence, an intersection that is gaining increasing relevance across the physical sciences.

\subsection{A Blueprint for Modernizing Legacy Codes}

Pyroclast also demonstrates a practical strategy for modernizing established geodynamic solvers. Many widely used codes rely on software architectures that were appropriate at the time of their creation but are now difficult to adapt to new hardware and evolving research needs. Pyroclast shows that it is possible to design simulation tools that are modular, maintainable, and performant while still meeting the numerical demands of geodynamics. The approach presented here provides a template for refactoring older tools toward hybrid CPU and GPU execution, matrix-free algorithms, and more modern software engineering practices.

\section{Future Work}

Although this work establishes a solid foundation, Pyroclast's long-term vision extends further. Several important directions for future development arise directly from the work presented here.

\subsection{Full Thermomechanical Coupling}

A major next step is the integration of the energy equation and of temperature-dependent rheology. This addition will enable simulations of visco-elasto-plastic deformation and will make it possible to reproduce the complete range of processes handled by models such as I2ELVIS and I3ELVIS.

\subsection{Three-Dimensional Models}

Extending Pyroclast to three dimensions represents a significant challenge and a major opportunity. The core algorithms described in this work generalize naturally, but practical implementation in three-dimensional domains requires careful optimization of data movement, cache optimization, and GPU kernel tuning. A three-dimensional version of Pyroclast will allow the study of more realistic geological scenarios, and it will provide a more demanding benchmark for hybrid CPU and GPU execution. It will also enable scaling to larger multi-GPU environments.

\subsection{Distributed Multigrid Support}

To fully realize Pyroclast's distributed potential, MPI support must be implemented in the Stokes multigrid solver. Domain decomposition methods such as Restricted Additive Schwarz preconditioning are particularly promising for this purpose. Completing this component will enable full multilevel scaling across large clusters and will make high-resolution three-dimensional simulations feasible.

\subsection{Numerical and Performance Enhancements}

Several promising avenues exist for further refinement of numerical components. Improving GPU kernel occupancy and reducing memory bandwidth bottlenecks would provide immediate performance gains. Optimizing smoother strategies, refining the construction of multigrid hierarchies, and improving marker to grid transfers would strengthen the robustness of the method, especially for heterogeneous materials. Exploring advanced advection methods such as Locally Polynomial Integration may further increase accuracy without compromising performance, especially in a distributed setting.

\subsection{Hybrid Physics and Machine Learning Methods}

Given Pyroclast's compatibility with machine learning frameworks, an exciting direction for future research involves integrating neural surrogate models, differentiable operators, and learned physics. These methods could accelerate simulations, improve parameter estimation, and support adaptive strategies guided by data. The combination of traditional physical modeling and modern data-driven approaches could open new directions in geodynamic research.

\section{Online Resources}

The Pyroclast framework is available as an open online repository. The source code, documentation, and example problems can be found on GitHub at \url{https://github.com/MarcelFerrari/Pyroclast}.

\section{Final Remarks}

Pyroclast demonstrates that high-performance geodynamic simulation can be both powerful and accessible. It bridges the gap between traditional high-performance computing practices and modern scientific workflows. The framework is flexible, GPU-ready, and designed to support experimentation, extensibility, and integration with machine learning.

The developments presented in this work form the basis of a next-generation geodynamics solver. The framework is modular, scalable, and aligned with emerging computational trends. Pyroclast is both a practical research tool and a vision for how next-generation geodynamic modeling software can evolve.


\newpage

\section*{Acknowledgements}
\addcontentsline{toc}{section}{Acknowledgements}

I would like to sincerely thank \textbf{Professor Dr.\ Taras Gerya (ETH Zurich)} for his unwavering support and guidance, not only during the development of this work but throughout my entire journey working on geodynamics at ETH Zurich. His mentorship has had a profound impact on the way I think about science, research, and the Earth itself.

A special thanks goes to \textbf{Alexander Sotoudeh (ETH Zurich)}, who worked alongside me during this project. Our countless discussions, shared frustrations, and moments of discovery made this experience not only more productive but genuinely enjoyable. His contributions helped shape many aspects of the work, and I’m grateful for the collaboration and friendship that grew from it.

I am also incredibly grateful to \textbf{Hussein Harake (Swiss National Supercomputing Center)} for the many insightful discussions, his technical expertise, and his last minute help during critical phases of benchmarking. His willingness to jump in and solve hardware issues was essential to getting everything across the finish line.

\bibliographystyle{IEEEtran}
\begingroup
\sloppy
\bibliography{refs}

@misc{rdma,
	title        = {Efficient RDMA Communication Protocols},
	author       = {Konstantin Taranov and Fabian Fischer and Torsten Hoefler},
	year         = 2022,
	url          = {https://arxiv.org/abs/2212.09134},
	eprint       = {2212.09134},
	archiveprefix = {arXiv},
	primaryclass = {cs.NI}
}

@misc{nvidia_gpudirect,
	title        = {{NVIDIA GPUDirect}: Enhancing Data Movement and Access for GPUs},
	author       = {{NVIDIA Corporation}},
	year         = 2025,
	note         = {Accessed: 2025-11-12},
	howpublished = {\url{https://developer.nvidia.com/gpudirect}}
}

@techreport{MPI-5.0,
	title        = {{MPI}: A Message-Passing Interface Standard, Version 5.0},
	author       = {{Message Passing Interface Forum}},
	year         = 2025,
	month        = jun,
	day          = 5,
	url          = {https://www.mpi-forum.org/docs/mpi-5.0/mpi50-report.pdf},
	note         = {Accessed: 2025-11-12},
	institution  = {Message Passing Interface Forum}
}

@manual{cupy_gpu_agnostic,
	title        = {{CuPy} User Guide: How to write CPU/GPU agnostic code},
	author       = {{CuPy Development Team}},
	year         = 2025,
	url          = {https://docs.cupy.dev/en/stable/user_guide/basic.html#how-to-write-cpu-gpu-agnostic-code},
	note         = {Accessed: 2025-11-10},
	version      = {13.6.0}
}

@article{mpi4py,
	title        = {{mpi4py}: Status Update After 12 Years of Development},
	author       = {Dalcin, Lisandro and Fang, Yao-Lung L.},
	year         = 2021,
	journal      = {Computing in Science \& Engineering},
	volume       = 23,
	number       = 4,
	pages        = {47--54},
	doi          = {10.1109/MCSE.2021.3083216}
}

@manual{cupy_memory_management,
	title        = {Memory Management — {CuPy} User Guide},
	author       = {{CuPy Development Team}},
	year         = 2025,
	url          = {https://docs.cupy.dev/en/stable/user_guide/memory.html},
	note         = {Accessed: 2025-11-10}
}

@manual{dlpack_v1.2,
	title        = {{DLPack}: Common in-memory Tensor Structure, Version 1.2},
	author       = {{DMLC Project}},
	year         = 2025,
	url          = {https://dmlc.github.io/dlpack/latest/},
	note         = {Version 1.2, released October 11 2025},
	organization = {DMLC}
}

@article{anderson_original,
	title        = {Iterative Procedures for Nonlinear Integral Equations},
	author       = {Anderson, Donald G.},
	year         = 1965,
	month        = oct,
	journal      = {J. ACM},
	publisher    = {Association for Computing Machinery},
	address      = {New York, NY, USA},
	volume       = 12,
	number       = 4,
	pages        = {547–560},
	doi          = {10.1145/321296.321305},
	issn         = {0004-5411},
	url          = {https://doi.org/10.1145/321296.321305},
	issue_date   = {Oct. 1965},
	numpages     = 14
}

@article{anderson_for_fixed_point,
	title        = {Anderson Acceleration for Fixed-Point Iterations},
	author       = {Walker, Homer F. and Ni, Peng},
	year         = 2011,
	journal      = {SIAM Journal on Numerical Analysis},
	volume       = 49,
	number       = 4,
	pages        = {1715--1735},
	doi          = {10.1137/10078356X},
	url          = {https://doi.org/10.1137/10078356X},
	eprint       = {https://doi.org/10.1137/10078356X}
}

@article{eisenstat_GCR,
	title        = {Variational Iterative Methods for Nonsymmetric Systems of Linear Equations},
	author       = {Eisenstat, Stanley C. and Elman, Howard C. and Schultz, Martin H.},
	year         = 1983,
	journal      = {SIAM Journal on Numerical Analysis},
	volume       = 20,
	number       = 2,
	pages        = {345--357},
	doi          = {10.1137/0720023},
	url          = {https://doi.org/10.1137/0720023},
	eprint       = {https://doi.org/10.1137/0720023}
}

@article{pyamg2023,
	title        = {{PyAMG}: Algebraic Multigrid Solvers in Python},
	author       = {Nathan Bell and Luke N. Olson and Jacob Schroder and Ben Southworth},
	year         = 2023,
	journal      = {Journal of Open Source Software},
	publisher    = {The Open Journal},
	volume       = 8,
	number       = 87,
	pages        = 5495,
	doi          = {10.21105/joss.05495},
	url          = {https://doi.org/10.21105/joss.05495}
}

@article{tackley_fast_stokes,
	title        = {Development of a Stokes flow solver robust to large viscosity jumps using a Schur complement approach with mixed precision arithmetic},
	author       = {Mikito Furuichi and Dave A. May and Paul J. Tackley},
	year         = 2011,
	journal      = {Journal of Computational Physics},
	volume       = 230,
	number       = 24,
	pages        = {8835--8851},
	doi          = {https://doi.org/10.1016/j.jcp.2011.09.007},
	issn         = {0021-9991},
	url          = {https://www.sciencedirect.com/science/article/pii/S0021999111005274}
}

@software{amgx,
	title        = {{AmgX}: GPU Accelerated Algebraic Multigrid Solvers},
	author       = {NVIDIA Corporation},
	year         = 2025,
	url          = {https://developer.nvidia.com/amgx},
	note         = {Accessed: 8 Nov 2025}
}

@inbook{amg_book,
	title        = {4. Algebraic Multigrid},
	author       = {J. W. Ruge and K. Stüben},
	booktitle    = {Multigrid Methods},
	pages        = {73--130},
	doi          = {10.1137/1.9781611971057.ch4},
	url          = {https://epubs.siam.org/doi/abs/10.1137/1.9781611971057.ch4},
	chapter      = {},
	eprint       = {https://epubs.siam.org/doi/pdf/10.1137/1.9781611971057.ch4}
}

@book{hackbusch_book,
	title        = {Multi-Grid Methods and Applications},
	author       = {Hackbusch, Wolfgang},
	year         = 1985,
	publisher    = {Springer Berlin Heidelberg},
	address      = {Berlin, Heidelberg},
	doi          = {10.1007/978-3-662-02427-0},
	isbn         = {978-3-662-02427-0},
	url          = {https://doi.org/10.1007/978-3-662-02427-0}
}

@inbook{hackbusch_book_intro,
	title        = {Preliminaries},
	author       = {Hackbusch, Wolfgang},
	year         = 1985,
	booktitle    = {Multi-Grid Methods and Applications},
	publisher    = {Springer Berlin Heidelberg},
	address      = {Berlin, Heidelberg},
	pages        = {1--16},
	doi          = {10.1007/978-3-662-02427-0_1},
	isbn         = {978-3-662-02427-0},
	url          = {https://doi.org/10.1007/978-3-662-02427-0_1}
}

@article{uzawa_book,
	title        = {Studies in linear and non-linear programming, by K. J. Arrow, L. Hurwicz and H. Uzawa. Stanford University Press, 1958. 229 pages. \$7.50.},
	author       = {Duguid, A.M.},
	year         = 1960,
	journal      = {Canadian Mathematical Bulletin},
	volume       = 3,
	number       = 3,
	pages        = {196–198},
	doi          = {10.1017/S0008439500025522}
}

@misc{ferrari_3d_blocking,
	title        = {3D Blocking for Matrix-free Smoothers in 2D Variable-Viscosity Stokes Equations with Applications to Geodynamics},
	author       = {Marcel Ferrari and Cyrill Püntener and Alexander Sotoudeh and Niklas Viebig},
	year         = 2025,
	url          = {https://arxiv.org/abs/2509.19061},
	eprint       = {2509.19061},
	archiveprefix = {arXiv},
	primaryclass = {physics.comp-ph}
}

@inbook{gerya_numerical_solution_of_stokes,
	title        = {Numerical solutions of the momentum and continuity equations},
	author       = {Gerya, Taras},
	year         = 2019,
	booktitle    = {Introduction to Numerical Geodynamic Modelling},
	publisher    = {Cambridge University Press},
	pages        = {82–104},
	place        = {Cambridge}
}

@inbook{butcher_runge_kutta,
	title        = {Runge–Kutta Methods},
	year         = 2016,
	booktitle    = {Numerical Methods for Ordinary Differential Equations},
	publisher    = {John Wiley \& Sons, Ltd},
	pages        = {143--331},
	doi          = {https://doi.org/10.1002/9781119121534.ch3},
	isbn         = 9781119121534,
	url          = {https://onlinelibrary.wiley.com/doi/abs/10.1002/9781119121534.ch3},
	chapter      = 3,
	eprint       = {https://onlinelibrary.wiley.com/doi/pdf/10.1002/9781119121534.ch3}
}

@inbook{gerya_mic,
	title        = {The advection equation and marker-in-cell method},
	author       = {Gerya, Taras},
	year         = 2019,
	booktitle    = {Introduction to Numerical Geodynamic Modelling},
	publisher    = {Cambridge University Press},
	pages        = {105–127},
	place        = {Cambridge}
}

@inbook{splitting_methods,
	title        = {Operator Splitting},
	author       = {MacNamara, Shev and Strang, Gilbert},
	year         = 2016,
	booktitle    = {Splitting Methods in Communication, Imaging, Science, and Engineering},
	publisher    = {Springer International Publishing},
	address      = {Cham},
	pages        = {95--114},
	doi          = {10.1007/978-3-319-41589-5_3},
	isbn         = {978-3-319-41589-5},
	url          = {https://doi.org/10.1007/978-3-319-41589-5_3},
	editor       = {Glowinski, Roland and Osher, Stanley J. and Yin, Wotao}
}

@inbook{gerya_boundary_conditions,
	title        = {Numerical solutions of the momentum and continuity equations},
	author       = {Gerya, Taras},
	year         = 2019,
	booktitle    = {Introduction to Numerical Geodynamic Modelling},
	publisher    = {Cambridge University Press},
	pages        = {82–104},
	place        = {Cambridge}
}

@book{gerya_book,
	title        = {Introduction to Numerical Geodynamic Modelling},
	author       = {Gerya, Taras},
	year         = 2019,
	publisher    = {Cambridge University Press},
	place        = {Cambridge},
	edition      = 2
}

@article{aurora,
	title        = {A foundation model for the Earth system},
	author       = {Bodnar, Cristian and Bruinsma, Wessel P. and Lucic, Ana and Stanley, Megan and Allen, Anna and Brandstetter, Johannes and Garvan, Patrick and Riechert, Maik and Weyn, Jonathan A. and Dong, Haiyu and Gupta, Jayesh K. and Thambiratnam, Kit and Archibald, Alexander T. and Wu, Chun-Chieh and Heider, Elizabeth and Welling, Max and Turner, Richard E. and Perdikaris, Paris},
	year         = 2025,
	journal      = {Nature},
	volume       = 641,
	number       = 8065,
	pages        = {1180--1187},
	doi          = {10.1038/s41586-025-09005-y},
	isbn         = {1476-4687},
	url          = {https://doi.org/10.1038/s41586-025-09005-y},
	date         = {2025-05-01},
	id           = {Bodnar2025}
}

@misc{poseidon,
	title        = {Poseidon: Efficient Foundation Models for PDEs},
	author       = {Maximilian Herde and Bogdan Raonić and Tobias Rohner and Roger Käppeli and Roberto Molinaro and Emmanuel de Bézenac and Siddhartha Mishra},
	year         = 2024,
	url          = {https://arxiv.org/abs/2405.19101},
	eprint       = {2405.19101},
	archiveprefix = {arXiv},
	primaryclass = {cs.LG}
}

@misc{diff_taichi,
	title        = {DiffTaichi: Differentiable Programming for Physical Simulation},
	author       = {Yuanming Hu and Luke Anderson and Tzu-Mao Li and Qi Sun and Nathan Carr and Jonathan Ragan-Kelley and Frédo Durand},
	year         = 2020,
	url          = {https://arxiv.org/abs/1910.00935},
	eprint       = {1910.00935},
	archiveprefix = {arXiv},
	primaryclass = {cs.LG}
}

@article{diff_phys_materials,
	title        = {End-to-end differentiability and tensor processing unit computing to accelerate materials'inverse design},
	author       = {Liu, Han and Liu, Yuhan and Li, Kevin and Zhao, Zhangji and Schoenholz, Samuel S. and Cubuk, Ekin D. and Gupta, Puneet and Bauchy, Mathieu},
	year         = 2023,
	journal      = {npj Computational Materials},
	volume       = 9,
	number       = 1,
	pages        = 121,
	doi          = {10.1038/s41524-023-01080-x},
	isbn         = {2057-3960},
	url          = {https://doi.org/10.1038/s41524-023-01080-x},
	date         = {2023-07-13},
	id           = {Liu2023}
}

@inproceedings{cno,
	title        = {Convolutional neural operators for robust and accurate learning of PDEs},
	author       = {Raoni\'{c}, Bogdan and Molinaro, Roberto and De Ryck, Tim and Rohner, Tobias and Bartolucci, Francesca and Alaifari, Rima and Mishra, Siddhartha and de B\'{e}zenac, Emmanuel},
	year         = 2023,
	booktitle    = {Proceedings of the 37th International Conference on Neural Information Processing Systems},
	location     = {New Orleans, LA, USA},
	publisher    = {Curran Associates Inc.},
	address      = {Red Hook, NY, USA},
	series       = {NIPS '23},
	articleno    = 3376,
	numpages     = 14
}

@misc{sfno,
	title        = {Spherical Fourier Neural Operators: Learning Stable Dynamics on the Sphere},
	author       = {Boris Bonev and Thorsten Kurth and Christian Hundt and Jaideep Pathak and Maximilian Baust and Karthik Kashinath and Anima Anandkumar},
	year         = 2023,
	url          = {https://arxiv.org/abs/2306.03838},
	eprint       = {2306.03838},
	archiveprefix = {arXiv},
	primaryclass = {cs.LG}
}

@misc{fno_paper,
	title        = {Fourier Neural Operator for Parametric Partial Differential Equations},
	author       = {Zongyi Li and Nikola Kovachki and Kamyar Azizzadenesheli and Burigede Liu and Kaushik Bhattacharya and Andrew Stuart and Anima Anandkumar},
	year         = 2021,
	url          = {https://arxiv.org/abs/2010.08895},
	eprint       = {2010.08895},
	archiveprefix = {arXiv},
	primaryclass = {cs.LG}
}

@article{pinn_geoscience,
	title        = {Applications of physics-informed neural networks in geosciences: From basic seismology to comprehensive environmental studies},
	title        = {},
	author       = {Maan Habib and Ahed Habib and Bashar Alibrahim},
	year         = 2025,
	journal      = {Open Geosciences},
	volume       = 17,
	number       = 1,
	pages        = 20250853,
	doi          = {doi:10.1515/geo-2025-0853},
	url          = {https://doi.org/10.1515/geo-2025-0853},
	lastchecked  = {2025-11-03}
}

@article{pinn_fluid_mechanics,
	title        = {Physics-informed neural networks (PINNs) for fluid mechanics: a review},
	author       = {Cai, Shengze and Mao, Zhiping and Wang, Zhicheng and Yin, Minglang and Karniadakis, George Em},
	year         = 2021,
	journal      = {Acta Mechanica Sinica},
	volume       = 37,
	number       = 12,
	pages        = {1727--1738},
	doi          = {10.1007/s10409-021-01148-1},
	isbn         = {1614-3116},
	url          = {https://doi.org/10.1007/s10409-021-01148-1},
	date         = {2021-12-01},
	id           = {Cai2021}
}

@misc{pinn_original_paper_2,
	title        = {Physics Informed Deep Learning (Part II): Data-driven Discovery of Nonlinear Partial Differential Equations},
	author       = {Maziar Raissi and Paris Perdikaris and George Em Karniadakis},
	year         = 2017,
	url          = {https://arxiv.org/abs/1711.10566},
	eprint       = {1711.10566},
	archiveprefix = {arXiv},
	primaryclass = {cs.AI}
}

@misc{pinn_original_paper_1,
	title        = {Physics Informed Deep Learning (Part I): Data-driven Solutions of Nonlinear Partial Differential Equations},
	author       = {Maziar Raissi and Paris Perdikaris and George Em Karniadakis},
	year         = 2017,
	url          = {https://arxiv.org/abs/1711.10561},
	eprint       = {1711.10561},
	archiveprefix = {arXiv},
	primaryclass = {cs.AI}
}

@article{devito-compiler,
	title        = {Architecture and Performance of Devito, a System for Automated Stencil Computation},
	author       = {Luporini, Fabio and Louboutin, Mathias and Lange, Michael and Kukreja, Navjot and Witte, Philipp and H\"{u}ckelheim, Jan and Yount, Charles and Kelly, Paul H. J. and Herrmann, Felix J. and Gorman, Gerard J.},
	year         = 2020,
	month        = 4,
	journal      = {ACM Trans. Math. Softw.},
	publisher    = {Association for Computing Machinery},
	address      = {New York, NY, USA},
	volume       = 46,
	number       = 1,
	doi          = {10.1145/3374916},
	issn         = {0098-3500},
	url          = {https://doi.org/10.1145/3374916},
	issue_date   = {March 2020},
	articleno    = 6,
	numpages     = 28
}

@article{devito-api,
	title        = {Devito (v3.1.0): an embedded domain-specific language for finite differences and geophysical exploration},
	author       = {Louboutin, M. and Lange, M. and Luporini, F. and Kukreja, N. and Witte, P. A. and Herrmann, F. J. and Velesko, P. and Gorman, G. J.},
	year         = 2019,
	journal      = {Geoscientific Model Development},
	volume       = 12,
	number       = 3,
	pages        = {1165--1187},
	doi          = {10.5194/gmd-12-1165-2019},
	url          = {https://www.geosci-model-dev.net/12/1165/2019/}
}

@misc{bisbas2024automatedmpixcodegeneration,
	title        = {Automated MPI-X code generation for scalable finite-difference solvers},
	author       = {George Bisbas and Rhodri Nelson and Mathias Louboutin and Fabio Luporini and Paul H. J. Kelly and Gerard Gorman},
	year         = 2024,
	url          = {https://arxiv.org/abs/2312.13094},
	eprint       = {2312.13094},
	archiveprefix = {arXiv},
	primaryclass = {cs.DC}
}

@article{sympy,
	title        = {SymPy: symbolic computing in Python},
	author       = {Meurer, Aaron and Smith, Christopher P. and Paprocki, Mateusz and \v{C}ert\'{i}k, Ond\v{r}ej and Kirpichev, Sergey B. and Rocklin, Matthew and Kumar, AMiT and Ivanov, Sergiu and Moore, Jason K. and Singh, Sartaj and Rathnayake, Thilina and Vig, Sean and Granger, Brian E. and Muller, Richard P. and Bonazzi, Francesco and Gupta, Harsh and Vats, Shivam and Johansson, Fredrik and Pedregosa, Fabian and Curry, Matthew J. and Terrel, Andy R. and Rou\v{c}ka, \v{S}t\v{e}p\'{a}n and Saboo, Ashutosh and Fernando, Isuru and Kulal, Sumith and Cimrman, Robert and Scopatz, Anthony},
	year         = 2017,
	month        = jan,
	journal      = {PeerJ Computer Science},
	volume       = 3,
	pages        = {e103},
	doi          = {10.7717/peerj-cs.103},
	issn         = {2376-5992},
	url          = {https://doi.org/10.7717/peerj-cs.103}
}

@misc{gt4py,
	title        = {GT4Py: High Performance Stencils for Weather and Climate Applications using Python},
	author       = {Enrique G. Paredes and Linus Groner and Stefano Ubbiali and Hannes Vogt and Alberto Madonna and Kean Mariotti and Felipe Cruz and Lucas Benedicic and Mauro Bianco and Joost VandeVondele and Thomas C. Schulthess},
	year         = 2023,
	url          = {https://arxiv.org/abs/2311.08322},
	eprint       = {2311.08322},
	archiveprefix = {arXiv},
	primaryclass = {cs.DC}
}

@misc{ICON2025.04,
	title        = {ICON release 2025.04},
	author       = {{ICON partnership (MPI-M; DWD; DKRZ; KIT; C2SM)}},
	year         = 2024,
	doi          = {10.35089/WDCC/IconRelease2025.04},
	note         = {World Data Center for Climate (WDCC) at DKRZ},
	howpublished = {\url{https://doi.org/10.35089/WDCC/IconRelease2025.04}}
}

@online{Russo2025NanoribbonTransistor,
	title        = {ETH Zürich team simulates a full nanoribbon transistor and earns a Gordon Bell Prize nomination},
	author       = {Santina Russo},
	year         = 2025,
	month        = 10,
	day          = 14,
	url          = {https://www.cscs.ch/science/computer-science-hpc/2025/eth-zuerich-team-simulates-a-full-nanoribbon-transistor-and-earns-a-gordon-bell-prize-nomination},
	note         = {Accessed: 2025-11-01},
	organization = {Swiss National Supercomputing Centre (CSCS)}
}

@misc{quantum_transport,
	title        = {Ab-initio Quantum Transport with the GW Approximation, 42,240 Atoms, and Sustained Exascale Performance},
	author       = {Nicolas Vetsch and Alexander Maeder and Vincent Maillou and Anders Winka and Jiang Cao and Grzegorz Kwasniewski and Leonard Deuschle and Torsten Hoefler and Alexandros Nikolaos Ziogas and Mathieu Luisier},
	year         = 2025,
	url          = {https://arxiv.org/abs/2508.19138},
	eprint       = {2508.19138},
	archiveprefix = {arXiv},
	primaryclass = {cs.DC}
}

@software{nccl,
	title        = {NVIDIA Collective Communication Library (NCCL)},
	author       = {{NVIDIA Corporation}},
	year         = 2025,
	url          = {https://developer.nvidia.com/nccl},
	note         = {Accessed: 31 Oct 2025},
	version      = {2.28.7},
	organization = {NVIDIA Corporation}
}

@software{nvshmem,
	title        = {NVIDIA NVSHMEM: OpenSHMEM Library for NVIDIA GPUs},
	author       = {{NVIDIA Corporation}},
	year         = 2025,
	url          = {https://docs.nvidia.com/nvshmem},
	note         = {Accessed: 31 Oct 2025},
	version      = {3.4.5},
	organization = {NVIDIA Corporation}
}

@software{nvidia_pytorch_container,
	title        = {NVIDIA PyTorch Container},
	author       = {{NVIDIA Corporation}},
	year         = 2025,
	url          = {https://catalog.ngc.nvidia.com/orgs/nvidia/containers/pytorch},
	note         = {Accessed: 31 Oct 2025},
	version      = {25.06},
	organization = {NVIDIA Corporation}
}

@software{enroot,
	title        = {Enroot: A simple yet powerful tool to turn traditional container/OS images into unprivileged sandboxes},
	author       = {{NVIDIA}},
	year         = 2025,
	url          = {https://github.com/NVIDIA/enroot},
	note         = {Accessed: 2025-10-31},
	version      = {4.0.1},
	license      = {Apache‐2.0}
}

@article{apptainer_2,
	title        = {Singularity: Scientific containers for mobility of compute},
	author       = {Kurtzer, Gregory M. AND Sochat, Vanessa AND Bauer, Michael W.},
	year         = 2017,
	month        = {05},
	journal      = {PLOS ONE},
	publisher    = {Public Library of Science},
	volume       = 12,
	number       = 5,
	pages        = {1--20},
	doi          = {10.1371/journal.pone.0177459},
	url          = {https://doi.org/10.1371/journal.pone.0177459}
}

@software{apptainer_1,
	title        = {hpcng/singularity: Singularity 3.7.3},
	author       = {Gregory M. Kurtzer and cclerget and Michael Bauer and Ian Kaneshiro and David Trudgian and David Godlove},
	year         = 2021,
	month        = apr,
	publisher    = {Zenodo},
	doi          = {10.5281/zenodo.4667718},
	url          = {https://doi.org/10.5281/zenodo.4667718},
	version      = {v3.7.3}
}

@misc{nanobind,
	title        = {nanobind: tiny and efficient C++/Python bindings},
	author       = {Wenzel Jakob},
	year         = 2022,
	note         = {https://github.com/wjakob/nanobind}
}

@software{numba_cuda,
	title        = {{Numba-CUDA}: A CUDA target for the Numba JIT compiler},
	author       = {{NVIDIA Corporation}},
	year         = 2024,
	url          = {https://nvidia.github.io/numba-cuda/},
	note         = {Python JIT compiler for NVIDIA GPU kernels via Numba.}
}

@software{numba_hip,
	title        = {{Numba-HIP}: A ROCm/HIP backend for the Numba JIT compiler},
	author       = {{AMD Inc.}},
	year         = 2024,
	url          = {https://github.com/ROCm/numba-hip},
	note         = {Experimental AMD HIP/ROCm GPU backend for Numba.}
}

@inproceedings{dace,
	title        = {Stateful dataflow multigraphs: a data-centric model for performance portability on heterogeneous architectures},
	author       = {Ben-Nun, Tal and de Fine Licht, Johannes and Ziogas, Alexandros N. and Schneider, Timo and Hoefler, Torsten},
	year         = 2019,
	booktitle    = {Proceedings of the International Conference for High Performance Computing, Networking, Storage and Analysis},
	location     = {Denver, Colorado},
	publisher    = {Association for Computing Machinery},
	address      = {New York, NY, USA},
	series       = {SC '19},
	doi          = {10.1145/3295500.3356173},
	isbn         = 9781450362290,
	url          = {https://doi.org/10.1145/3295500.3356173},
	articleno    = 81,
	numpages     = 14
}

@manual{cublas,
	title        = {cuBLAS: NVIDIA CUDA Basic Linear Algebra Subprograms Library},
	author       = {{NVIDIA Corporation}},
	year         = 2025,
	url          = {https://developer.nvidia.com/cublas},
	note         = {GPU-accelerated BLAS library for NVIDIA GPUs}
}

@manual{cutensor,
	title        = {cuTENSOR: NVIDIA Tensor Linear Algebra Library},
	author       = {{NVIDIA Corporation}},
	year         = 2025,
	url          = {https://developer.nvidia.com/cutensor},
	note         = {GPU-accelerated tensor primitives library for NVIDIA GPUs}
}

@manual{hipblas,
	title        = {hipBLAS: ROCm/HIP Basic Linear Algebra Subprograms Library},
	author       = {{AMD Inc.}},
	year         = 2025,
	url          = {https://rocm.docs.amd.com/projects/hipBLAS/en/latest/},
	note         = {BLAS-marshalling library supporting multiple backends (rocBLAS, cuBLAS)}
}

@manual{hiptensor,
	title        = {hipTensor: ROCm/HIP Tensor Primitives Library},
	author       = {{AMD Inc.}},
	year         = 2025,
	url          = {https://rocm.docs.amd.com/projects/hipTensor/en/latest/},
	note         = {C++ library for tensor primitives on AMD GPUs}
}

@manual{openacc-spec,
	title        = {The OpenACC Application Programming Interface},
	author       = {{OpenACC Organization}},
	year         = 2025,
	month        = jun,
	url          = {https://www.openacc.org/specification},
	version      = {3.4}
}

@inproceedings{raja_1,
	title        = {RAJA: Portable Performance for Large-Scale Scientific Applications},
	author       = {Beckingsale, David A. and Burmark, Jason and Hornung, Rich and Jones, Holger and Killian, William and Kunen, Adam J. and Pearce, Olga and Robinson, Peter and Ryujin, Brian S. and Scogland, Thomas RW},
	year         = 2019,
	booktitle    = {2019 IEEE/ACM International Workshop on Performance, Portability and Productivity in HPC (P3HPC)},
	volume       = {},
	number       = {},
	pages        = {71--81},
	doi          = {10.1109/P3HPC49587.2019.00012}
}

@inproceedings{raja_2,
	title        = {Performance portable C++ programming with RAJA},
	author       = {Beckingsale, David and Hornung, Richard and Scogland, Tom and Vargas, Arturo},
	year         = 2019,
	booktitle    = {Proceedings of the 24th Symposium on Principles and Practice of Parallel Programming},
	location     = {Washington, District of Columbia},
	publisher    = {Association for Computing Machinery},
	address      = {New York, NY, USA},
	series       = {PPoPP '19},
	pages        = {455–456},
	doi          = {10.1145/3293883.3302577},
	isbn         = 9781450362252,
	url          = {https://doi.org/10.1145/3293883.3302577},
	numpages     = 2
}

@article{kokkos_1,
	title        = {Kokkos 3: Programming Model Extensions for the Exascale Era},
	author       = {Trott, Christian R. and Lebrun-Grandié, Damien and Arndt, Daniel and Ciesko, Jan and Dang, Vinh and Ellingwood, Nathan and Gayatri, Rahulkumar and Harvey, Evan and Hollman, Daisy S. and Ibanez, Dan and Liber, Nevin and Madsen, Jonathan and Miles, Jeff and Poliakoff, David and Powell, Amy and Rajamanickam, Sivasankaran and Simberg, Mikael and Sunderland, Dan and Turcksin, Bruno and Wilke, Jeremiah},
	year         = 2022,
	journal      = {IEEE Transactions on Parallel and Distributed Systems},
	volume       = 33,
	number       = 4,
	pages        = {805--817},
	doi          = {10.1109/TPDS.2021.3097283}
}

@article{kokkos_2,
	title        = {Kokkos: Enabling manycore performance portability through polymorphic memory access patterns},
	author       = {H. Carter Edwards and Christian R. Trott and Daniel Sunderland},
	year         = 2014,
	journal      = {Journal of Parallel and Distributed Computing},
	volume       = 74,
	number       = 12,
	pages        = {3202--3216},
	doi          = {https://doi.org/10.1016/j.jpdc.2014.07.003},
	issn         = {0743-7315},
	url          = {http://www.sciencedirect.com/science/article/pii/S0743731514001257},
	note         = {Domain-Specific Languages and High-Level Frameworks for High-Performance Computing}
}

@misc{petsc_changes,
	title        = {Changes for each release},
	author       = {PETSc Development Team},
	year         = 2025,
	note         = {Accessed: 2025-10-31},
	howpublished = {\url{https://petsc.org/release/changes/}}
}

@misc{petsc_gpu_roadmap,
	title        = {GPU Support Roadmap},
	author       = {{PETSc Development Team}},
	year         = 2025,
	note         = {Last updated 2025-10-31 (v3.24.1)},
	howpublished = {\url{https://petsc.org/release/overview/gpu_roadmap/}}
}

@article{pylith,
	title        = {A domain decomposition approach to implementing fault slip in finite-element models of quasi-static and dynamic crustal deformation},
	author       = {Aagaard, B. T. and Knepley, M. G. and Williams, C. A.},
	year         = 2013,
	journal      = {Journal of Geophysical Research: Solid Earth},
	volume       = 118,
	number       = 6,
	pages        = {3059--3079},
	doi          = {https://doi.org/10.1002/jgrb.50217},
	url          = {https://agupubs.onlinelibrary.wiley.com/doi/abs/10.1002/jgrb.50217},
	eprint       = {https://agupubs.onlinelibrary.wiley.com/doi/pdf/10.1002/jgrb.50217}
}

@software{sw4_1,
	title        = {geodynamics/sw4: SW4, version 3.0},
	author       = {Petersson, N. Anders and Sjogreen, Bjorn and Tang, Houjun and Pankajakshan, Ramesh},
	year         = 2023,
	month        = sep,
	publisher    = {Zenodo},
	doi          = {10.5281/zenodo.8322590},
	url          = {https://doi.org/10.5281/zenodo.8322590},
	version      = {v3.0}
}

@article{sw4_2,
	title        = {A Fourth Order Accurate Finite Difference Scheme for the Elastic Wave Equation in Second Order Formulation},
	author       = {Sj{\"o}green, Bj{\"o}rn and Petersson, N. Anders},
	year         = 2012,
	journal      = {Journal of Scientific Computing},
	volume       = 52,
	number       = 1,
	pages        = {17--48},
	doi          = {10.1007/s10915-011-9531-1},
	isbn         = {1573-7691},
	url          = {https://doi.org/10.1007/s10915-011-9531-1},
	date         = {2012-07-01},
	id           = {Sj{\"o}green2012}
}

@article{sw4_3,
	title        = {Porting a 3D seismic modeling code (SW4) to CORAL machines},
	author       = {Pankajakshan, R. and Lin, P.-H. and Sjögreen, B.},
	year         = 2020,
	journal      = {IBM Journal of Research and Development},
	volume       = 64,
	number       = {3/4},
	pages        = {17:1--17:11},
	doi          = {10.1147/JRD.2019.2960218}
}

@inproceedings{awp_odc_1,
	title        = {Scalable Earthquake Simulation on Petascale Supercomputers},
	author       = {Cui, Yifeng and Olsen, Kim B. and Jordan, Thomas H. and Lee, Kwangyoon and Zhou, Jun and Small, Patrick and Roten, Daniel and Ely, Geoffrey and Panda, Dhabaleswar K. and Chourasia, Amit and Levesque, John and Day, Steven M. and Maechling, Philip},
	year         = 2010,
	booktitle    = {SC '10: Proceedings of the 2010 ACM/IEEE International Conference for High Performance Computing, Networking, Storage and Analysis},
	volume       = {},
	number       = {},
	pages        = {1--20},
	doi          = {10.1109/SC.2010.45}
}

@article{awp_odc_2,
	title        = {Hands-on Performance Tuning of 3D Finite Difference Earthquake Simulation on GPU Fermi Chipset},
	author       = {Jun Zhou and Didem Unat and Dong Ju Choi and Clark C. Guest and Yifeng Cui},
	year         = 2012,
	journal      = {Procedia Computer Science},
	volume       = 9,
	pages        = {976--985},
	doi          = {https://doi.org/10.1016/j.procs.2012.04.104},
	issn         = {1877-0509},
	url          = {https://www.sciencedirect.com/science/article/pii/S1877050912002256},
	note         = {Proceedings of the International Conference on Computational Science, ICCS 2012}
}

@software{specfem_1,
	title        = {SPECFEM/specfem3d: SPECFEM3D v4.1.1},
	author       = {Komatitsch, Dimitri and Tromp, Jeroen and Garg, Rahul and Gharti, Hom Nath and Nagaso, Masaru and Oral, Elif and Peter, Daniel and Afanasiev, Michael and Almada, Rafael and Ampuero, Jean-Paul and Bachmann, Etienne and Bai, Kangchen and Basini, Piero and Beller, Stephen and Bishop, Jordan and Bissey, Francois and Blitz, Celine and Bottero, Alexis and Bozdag, Ebru and Casarotti, Emanuele and Charles, Joseph and Chen, Min and Cristini, Paul and Durochat, Clement and Galvez Barron, Percy and Gassmoeller, Rene and Goeddeke, Dominik and Grinberg, Leopold and Gupta, Aakash and Heien, Eric and Hjoerleifsdottir, Vala and Karakostas, Foivos and Kientz, Sue and Labarta, Jesus and Le Goff, Nicolas and Le Loher, Pieyre and Lefebvre, Matthieu and Liu, Qinya and Liu, Youshan and Luet, David and Luo, Yang and Maggi, Alessia and Magnoni, Federica and Martin, Roland and Matzen, Rene and McBain, G. D. and McRitchie, Dennis and Meschede, Matthias and Messmer, Peter and Michea, David and Miller, David and Modrak, Ryan and Monteiller, Vadim and Morency, Christina and Nadh Somala, Surendra and Nissen-Meyer, Tarje and Pouget, Kevin and Rietmann, Max and Sales de Andrade, Elliott and Savage, Brian and Schuberth, Bernhard and Sieminski, Anne and Smith, James and Strand, Leif and Tape, Carl and Valero Cano, Eduardo and Videau, Brice and Vilotte, Jean-Pierre and Weng, Huihui and Xie, Zhinan and Zhang, Chang-Hua and Zhu, Hejun},
	year         = 2024,
	month        = mar,
	publisher    = {Zenodo},
	doi          = {10.5281/zenodo.10823181},
	url          = {https://doi.org/10.5281/zenodo.10823181},
	version      = {v4.1.1}
}

@article{specfem_2,
	title        = {Spectral-element simulations of global seismic wave propagation—I. Validation},
	author       = {Komatitsch, Dimitri and Tromp, Jeroen},
	year         = 2002,
	journal      = {Geophysical Journal International},
	volume       = 149,
	number       = 2,
	pages        = {390--412},
	doi          = {https://doi.org/10.1046/j.1365-246X.2002.01653.x},
	url          = {https://onlinelibrary.wiley.com/doi/abs/10.1046/j.1365-246X.2002.01653.x},
	eprint       = {https://onlinelibrary.wiley.com/doi/pdf/10.1046/j.1365-246X.2002.01653.x}
}

@article{specfem_3,
	title        = {Fluid–solid coupling on a cluster of GPU graphics cards for seismic wave propagation},
	author       = {Dimitri Komatitsch},
	year         = 2011,
	journal      = {Comptes Rendus Mécanique},
	volume       = 339,
	number       = 2,
	pages        = {125--135},
	doi          = {https://doi.org/10.1016/j.crme.2010.11.007},
	issn         = {1631-0721},
	url          = {https://www.sciencedirect.com/science/article/pii/S1631072110002081},
	note         = {High Performance Computing}
}

@article{seissol_1,
	title        = {An arbitrary high-order discontinuous Galerkin method for elastic waves on unstructured meshes — I. The two-dimensional isotropic case with external source terms},
	author       = {Käser, Martin and Dumbser, Michael},
	year         = 2006,
	month        = {08},
	journal      = {Geophysical Journal International},
	volume       = 166,
	number       = 2,
	pages        = {855--877},
	doi          = {10.1111/j.1365-246X.2006.03051.x},
	issn         = {0956-540X},
	url          = {https://doi.org/10.1111/j.1365-246X.2006.03051.x},
	eprint       = {https://academic.oup.com/gji/article-pdf/166/2/855/1514333/166-2-855.pdf}
}

@article{seissol_2,
	title        = {An arbitrary high-order discontinuous Galerkin method for elastic waves on unstructured meshes — II. The three-dimensional isotropic case},
	author       = {Dumbser, Michael and Käser, Martin},
	year         = 2006,
	month        = 10,
	journal      = {Geophysical Journal International},
	volume       = 167,
	number       = 1,
	pages        = {319--336},
	doi          = {10.1111/j.1365-246X.2006.03120.x},
	issn         = {0956-540X},
	url          = {https://doi.org/10.1111/j.1365-246X.2006.03120.x},
	eprint       = {https://academic.oup.com/gji/article-pdf/167/1/319/6021938/167-1-319.pdf}
}

@article{seissol_3,
	title        = {Fused GEMMs towards an efficient GPU implementation of the ADER-DG method in SeisSol},
	author       = {Dorozhinskii, Ravil and Gadeschi, Gonzalo Brito and Bader, Michael},
	year         = 2024,
	journal      = {Concurrency and Computation: Practice and Experience},
	volume       = 36,
	number       = 12,
	pages        = {e8037},
	doi          = {https://doi.org/10.1002/cpe.8037},
	url          = {https://onlinelibrary.wiley.com/doi/abs/10.1002/cpe.8037},
	eprint       = {https://onlinelibrary.wiley.com/doi/pdf/10.1002/cpe.8037}
}

@inproceedings{seissol_4,
	title        = {Petascale High Order Dynamic Rupture Earthquake Simulations on Heterogeneous Supercomputers},
	author       = {Heinecke, Alexander and Breuer, Alexander and Rettenberger, Sebastian and Bader, Michael and Gabriel, Alice-Agnes and Pelties, Christian and Bode, Arndt and Barth, William and Liao, Xiang-Ke and Vaidyanathan, Karthikeyan and Smelyanskiy, Mikhail and Dubey, Pradeep},
	year         = 2014,
	booktitle    = {SC '14: Proceedings of the International Conference for High Performance Computing, Networking, Storage and Analysis},
	volume       = {},
	number       = {},
	pages        = {3--14},
	doi          = {10.1109/SC.2014.6}
}

@article{underworld_1,
	title        = {Underworld2: Python Geodynamics Modelling for Desktop, HPC and Cloud},
	author       = {Mansour, John and Giordani, Julian and Moresi, Louis and Beucher, Romain and Kaluza, Owen and Velic, Mirko and Farrington, Rebecca and Quenette, Steve and Beall, Adam},
	year         = 2020,
	journal      = {Journal of Open Source Software},
	publisher    = {The Open Journal},
	volume       = 5,
	number       = 47,
	pages        = 1797,
	doi          = {10.21105/joss.01797},
	url          = {https://doi.org/10.21105/joss.01797}
}

@article{underworld_2,
	title        = {UWGeodynamics: A teaching and research tool for numerical geodynamic modelling},
	author       = {Beucher, Romain and Moresi, Louis and Giordani, Julian and Mansour, John and Sandiford, Dan and Farrington, Rebecca and Mondy, Luke and Mallard, Claire and Rey, Patrice and Duclaux, Guillaume and Kaluza, Owen and Laik, Arijit and Morón, Sara},
	year         = 2019,
	journal      = {Journal of Open Source Software},
	publisher    = {The Open Journal},
	volume       = 4,
	number       = 36,
	pages        = 1136,
	doi          = {10.21105/joss.01136},
	url          = {https://doi.org/10.21105/joss.01136}
}

@article{underworld_3,
	title        = {Mantle Convection Modeling with Viscoelastic/Brittle Lithosphere: Numerical Methodology and Plate Tectonic Modeling},
	author       = {Moresi, L. and Dufour, F. and M{\"u}hlhaus, H. -B.},
	year         = 2002,
	journal      = {pure and applied geophysics},
	volume       = 159,
	number       = 10,
	pages        = {2335--2356},
	doi          = {10.1007/s00024-002-8738-3},
	isbn         = {1420-9136},
	url          = {https://doi.org/10.1007/s00024-002-8738-3},
	date         = {2002-08-01},
	id           = {Moresi2002}
}

@article{underworld_4,
	title        = {A Lagrangian integration point finite element method for large deformation modeling of viscoelastic geomaterials},
	author       = {L. Moresi and F. Dufour and H.-B. Mühlhaus},
	year         = 2003,
	journal      = {Journal of Computational Physics},
	volume       = 184,
	number       = 2,
	pages        = {476--497},
	doi          = {https://doi.org/10.1016/S0021-9991(02)00031-1},
	issn         = {0021-9991},
	url          = {https://www.sciencedirect.com/science/article/pii/S0021999102000311}
}

@article{underworld_5,
	title        = {Computational approaches to studying non-linear dynamics of the crust and mantle},
	author       = {L. Moresi and S. Quenette and V. Lemiale and C. Mériaux and B. Appelbe and H.-B. Mühlhaus},
	year         = 2007,
	journal      = {Physics of the Earth and Planetary Interiors},
	volume       = 163,
	number       = 1,
	pages        = {69--82},
	doi          = {https://doi.org/10.1016/j.pepi.2007.06.009},
	issn         = {0031-9201},
	url          = {https://www.sciencedirect.com/science/article/pii/S0031920107001446},
	note         = {Computational Challenges in the Earth Sciences}
}

@misc{stagbl,
	title        = {StagBL: A scalable, portable, high-performance discretization and solver layer for geodynamic simulation},
	author       = {StagBL contributors},
	year         = 2025,
	month        = 10,
	note         = {Accessed: 2025-10-31},
	howpublished = {\url{https://github.com/stagbl/stagbl}}
}

@misc{gaia_software,
	title        = {GAIA: A framework to study the thermo-chemical evolution of rocky and icy bodies},
	author       = {Hüttig, Christian and Plesa, Ana-Catalina and Tosi, Nicola},
	year         = 2023,
	month        = sep,
	note         = {Accessed: 2025-10-31},
	howpublished = {\url{https://helmholtz.software/software/gaia}}
}

@article{gaia_article_1,
	title        = {The spiral grid: A new approach to discretize the sphere and its application to mantle convection},
	author       = {Hüttig, Christian and Stemmer, Kai},
	year         = 2008,
	journal      = {Geochemistry, Geophysics, Geosystems},
	volume       = 9,
	number       = 2,
	pages        = {},
	doi          = {https://doi.org/10.1029/2007GC001581},
	url          = {https://agupubs.onlinelibrary.wiley.com/doi/abs/10.1029/2007GC001581},
	eprint       = {https://agupubs.onlinelibrary.wiley.com/doi/pdf/10.1029/2007GC001581}
}

@article{gaia_article_2,
	title        = {Finite volume discretization for dynamic viscosities on Voronoi grids},
	author       = {Christian Hüttig and Kai Stemmer},
	year         = 2008,
	journal      = {Physics of the Earth and Planetary Interiors},
	volume       = 171,
	number       = 1,
	pages        = {137--146},
	doi          = {https://doi.org/10.1016/j.pepi.2008.07.007},
	issn         = {0031-9201},
	url          = {https://www.sciencedirect.com/science/article/pii/S0031920108001659},
	note         = {Recent Advances in Computational Geodynamics: Theory, Numerics and Applications}
}

@article{gaia_article_3,
	title        = {An improved formulation of the incompressible Navier–Stokes equations with variable viscosity},
	author       = {Christian Hüttig and Nicola Tosi and William B. Moore},
	year         = 2013,
	journal      = {Physics of the Earth and Planetary Interiors},
	volume       = 220,
	pages        = {11--18},
	doi          = {https://doi.org/10.1016/j.pepi.2013.04.002},
	issn         = {0031-9201},
	url          = {https://www.sciencedirect.com/science/article/pii/S0031920113000459}
}

@inbook{gerya_intro_book,
	title        = {Introduction},
	author       = {Gerya, Taras},
	year         = 2019,
	booktitle    = {Introduction to Numerical Geodynamic Modelling},
	publisher    = {Cambridge University Press},
	pages        = {1–11},
	place        = {Cambridge}
}

@inbook{turcotte_plate_tectonics,
	title        = {Plate Tectonics},
	author       = {Turcotte, Donald and Schubert, Gerald},
	year         = 2014,
	booktitle    = {Geodynamics},
	publisher    = {Cambridge University Press},
	pages        = {1–91},
	place        = {Cambridge}
}

@inbook{turcotte_heat,
	title        = {Heat Transfer},
	author       = {Turcotte, Donald and Schubert, Gerald},
	year         = 2014,
	booktitle    = {Geodynamics},
	publisher    = {Cambridge University Press},
	pages        = {160–229},
	place        = {Cambridge}
}

@inbook{gerya_heat,
	title        = {The heat conservation equation},
	author       = {Gerya, Taras},
	year         = 2019,
	booktitle    = {Introduction to Numerical Geodynamic Modelling},
	publisher    = {Cambridge University Press},
	pages        = {128–138},
	place        = {Cambridge}
}

@book{schubert_book,
	title        = {Mantle Convection in the Earth and Planets},
	author       = {Schubert, Gerald and Turcotte, Donald L. and Olson, Peter},
	year         = 2001,
	publisher    = {Cambridge University Press},
	place        = {Cambridge}
}

@inbook{gerya_rheology,
	title        = {Viscous rheology of rocks},
	author       = {Gerya, Taras},
	year         = 2019,
	booktitle    = {Introduction to Numerical Geodynamic Modelling},
	publisher    = {Cambridge University Press},
	pages        = {73–81},
	place        = {Cambridge}
}

@inbook{turcotte_rheology,
	title        = {Rock Rheology},
	author       = {Turcotte, Donald and Schubert, Gerald},
	year         = 2014,
	booktitle    = {Geodynamics},
	publisher    = {Cambridge University Press},
	pages        = {336–385},
	place        = {Cambridge}
}

@inbook{gerya_seismo_modeling,
	title        = {Seismo-thermomechanical modelling},
	author       = {Gerya, Taras},
	year         = 2019,
	booktitle    = {Introduction to Numerical Geodynamic Modelling},
	publisher    = {Cambridge University Press},
	pages        = {224–239},
	place        = {Cambridge}
}

@inbook{gerya_hydro_modeling,
	title        = {Hydro-thermomechanical modelling},
	author       = {Gerya, Taras},
	year         = 2019,
	booktitle    = {Introduction to Numerical Geodynamic Modelling},
	publisher    = {Cambridge University Press},
	pages        = {240–276},
	place        = {Cambridge}
}

@inbook{gerya_amr,
	title        = {Adaptive mesh refinement},
	author       = {Gerya, Taras},
	year         = 2019,
	booktitle    = {Introduction to Numerical Geodynamic Modelling},
	publisher    = {Cambridge University Press},
	pages        = {277–291},
	place        = {Cambridge}
}

@inproceedings{pytorch,
	title        = {{PyTorch 2: Faster Machine Learning Through Dynamic Python Bytecode Transformation and Graph Compilation}},
	author       = {Ansel, Jason and Yang, Edward and He, Horace and Gimelshein, Natalia and Jain, Animesh and Voznesensky, Michael and Bao, Bin and Bell, Peter and Berard, David and Burovski, Evgeni and Chauhan, Geeta and Chourdia, Anjali and Constable, Will and Desmaison, Alban and DeVito, Zachary and Ellison, Elias and Feng, Will and Gong, Jiong and Gschwind, Michael and Hirsh, Brian and Huang, Sherlock and Kalambarkar, Kshiteej and Kirsch, Laurent and Lazos, Michael and Lezcano, Mario and Liang, Yanbo and Liang, Jason and Lu, Yinghai and Luk, CK and Maher, Bert and Pan, Yunjie and Puhrsch, Christian and Reso, Matthias and Saroufim, Mark and Siraichi, Marcos Yukio and Suk, Helen and Suo, Michael and Tillet, Phil and Wang, Eikan and Wang, Xiaodong and Wen, William and Zhang, Shunting and Zhao, Xu and Zhou, Keren and Zou, Richard and Mathews, Ajit and Chanan, Gregory and Wu, Peng and Chintala, Soumith},
	year         = 2024,
	month        = apr,
	booktitle    = {29th ACM International Conference on Architectural Support for Programming Languages and Operating Systems, Volume 2 (ASPLOS '24)},
	publisher    = {ACM},
	doi          = {10.1145/3620665.3640366},
	url          = {https://docs.pytorch.org/assets/pytorch2-2.pdf}
}

@misc{tensorflow,
	title        = {{TensorFlow}: Large-Scale Machine Learning on Heterogeneous Systems},
	author       = {Mart\'{i}n~Abadi and Ashish~Agarwal and Paul~Barham and Eugene~Brevdo and Zhifeng~Chen and Craig~Citro and Greg~S.~Corrado and Andy~Davis and Jeffrey~Dean and Matthieu~Devin and Sanjay~Ghemawat and Ian~Goodfellow and Andrew~Harp and Geoffrey~Irving and Michael~Isard and Yangqing Jia and Rafal~Jozefowicz and Lukasz~Kaiser and Manjunath~Kudlur and Josh~Levenberg and Dandelion~Man\'{e} and Rajat~Monga and Sherry~Moore and Derek~Murray and Chris~Olah and Mike~Schuster and Jonathon~Shlens and Benoit~Steiner and Ilya~Sutskever and Kunal~Talwar and Paul~Tucker and Vincent~Vanhoucke and Vijay~Vasudevan and Fernanda~Vi\'{e}gas and Oriol~Vinyals and Pete~Warden and Martin~Wattenberg and Martin~Wicke and Yuan~Yu and Xiaoqiang~Zheng},
	year         = 2015,
	url          = {https://www.tensorflow.org/},
	note         = {Software available from tensorflow.org}
}

@misc{icon4py,
	title        = {{ICON4Py}: Python implementations of the ICON climate and weather model},
	author       = {{C2SM}},
	year         = 2024,
	note         = {GitHub repository, accessed 2025-10-30},
	howpublished = {\url{https://github.com/C2SM/icon4py}}
}

@misc{gpu4geo,
	title        = {{GPU4GEO}: Frontier GPU Multi-Physics Solvers},
	author       = {{GPU4GEO Research Team}},
	year         = 2025,
	url          = {https://gpu4geo.org},
	howpublished = {\url{https://gpu4geo.org}}
}

@misc{gpu4geo_software,
	title        = {{GPU4GEO}: Software},
	author       = {{GPU4GEO Research Team}},
	year         = 2025,
	url          = {https://gpu4geo.org/software/},
	howpublished = {\url{https://gpu4geo.org/software/}}
}

@misc{ferrari_sparse_solvers,
	title        = {A Comparison of Sparse Solvers for Severely Ill-Conditioned Linear Systems in Geophysical Marker-In-Cell Simulations},
	author       = {Marcel Ferrari},
	year         = 2024,
	url          = {https://arxiv.org/abs/2409.11515},
	eprint       = {2409.11515},
	archiveprefix = {arXiv},
	primaryclass = {math.NA}
}

@misc{ho2016acceleratinguzawaalgorithm,
	title        = {Accelerating the Uzawa Algorithm},
	author       = {Nguyenho Ho and Sarah D. Olson and Homer F. Walker},
	year         = 2016,
	url          = {https://arxiv.org/abs/1510.04246},
	eprint       = {1510.04246},
	archiveprefix = {arXiv},
	primaryclass = {math.NA}
}

@article{heister:etal:2017,
	title        = {High Accuracy Mantle Convection Simulation through Modern Numerical Methods. {II}: {R}ealistic Models and Problems},
	author       = {Heister, Timo and Dannberg, Juliane and Gassm{\"o}ller, Rene and Bangerth, Wolfgang},
	year         = 2017,
	journal      = {Geophysical Journal International},
	volume       = 210,
	number       = 2,
	pages        = {833--851},
	doi          = {10.1093/gji/ggx195},
	url          = {https://doi.org/10.1093/gji/ggx195}
}

@article{kronbichler:etal:2012,
	title        = {High Accuracy Mantle Convection Simulation through Modern Numerical Methods},
	author       = {M. Kronbichler and T. Heister and W. Bangerth},
	year         = 2012,
	journal      = {Geophysical Journal International},
	volume       = 191,
	pages        = {12--29},
	doi          = {10.1111/j.1365-246X.2012.05609.x},
	url          = {http://dx.doi.org/10.1111/j.1365-246X.2012.05609.x}
}

@software{aspect-doi-v3.0.0,
	title        = {ASPECT v3.0.0},
	author       = {Wolfgang Bangerth and Juliane Dannberg and Menno Fraters and Rene Gassmoeller and Anne Glerum and Timo Heister and Robert Myhill and John Naliboff},
	year         = 2024,
	month        = dec,
	publisher    = {Zenodo},
	doi          = {10.5281/zenodo.14371679},
	url          = {https://doi.org/10.5281/zenodo.14371679},
	version      = {v3.0.0}
}

@misc{aspectmanual,
	title        = {{{ASPECT}: Advanced Solver for Planetary Evolution, Convection, and Tectonics, User Manual}},
	author       = {Bangerth, Wolfgang and Dannberg, Juliane and Fraters, Menno and Gassmoeller, Rene and Glerum, Anne and Heister, Timo and Myhill, Robert and Naliboff, John},
	year         = 2024,
	month        = dec,
	doi          = {10.6084/m9.figshare.4865333},
	url          = {https://doi.org/10.6084/m9.figshare.4865333}
}

@article{citcoms_gpu,
	title        = {Accelerating Geophysics Simulation using CUDA},
	author       = {Holt, Brandon and Ernst, Daniel},
	year         = 2011,
	month        = 12,
	journal      = {The Journal of Computational Science Education},
	volume       = 2,
	pages        = {21--27},
	doi          = {10.22369/issn.2153-4136/2/1/4}
}

@software{citcoms_software,
	title        = {CitcomS v3.3.1},
	author       = {Moresi, Louis and Zhong, Shijie and Han, Lijie and Conrad, Clint and Tan, Eh and Gurnis, Michael and Choi, Eunseo and Thoutireddy, Pururav and Manea, Vlad and McNamara, Allen and Becker, Thorsten and Leng, Wei and Armendariz, Luis},
	year         = 2014,
	month        = nov,
	publisher    = {Zenodo},
	doi          = {10.5281/zenodo.7271920},
	url          = {https://doi.org/10.5281/zenodo.7271920},
	version      = {v3.3.1}
}

@article{citcoms_article_1,
	title        = {GeoFramework: Coupling multiple models of mantle convection within a computational framework},
	author       = {Tan, E. and Choi, E. and Thoutireddy, P. and Gurnis, M. and Aivazis, M.},
	year         = 2006,
	journal      = {Geochemistry, Geophysics, Geosystems},
	volume       = 7,
	number       = 6,
	pages        = {Q06001},
	doi          = {10.1029/2005GC00115},
	issn         = {1525-2027},
	optkeywords  = {CitcomS},
	opturl       = {http://doi.wiley.com/10.1029/2005GC001155}
}

@article{citcoms_article_2,
	title        = {Role of temperature-dependent viscosity and surface plates in spherical shell models of mantle convection},
	author       = {Zhong, S. and Zuber, M.T. and Moresi, L. and Gurnis, M.},
	year         = 2000,
	journal      = {Journal of Geophysical Research: Solid Earth},
	volume       = 105,
	number       = {B5},
	pages        = {11063--11082},
	doi          = {10.1029/2000JB900003},
	issn         = {1525-2027},
	optkeywords  = {CitcomS},
	opturl       = {http://doi.wiley.com/10.1029/2000JB900003}
}

@article{citcoms_article_3,
	title        = {Thermochemical structures within a spherical mantle: Superplumes or piles?},
	author       = {McNamara, A.K. and Zhong, S.},
	year         = 2004,
	journal      = {Journal of Geophysical Research: Solid Earth},
	volume       = 109,
	number       = {B7},
	pages        = {B07402},
	doi          = {10.1029/2003JB002847},
	issn         = {0148-0227},
	optkeywords  = {CitcomS},
	opturl       = {http://doi.wiley.com/10.1029/2003JB002847}
}

@incollection{Gerya_stokes_flow,
	title        = {The momentum equation},
	author       = {Gerya, Taras},
	year         = 2019,
	booktitle    = {Introduction to Numerical Geodynamic Modelling},
	publisher    = {Cambridge University Press},
	pages        = {60–72},
	place        = {Cambridge}
}

@inproceedings{ptatin3d,
	title        = {pTatin3D: high-performance methods for long-term lithospheric dynamics},
	author       = {May, Dave A. and Brown, Jed and Le Pourhiet, Laetitia},
	year         = 2014,
	booktitle    = {Proceedings of the International Conference for High Performance Computing, Networking, Storage and Analysis},
	location     = {New Orleans, Louisana},
	publisher    = {IEEE Press},
	series       = {SC '14},
	pages        = {274–284},
	doi          = {10.1109/SC.2014.28},
	isbn         = 9781479955008,
	url          = {https://doi.org/10.1109/SC.2014.28},
	numpages     = 11
}

@article{gerya_i2elvis_old,
	title        = {Characteristics-based marker-in-cell method with conservative finite-differences schemes for modeling geological flows with strongly variable transport properties},
	author       = {Taras V. Gerya and David A. Yuen},
	year         = 2003,
	journal      = {Physics of the Earth and Planetary Interiors},
	volume       = 140,
	number       = 4,
	pages        = {293--318},
	doi          = {https://doi.org/10.1016/j.pepi.2003.09.006},
	issn         = {0031-9201},
	url          = {https://www.sciencedirect.com/science/article/pii/S0031920103001900}
}

@article{gerya_i2elvis,
	title        = {Robust characteristics method for modelling multiphase visco-elasto-plastic thermo-mechanical problems},
	author       = {Taras V. Gerya and David A. Yuen},
	year         = 2007,
	journal      = {Physics of the Earth and Planetary Interiors},
	volume       = 163,
	number       = 1,
	pages        = {83--105},
	doi          = {https://doi.org/10.1016/j.pepi.2007.04.015},
	issn         = {0031-9201},
	url          = {https://www.sciencedirect.com/science/article/pii/S0031920107000969},
	note         = {Computational Challenges in the Earth Sciences}
}

@article{tackley_stagyy,
	title        = {Modelling compressible mantle convection with large viscosity contrasts in a three-dimensional spherical shell using the yin-yang grid},
	author       = {Paul J. Tackley},
	year         = 2008,
	journal      = {Physics of the Earth and Planetary Interiors},
	volume       = 171,
	number       = 1,
	pages        = {7--18},
	doi          = {https://doi.org/10.1016/j.pepi.2008.08.005},
	issn         = {0031-9201},
	url          = {https://www.sciencedirect.com/science/article/pii/S0031920108002276},
	note         = {Recent Advances in Computational Geodynamics: Theory, Numerics and Applications}
}

@incollection{Gerya_viscoelastoplastic,
	title        = {2D implementation of visco-elasto-plasticity},
	author       = {Gerya, Taras},
	year         = 2019,
	booktitle    = {Introduction to Numerical Geodynamic Modelling},
	publisher    = {Cambridge University Press},
	pages        = {188–208},
	place        = {Cambridge}
}

@incollection{gerya_multigrid,
	title        = {The multigrid method},
	author       = {Gerya, Taras},
	year         = 2019,
	booktitle    = {Introduction to Numerical Geodynamic Modelling},
	publisher    = {Cambridge University Press},
	pages        = {292–318},
	place        = {Cambridge}
}

@book{yousef_saad_iterative_methods_linear_systems,
	title        = {Iterative Methods for Sparse Linear Systems},
	author       = {Saad, Yousef},
	year         = 2003,
	publisher    = {Society for Industrial and Applied Mathematics},
	address      = {},
	doi          = {10.1137/1.9780898718003},
	url          = {https://epubs.siam.org/doi/abs/10.1137/1.9780898718003},
	edition      = {Second},
	eprint       = {https://epubs.siam.org/doi/pdf/10.1137/1.9780898718003}
}

@book{tim_davis_direct_methods_linear_systems,
	title        = {Direct Methods for Sparse Linear Systems},
	author       = {Davis, Timothy A.},
	year         = 2006,
	publisher    = {Society for Industrial and Applied Mathematics},
	address      = {},
	doi          = {10.1137/1.9780898718881},
	url          = {https://epubs.siam.org/doi/abs/10.1137/1.9780898718881},
	edition      = {},
	eprint       = {https://epubs.siam.org/doi/pdf/10.1137/1.9780898718881}
}

@misc{petsc-web-page,
	title        = {{PETS}c {W}eb page},
	author       = {Satish Balay and Shrirang Abhyankar and Mark~F. Adams and Steven Benson and Jed Brown and Peter Brune and Kris Buschelman and Emil~M. Constantinescu and Lisandro Dalcin and Alp Dener and Victor Eijkhout and Jacob Faibussowitsch and William~D. Gropp and V\'{a}clav Hapla and Tobin Isaac and Pierre Jolivet and Dmitry Karpeev and Dinesh Kaushik and Matthew~G. Knepley and Fande Kong and Scott Kruger and Dave~A. May and Lois Curfman McInnes and Richard Tran Mills and Lawrence Mitchell and Todd Munson and Jose~E. Roman and Karl Rupp and Patrick Sanan and Jason Sarich and Barry~F. Smith and Stefano Zampini and Hong Zhang and Hong Zhang and Junchao Zhang},
	year         = 2024,
	url          = {https://petsc.org/},
	howpublished = {\url{https://petsc.org/}}
}

@techreport{petsc-user-ref,
	title        = {{PETSc/TAO} Users Manual},
	author       = {Satish Balay and Shrirang Abhyankar and Mark~F. Adams and Steven Benson and Jed Brown and Peter Brune and Kris Buschelman and Emil Constantinescu and Lisandro Dalcin and Alp Dener and Victor Eijkhout and Jacob Faibussowitsch and William~D. Gropp and V\'{a}clav Hapla and Tobin Isaac and Pierre Jolivet and Dmitry Karpeev and Dinesh Kaushik and Matthew~G. Knepley and Fande Kong and Scott Kruger and Dave~A. May and Lois Curfman McInnes and Richard Tran Mills and Lawrence Mitchell and Todd Munson and Jose~E. Roman and Karl Rupp and Patrick Sanan and Jason Sarich and Barry~F. Smith and Hansol Suh and Stefano Zampini and Hong Zhang and Hong Zhang and Junchao Zhang},
	year         = 2025,
	number       = {ANL-21/39 - Revision 3.24},
	doi          = {10.2172/2998643},
	institution  = {Argonne National Laboratory}
}

@inproceedings{petsc-efficient,
	title        = {Efficient Management of Parallelism in Object Oriented Numerical Software Libraries},
	author       = {Satish Balay and William~D. Gropp and Lois Curfman McInnes and Barry~F. Smith},
	year         = 1997,
	booktitle    = {Modern Software Tools in Scientific Computing},
	publisher    = {Birkh{\"{a}}user Press},
	pages        = {163--202},
	editor       = {E. Arge and A.~M. Bruaset and H.~P. Langtangen}
}

@article{petsc-gpu,
	title        = {Toward performance-portable {PETS}c for {GPU}-based exascale systems},
	author       = {Richard Tran Mills and Mark F. Adams and Satish Balay and Jed Brown and Alp Dener},
	year         = 2021,
	journal      = {Parallel Computing},
	volume       = 108,
	pages        = 102831,
	doi          = {https://doi.org/10.1016/j.parco.2021.102831},
	issn         = {0167-8191},
	url          = {https://www.sciencedirect.com/science/article/pii/S016781912100079X}
}

@inproceedings{numba,
	title        = {Numba: a LLVM-based Python JIT compiler},
	author       = {Lam, Siu Kwan and Pitrou, Antoine and Seibert, Stanley},
	year         = 2015,
	booktitle    = {Proceedings of the Second Workshop on the LLVM Compiler Infrastructure in HPC},
	location     = {Austin, Texas},
	publisher    = {Association for Computing Machinery},
	address      = {New York, NY, USA},
	series       = {LLVM '15},
	doi          = {10.1145/2833157.2833162},
	isbn         = 9781450340052,
	url          = {https://doi.org/10.1145/2833157.2833162},
	articleno    = 7,
	numpages     = 6
}

@article{numpy,
	title        = {Array programming with {NumPy}},
	author       = {Charles R. Harris and K. Jarrod Millman and St{\'{e}}fan J. van der Walt and Ralf Gommers and Pauli Virtanen and David Cournapeau and Eric Wieser and Julian Taylor and Sebastian Berg and Nathaniel J. Smith and Robert Kern and Matti Picus and Stephan Hoyer and Marten H. van Kerkwijk and Matthew Brett and Allan Haldane and Jaime Fern{\'{a}}ndez del R{\'{i}}o and Mark Wiebe and Pearu Peterson and Pierre G{\'{e}}rard-Marchant and Kevin Sheppard and Tyler Reddy and Warren Weckesser and Hameer Abbasi and Christoph Gohlke and Travis E. Oliphant},
	year         = 2020,
	month        = sep,
	journal      = {Nature},
	publisher    = {Springer Science and Business Media {LLC}},
	volume       = 585,
	number       = 7825,
	pages        = {357--362},
	doi          = {10.1038/s41586-020-2649-2},
	url          = {https://doi.org/10.1038/s41586-020-2649-2}
}

@inproceedings{cupy,
	title        = {{CuPy}: A NumPy-Compatible Library for NVIDIA GPU Calculations},
	author       = {Okuta, Ryosuke and Unno, Yuya and Nishino, Daisuke and Hido, Shohei and Loomis, Crissman},
	year         = 2017,
	booktitle    = {Proceedings of Workshop on Machine Learning Systems (LearningSys) in The Thirty-first Annual Conference on Neural Information Processing Systems (NIPS)},
	url          = {http://learningsys.org/nips17/assets/papers/paper_16.pdf}
}

@article{amgcl,
	title        = {AMGCL -- A C++ library for efficient solution of large sparse linear systems},
	author       = {Denis Demidov},
	year         = 2020,
	journal      = {Software Impacts},
	volume       = 6,
	pages        = 100037,
	doi          = {10.1016/j.simpa.2020.100037},
	issn         = {2665-9638},
	url          = {https://doi.org/10.1016/j.simpa.2020.100037}
}

@article{saddle_point,
	title        = {Numerical solution of saddle point problems},
	author       = {Benzi, Michele and Golub, Gene H. and Liesen, Jörg},
	year         = 2005,
	journal      = {Acta Numerica},
	volume       = 14,
	pages        = {1–137},
	doi          = {10.1017/S0962492904000212}
}
\endgroup

\end{document}